\documentclass[%
 reprint,
 amsmath,amssymb,
 aps,
]{revtex4-1}

\usepackage[english]{babel}
\usepackage[utf8]{inputenc} 
\usepackage{physics}
\usepackage{ dsfont }
\usepackage{enumerate}
\usepackage{verbatim}
\usepackage[normalem]{ulem}

\usepackage[colorlinks=false, linktocpage=true]{hyperref}

\usepackage{graphicx}
\graphicspath{{figures/}{../../figures/}}
\usepackage{tikz}

\usepackage{dcolumn} 
\usepackage{bm}

\begin{document}

\title{Readout of Majorana Qubits}

\author{Jacob F.\ Steiner and Felix von Oppen}
\affiliation{%
\mbox{Dahlem Center for Complex Quantum Systems and Fachbereich Physik, Freie Universität Berlin, 14195 Berlin, Germany}
}%

\date{\today}

\begin{abstract}
Schemes for topological quantum computation with Majorana bound states rely heavily on the ability to measure products of Majorana operators projectively. Here, we employ Markovian quantum measurement theory, including the readout device, to analyze such measurements. Specifically, we focus on the readout of Majorana qubits via continuous charge sensing of a tunnel-coupled quantum dot by a quantum point contact. We show that projective measurements of Majorana products $\prod_i\hat{\gamma}_i$ can be implemented by continuous charge sensing under quite general circumstances. Essential requirements are that a combined local parity $\hat{\pi}$, involving the quantum dot charge along with the Majorana product of interest, be conserved, and that the two eigenspaces of the combined parity $\hat{\pi}$ generate distinguishable measurement signals. We find that qubit readout may have to rely on measuring noise correlations of the quantum-point-contact current. The average current encodes the qubit readout only transiently for fine-tuned parameters or in the presence of relaxation processes. We also discuss the corresponding measurement and decoherence times and consider processes such as residual Majorana hybridizations which are detrimental to the measurement protocol. Finally, we emphasize that the underlying mechanism -- which we term symmetry-protected readout -- is quite general and has further implications for both, Majorana and non-Majorana systems. 

\end{abstract}


\maketitle

\section{Introduction}

Qubits based on Majorana bound states promise key advantages for quantum computing, including long intrinsic lifetimes deriving from the nonlocal encoding of quantum information \cite{kitaev_fault-tolerant_2003,nayak_non-abelian_2008,Oreg2020} and topologically protected single-qubit gates based on braiding or, alternatively, on exploiting measurements in all Pauli bases. A popular Majorana qubit -- known as Majorana box qubit \cite{plugge_majorana_2017} or tetron and hexon \cite{karzig_scalable_2017} -- is based on semiconductor quantum wires proximity coupled to a superconductor \cite{lutchyn_majorana_2010,oreg_helical_2010,alicea_non-abelian_2011}. These qubits are believed to be within experimental reach \cite{lutchyn_majorana_2018} and quantum computing architectures have been developed on their basis \cite{vijay_majorana_2015,plugge_roadmap_2016,karzig_scalable_2017,litinski_quantum_2018}. Quantum computation with Majorana qubits is expected to rely heavily on projective qubit measurements, with all Clifford gates implemented using single and two-qubit measurements \cite{Zilberberg2008,karzig_scalable_2017,litinski_quantum_2018}. Indeed, these schemes can be referred to as measurement-based topological quantum computing and good readout fidelities are absolutely central to their performance. This makes it essential to develop a detailed theoretical understanding of the proposed readout schemes for Majorana-based topological qubits.

\begin{figure}[b]
    \centering
    \includegraphics[width=0.9\columnwidth]{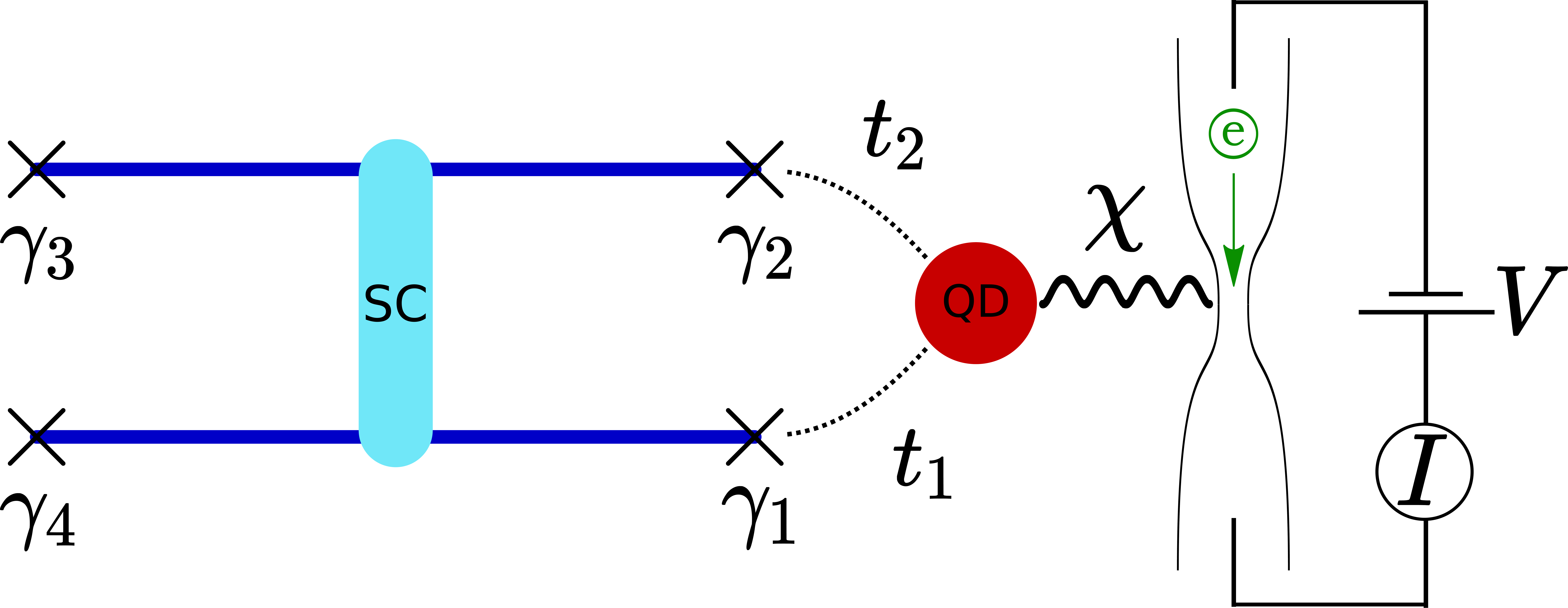}
    \caption{Setup for readout of Majorana qubit. A quantum dot (QD) is tunnel coupled to a Majorana qubit consisting of two topological superconducting wires (dark blue) with four Majorana bound states $\hat{\gamma}_1,\ldots,\hat{\gamma}_4$. The wires are connected by a conventional superconducting bridge (SC) allowing charge to move freely between the wires, so that only the overall charge of the Majorana qubit is fixed by the charging energy. The Majorana parity $\hat Z = - i\hat{\gamma}_1\hat{\gamma}_2$ defines the Pauli-$Z$ operator of the Majorana qubit, and can be read out by tunnel coupling the two Majoranas $\hat{\gamma}_1$ and $\hat{\gamma}_2$ to the quantum dot. The quantum dot charge is measured by capacitively coupling (with strength $k\propto |\chi|^2$) the dot to a quantum point contact.}
    \label{fig:tetron_quantum_dot_measurement_device}
\end{figure}

While a blessing for their characteristics as a quantum memory, the nonlocal nature of topological qubits complicates the readout of the encoded quantum information. Readout requires one to make the nonlocally encoded quantum information available locally. This can be achieved by exploiting interference effects which are sensitive to the Majorana parity operator of interest \cite{fu_electron_2010,plugge_majorana_2017,karzig_scalable_2017}. A schematic Majorana qubit is shown in Fig.\ \ref{fig:tetron_quantum_dot_measurement_device} and involves four Majorana bound states located at the ends of two proximity-coupled semiconductor quantum wires. The Pauli operators associated with the Majorana qubit are parity operators involving products of two Majorana operators. Here, we focus on a readout procedure which measures these Majorana parity operators by tunnel coupling a quantum dot to the relevant pair of Majoranas as shown in Fig.\ \ref{fig:tetron_quantum_dot_measurement_device}. 
Virtual tunneling processes between Majorana qubit and quantum dot shift the energy levels of the quantum dot in a manner that depends on the Majorana parity. As a result, the coupled time evolution of Majorana qubit and quantum dot entangles the two, and the charge state of the quantum dot becomes correlated with the Majorana parity. Measurements of the quantum dot charge, for instance by a nearby quantum point contact, can thus be used to read out the Majorana qubit. 

In principle, it is possible to design parity-to-charge conversion procedures which allow for a projective measurement of the Majorana parity based on a single-shot projective measurement of the charge $\hat{n}$ of the quantum dot \cite{plugge_majorana_2017}. However, these schemes are not robust and require fine tuning and rapid manipulation of system parameters. More generically, the charge state of the quantum dot becomes only weakly correlated with the state of the Majorana qubit, and qubit readout requires multiple measurements. This can be achieved by repeatedly coupling and decoupling qubit and quantum dot, with intervening projective measurements of $\hat{n}$ and resets of the qubit charge state. 

In practice this coupling, decoupling, and resetting is challenging and prone to errors. It would be preferable and more natural to keep Majorana qubit and quantum dot coupled during the entire readout procedure and to \textit{monitor} the charge of the quantum dot continuously. Here, we show that a projective readout of the Majorana qubit can indeed be robustly implemented in this manner. In particular, our strategy significantly relaxes the requirements on dynamical control over system parameters, and obviates the need for resets of the quantum dot charge state.

To describe the dynamics of the quantum measurement, we include the measurement device in the theoretical description. The continuous measurement decoheres the system in the basis of the quantum dot charge and outputs the noisy measurement signal $j(t)$ of the quantum point contact \cite{wiseman_quantum_2009}. The task is then twofold. First, one needs to show that the system of Majorana qubit and quantum dot decoheres in the actual basis of interest associated with the parity operator of the Majorana qubit. Second, one needs to ascertain that the measurement outcome can be extracted from the signal $j(t)$. Both criteria must be satisfied to effectively implement a projective (Born rule) readout of the Majorana qubit which can be employed for measurement-based topological quantum computation. 

In section \ref{sec:basics_majorana_qubits}, we introduce the system under consideration, a Majorana qubit with two of the four Majorana bound states tunnel coupled to a quantum dot as illustrated in Fig.\ \ref{fig:tetron_quantum_dot_measurement_device}, and discuss how under idealized assumptions, quantum dot charge measurements can be used for single-shot readout of the Majorana qubit. We then turn to more realistic readout protocols which rely on continuous monitoring of the quantum dot charge, collecting our central results in Sec.\ \ref{sec:qpc_readout}. The basic master-equation formalism describing weak measurements of the quantum dot charge by a quantum point contact is described in Sec.\ \ref{sec:contcharmeas}. As a backdrop, we first illustrate the formalism in Sec.\ \ref{sec:simple_charge_readout} by reviewing charge monitoring of a quantum dot in the absence of coupling to a Majorana qubit. We then include the coupling to the Majorana qubit in Sec.\ \ref{sec:majorana-qubit_readout} and show how a two-Majorana parity (Pauli operator) of the Majorana qubit can be read out. While it suffices to monitor the average quantum-point-contact current for charge readout of an uncoupled quantum dot, we find that  in general, readout of the Majorana qubit requires one to measure noise correlations of the current. Readout based on noise correlations can be avoided by tuning to a sweet spot in parameter space or, as shown in Sec.\ \ref{sec:with_relaxation}, by including additional processes which cause relaxation of the coupled Majorana qubit-quantum dot system to its ground state. In both of these cases, it suffices in principle to monitor the average quantum-point-contact current. In Sec.\ \ref{sec:imperfect}, we discuss various processes which are detrimental to the readout protocol. Most importantly, the previous sections assume that the residual Majorana hybridizations of the qubit are negligible, and we show here how these hybridizations affect the measurement protocol. Finally, Sec.\ \ref{sec:qpc_readout} closes with a discussion of alternative readout schemes which rely on coupling the Majorana qubit to double quantum dots, see Sec.\ \ref{sec:DQDreadout}. We find that this readout scheme adds flexibility in designing the coupling between Majorana qubit and quantum dots. We also discuss readout of Majorana parity operators involving more than two Majoranas, which represent two-qubit parities or stabilizer operators of topological quantum error correcting codes. While our results are mostly analytical in nature, we illustrate the various measurement protocols by simulations of the stochastic master equation. In addition to our analytical estimates throughout Sec.\ \ref{sec:qpc_readout}, these simulations also illustrate the required measurement times. Sec.\ \ref{sec:general_readout_analysis} discusses the measurement protocols from a more general point of view, not restricted to the readout of Majorana qubits. We finally summarize and conclude in Sec.\ \ref{sec:discussion}. Throughout the paper, we focus on the principal arguments and results. Explicit calculations and background material are relegated to a series of appendices. 

\section{Majorana Qubits and Quantum Dot Readout}
\label{sec:basics_majorana_qubits}\label{sec:majorana_qubits_and_quantum_dots}

\subsection{Majorana qubit coupled to quantum dot}

Majorana qubits are Coulomb-blockaded islands hosting $2m$ Majorana bound states $\hat\gamma_j$ as described by the Hamiltonian
\begin{equation}\label{eq:charging_ham}
\hat{H}_{\textrm{M}} = 
E_C\pqty{\hat{N}-N_g}^2 +\ i\sum_{i < j}^{2m} \varepsilon_{ij}\hat{\gamma}_i\hat{\gamma}_j.
\end{equation}
The first term reflects the charging energy $E_C$ of the device, which depends on the total charge $\hat{N}$ as well as a gate-controlled offset $N_g$. For a fixed charge and well-separated Majorana bound states, the ground state of the system is $2^{m-1}$-fold degenerate. Residual splittings are included in $H_M$ through the $\varepsilon_{ij}$. Above-gap excitations of the Majorana wires are ignored by virtue of a sufficiently large gap.

The minimal number of Majoranas required for a single qubit is four, in which case the Majorana island realizes a Majorana box qubit or tetron. Figure \ref{fig:tetron_quantum_dot_measurement_device} shows such a Majorana qubit assembled from a pair of topological superconducting quantum wires hosting two Majorana bound states each. The superconducting bridge between the quantum wires provides a sufficiently large mutual capacitance so that the charging energy depends on the charges of the individual wires only via the total charge of the device \cite{karzig_scalable_2017}. For definiteness, we choose $N_g = 0$, so that the ground-state manifold has even fermion parity, $\hat P= (i\hat\gamma_1\hat\gamma_4) (i\hat\gamma_2\hat\gamma_3)=1$. We can then define the Pauli operators 
\begin{subequations}
\begin{eqnarray}
\hat{Z} &=& -i\hat{\gamma}_1\hat{\gamma}_2 \\ 
\hat{X} &=& -i\hat{\gamma}_2\hat{\gamma}_3 \\ 
\hat{Y} &=& -i\hat{\gamma}_3 \hat{\gamma}_1
\end{eqnarray}
\end{subequations}
of the qubit. Fermion parity conservation implies that one can alternatively use the operators $\hat{Z}' = \hat{P} \hat{Z}$ (with $\hat{X}'$ and $\hat{Y}'$ defined analogously). 

Readout of the qubit operators (say, $\hat Z$ for definiteness) can be effected by connecting the Majorana island to a quantum dot via tunnel junctions as depicted in Fig.\ \ref{fig:tetron_quantum_dot_measurement_device} \cite{plugge_majorana_2017,karzig_scalable_2017}. We assume that the quantum dot has a single nondegenerate level $\epsilon$, which is spin resolved due to the magnetic field required for realizing topological superconductivity. Then, the quantum dot is described by the Hamiltonian 
\begin{equation}\label{eq:ham_qd}
\hat{H}_{\textrm{QD}} = \epsilon \hat{n},
\end{equation}
where the quantum dot occupation $\hat{n} =\hat{d}^{\dagger}\hat{d}$ involves the annihilation operator $\hat d$ of the gate-tunable dot level. We assume that the quantum dot is unoccupied in the ground state, $n=0$. 

Tunneling between quantum dot and Majorana qubit is described by
\begin{equation}\label{eq:tetron_qd}
\hat{H}_{\textrm{T}} = \pqty{t_1 \hat{\gamma}_1+t_2\hat{\gamma}_2}e^{i\hat{\phi}/2} \hat{d} + \textrm{h.c.}
\end{equation} 
Electrons tunneling into or out of the Majorana island affect the state of the system within the ground-state manifold and change its charge. We describe the first effect through the Majorana operators $\hat{\gamma}_j$ (which leave the charge state of the system unchanged, $[\hat{\gamma}_i,\hat{N}] = 0$), and the second through a charge shift operator $e^{i\hat{\phi}/2}$ (with $[\hat{\phi},\hat{N}] = 2i$ and $[\hat{\gamma}_i,\hat{\phi}] = 0$). In this formulation, physical states must satisfy the total parity constraint $(-i)^m\prod_{i=1}^{2m}\hat{\gamma}_{i} = (-1)^{\hat{N}}$.

For a topological qubit, we assume that the hybridizations $\varepsilon_{ij}$ are negligible, so that the Majorana bound states are true zero-energy modes of the Majorana island. This implies that in addition to the fermion parity $\hat P$ of the qubit, also all two-Majorana parities $i\hat\gamma_i\hat\gamma_j$ are good quantum numbers. In particular, this is the case for the Pauli-$\hat Z$ operator of the Majorana qubit. The resulting degeneracy is partially lifted when tunnel coupling the Majorana qubit to the quantum dot. It is important to notice, however, that unlike the Pauli-$\hat Z$ operator, the combined fermion parity operator 
\begin{equation}\label{eq:combined_parity}
    \hat{\pi} = \hat{Z}\pqty{-1}^{\hat{n}}
\end{equation}
remains a conserved quantity \cite{akhmerov_topological_2010}. Restricting ourselves to states which have total fermion parity $P(-1)^n=1$ and can thus be reached from the ground state with $n=N=0$ by tunneling, the Hamiltonian $\hat{H} = \hat{H}_{\textrm{M}} +  \hat{H}_{\textrm{QD}} + \hat{H}_{\textrm{T}} $ becomes block diagonal in the subspaces of the combined fermion parity $\hat\pi$,
\begin{equation}\label{eq:hamiltonian_parities}
    \hat{H} = \left(\begin{array}{cc} h_+ & 0 \\ 0 & h_- \end{array}\right).
\end{equation}
Here, we choose the basis $\{\ket{\uparrow,0},\ket{ \downarrow,1},\ket{\downarrow,0},\ket{\uparrow,1}\}$ with the first entry distinguishing the eigenstates of $\hat Z$ and the second entry denoting the occupation of the quantum dot level, see App.\ \ref{app:definitions} for details. We use lower-case letters without hats to denote operators within subspaces of fixed combined parity $\hat \pi$. The corresponding $2\times2$ blocks take the explicit form
\begin{equation}\label{eq:sector_hams}
h_{\pi} = 
\begin{pmatrix}
0  & t_1 - i\pi  t_2 \\ t^*_1 + i\pi  t^*_2 & \varepsilon
\end{pmatrix}
= 
\frac{\varepsilon}{2} + \Omega_{\pi} \bm{h}_{\pi}\cdot \bm{\sigma},
\end{equation}
where $\bm{\sigma}$ denotes a vector of Pauli matrices and $\varepsilon = \epsilon + E_C$ the detuning (which depends on both, the level energy $\epsilon$ of the quantum dot and the charging energy $E_C$ of the Majorana island). We also defined a Bloch vector
\begin{equation}\label{eq:bloch_vec}
\bm{h}_\pi = (\sin\theta_\pi \cos\phi_\pi, \sin\theta_\pi \sin\phi_\pi,\cos\theta_\pi)
\end{equation}
with $\cos\theta_\pi = -\varepsilon/2 \Omega_\pi$ and $\sin\theta_\pi e^{i\phi_\pi} = (t^*_1+ i\pi t^*_2)/\Omega_\pi$, as well as the Rabi frequency $\Omega_{\pi}^2 = \varepsilon^2/4 + \abs{t^*_1 +i\pi t^*_2}^2$. The eigenenergies and Rabi frequencies of the two subspaces are different provided that $\Im\pqty{t_1 t_2^*} \neq  0$.

\subsection{Basic measurement protocols}\label{sec:basic_protocols}

The tunnel couplings entangle the Pauli-$\hat Z$ operator of the Majorana qubit with the charge of the quantum dot. The entanglement emerges from processes in which an electron virtually occupies the Majorana island through one Majorana involved in $\hat Z$ and leaves it through the other. This makes the quantum information stored nonlocally in the Majorana qubit accessible locally in the quantum dot. We first discuss to which degree a single projective measurement of the charge of the quantum dot realizes a measurement of $\hat Z$. 

We begin with a protocol in which the tunnel coupling between dot and Majorana qubit is turned on adiabatically on the scale of the system dynamics. We assume that prior to turning on the tunnel couplings, the Majorana qubit is in an arbitrary qubit state and the quantum dot is initialized in the $n=0$ state,
\begin{equation}
    \ket{\psi}=\pqty{\alpha\ket{\uparrow}+\beta\ket{\downarrow}} \ket{0}.
    \label{initialstate}
\end{equation}
Adiabatically turning on the tunnel couplings $t_1$ and $t_2$, this state evolves into the corresponding eigenstate of the coupled system,
\begin{equation}\label{eq:adiabatic_turn_on}
\ket{\psi'}=\ \alpha\ket{g_+}+\beta e^{i\chi}\ket{ g_- }.
\end{equation}
Here, the relative phase $\chi$ depends on details of the protocol, and we defined the exact ground states
\begin{subequations}\label{eq:ground_states}
\begin{eqnarray}
    \ket{g_+} &=& \sin\frac{\theta_+}{2} \ket{\uparrow,0}  -\cos\frac{\theta_+}{2}e^{i\phi_+} 
    \ket{\downarrow,1}
    \\
    \ket{g_-} &=& \sin\frac{\theta_-}{2} \ket{\downarrow,0}  -\cos\frac{\theta_-}{2}e^{i\phi_-} 
    \ket{\uparrow,1}
\end{eqnarray}
\end{subequations}
of $h_\pm$. Notice that $\theta_\pm=\pi$ in the absence of the tunnel couplings. Then, a subsequent projective measurement of the charge $n$ of the quantum dot either yields $n=0$ and the state
\begin{equation}\label{eq:adiabatic_post_measurement_state_0}
\ket{\psi'_0}= \frac{1}{\sqrt{p'_0}}\pqty{ \alpha \sin\frac{\theta_+}{2} \ket{\uparrow} + \beta e^{i\chi} \sin\frac{\theta_-}{2}  \ket{\downarrow} } \ket{0},
\end{equation}
with probability 
\begin{equation}
p'_0=|\alpha|^2 \sin^2\frac{\theta_+}{2}
   +|\beta|^2 \sin^2\frac{\theta_-}{2},
\end{equation}
or $n=1$ and 
\begin{equation}\label{eq:adiabatic_post_measurement_state_1}
\ket{\psi'_1}= \frac{1}{\sqrt{p'_1}}\pqty{ \alpha \cos\frac{\theta_+}{2}e^{i\phi_+} \ket{\downarrow} + \beta e^{i\chi} \cos\frac{\theta_-}{2}e^{i\phi_-} \ket{\uparrow} } \ket{1},
\end{equation}
with probability $p'_1=1-p'_0$.

Unfortunately, such charge measurements provide only partial information on the qubit state. One can arrange perfect correlation between the measurement outcome $n=0$ and, say, the $\ket{\uparrow}$ state of the qubit by fine tuning $\sin\frac{\theta_-}{2}=0$. However, one readily proves that 
\begin{equation}
 \abs{\sin^2\frac{\theta_+}{2}-\sin^2\frac{\theta_-}{2}}\leq \frac{1}{2}.
\end{equation}
This implies that a measurement outcome of $n=1$ remains compatible with both qubit states even if $n=0$ is perfectly correlated with the $\ket{\uparrow}$ state.

In principle, a projective measurement can be implemented when turning on the quantum dot-Majorana qubit tunneling instantaneously. In this case, the initial state $\ket{\psi}$ is no longer an eigenstate, and its unitary evolution under the Hamiltonian $\hat H$ entangles qubit and quantum dot. Depending on the measurement outcome, a projective charge measurement after a waiting time $T$ yields the states
\begin{subequations}
\begin{align}
\ket{\psi''_0}=&\ \frac{1}{\sqrt{p''_0}}\left[ \alpha A_+(T) \ket{\uparrow}  + \beta A_-(T) \ket{\downarrow} \right] \ket{0}, \label{eq:sudden_post_measurement_state_0}\\
\ket{\psi''_1}=&\ \frac{1}{\sqrt{p''_1}}\left[ \alpha B_+(T) \ket{\downarrow}  + \beta B_-(T) \ket{\uparrow} \right] \ket{1}. \label{eq:sudden_post_measurement_state_1}
\end{align}
\end{subequations}
with probabilities $p''_0=|\alpha A_+|^2 + |\beta A_-|^2$ and $p''_1=1-p''_0$, respectively. Here, we defined the  amplitudes $A_{\pm}(T) = \cos\Omega_{\pm}T + i\cos\theta_\pm \sin\Omega_{\pm}T$ and $B_{\pm}(T) =  -i\sin\theta_\pm e^{i\phi_\pm} \sin\Omega_{\pm}T$, where $|A_\pm|^2+|B_\pm|^2=1$. This scheme implements a projective measurement of $\hat{Z}$, when these coefficients satisfy, say, $\vert A_- \vert^2 =\vert B_+ \vert^2=0$, with the remaining two coefficients being equal to unity. This can be realized for $t_1 = i t_2$, $\abs{t_{1,2}}^2 \gg \varepsilon^2/4$, and $T=\pi/2\Omega_-$. As such, the qubit is left in the $\ket{\uparrow}$ state for both measurement outcomes. Thus, for the outcome $n=1$, this protocol would need to be followed by another waiting period to rotate the state back to $\ket{\downarrow}$, before turning off the tunnel couplings. 

While in principle this allows for single-shot projective measurements of $\hat{Z}$, the protocol relies on fine-tuned waiting periods and a hierarchy of time scales which would be challenging to fulfill in experiment. Most importantly, we need to assume that the charge measurement is fast compared to internal time scales of the coupled Majorana qubit-quantum dot system. This requirement derives from the fact that the measurement operator $\hat n$ does not commute with the Hamiltonian of the system. Alternatively, we can abruptly turn off the tunnel couplings after entangling Majorana qubit and quantum dot state, and measure the quantum dot charge only subsequently. Then, the charge measurement is quantum nondemolition. Under realistic circumstances, a single projective charge measurement would presumably provide only partial information on $\hat{Z}$. This can be remedied by repeating the above protocol sufficiently many times until the measurement outcome is certain (see App.\ \ref{app:repeated_measurements}). 

Implementing this protocol is clearly challenging and requires detailed and fast control. One may also worry that the fast switching excites the qubit in unwanted ways. It would be preferable to implement a projective measurement of $\hat{Z}$ by monitoring the quantum dot charge {\em while} the quantum dot is coupled to the Majorana qubit. This obviates the need for repeated switching, ideally without compromising on the achievable measurement times. We now turn to describe such continuous measurement procedures.  

\section{Majorana qubit readout via quantum dot charge monitoring}
\label{sec:qpc_readout}

\subsection{Continuous charge measurements}
\label{sec:contcharmeas}

To describe the readout dynamics, we need to include a measurement device for the quantum dot charge in the microscopic description. We assume that the quantum dot is capacitively coupled to a voltage-biased quantum point contact in such a way that the transmission amplitude $\mathcal{T}$ of the quantum point contact depends on the charge state of the quantum dot \cite{korotkov_continuous_1999,goan_dynamics_2001},
\begin{equation}
    \hat{\mathcal{T}} = \tau + \chi \hat{n}.
\end{equation}
The current through the quantum point contact will then depend on the quantum dot charge, taking the values $I_0 \propto \abs{\tau}^2$ when the quantum dot is empty and $I_1 \propto \abs{\tau+\chi}^2$ when the quantum dot is occupied. (For simplicity, we assume that $\tau$ and $\chi$ have relative phase $\pi$.) Provided that the voltage applied to the quantum point contact is sufficiently large, $eV\gg \Omega_\pi$, this setup decoheres the system in the quantum dot charge basis (see App.\ \ref{app:derivation_unconditional}). 

When $I_0 \gg \abs{\delta I} = \abs{I_1 - I_0}$, the fluctuating current through the quantum point contact becomes a Gaussian random process (see App.\ \ref{app:derivation_conditional_diffusive}),
\begin{equation}
    I(t) = I_0 + \delta I \ev{\hat{n}(t)} + \sqrt{I_0} \xi(t),
\end{equation}
involving the Langevin current $\xi(t)$ with $\mathbb{E}\bqty{\xi(t)} = 0$ and $\mathbb{E}\bqty{\xi(t)\xi(t+\tau)} = \delta(\tau)$. Here, $\mathbb{E}[.]$ denotes an ensemble average over many realizations of the measurement procedure.
It will prove useful to work with a dimensionless measurement signal 
\begin{equation}\label{eq:meas_signal}
    j(t) = \frac{{I}(t) - I_0}{\delta I} = \ev{\hat{n}(t)} + \frac{1}{\sqrt{4k}}\xi(t)
\end{equation}
relative to the measured background current $I_0$, with $k\propto \abs{\chi}^2$. The first term contains information on the quantum dot charge and the second is the noise added by the quantum point contact. Within the Born-Markov approximation, the state of the system evolves according to the stochastic master equation \cite{korotkov_continuous_1999,goan_dynamics_2001} (see App.\ \ref{app:derivation} for a derivation)
\begin{multline}\label{eq:sme_mbq-qd-qpc}
    \frac{\textrm{d}}{\textrm{d}t}\hat{\rho}_c(t) = \underbrace{-i\bqty{\hat{H},\hat{\rho}_c(t)} + k \mathcal{D}\bqty{\hat{n}}\hat{\rho}_c(t)}_{\equiv  \mathcal{L} \hat{\rho}_c(t)} 
    \\
    + \sqrt{k} \xi(t) \mathcal{H}\bqty{\hat{n}}\hat{\rho}_c(t).
\end{multline}
The term involving the superoperator $\mathcal{D}[\hat{L}]\hat{\rho}=\hat{L}\hat{\rho}\hat{L}^{\dagger}-(\hat{L}^{\dagger}\hat{L}\hat{\rho}+\hat{\rho}\hat{L}^{\dagger}\hat{L})/2$ causes decoherence in the eigenbasis of $\hat{n}$, and the term with the superoperator $\mathcal{H}[\hat{L}]\hat{\rho} = \hat{L}\hat{\rho} + \hat{\rho} \hat{L}^{\dagger}- \langle \hat{L} + \hat{L}^{\dagger}\rangle \hat{\rho}$ describes the information gain by the measurement. The latter is a Langevin term due to the
stochastic nature of the measurement. The stochastic master equation (\ref{eq:sme_mbq-qd-qpc}) describes the state of the system \textit{conditioned} on the measurement signal $j(t)$, as denoted by the  subscript $c$. The ensemble-averaged -- and thus \textit{unconditioned} -- evolution of the system simply follows by dropping the stochastic term, $ \textrm{d}\hat{\rho}(t)/ \textrm{dt} = \mathcal{L} \hat{\rho}(t)$.

\subsection{Quantum dot charge measurement}\label{sec:simple_charge_readout}

To recall how this formalism describes standard projective measurements, consider first a simple quantum dot charge readout with $\hat{H}_T = 0$. The deterministic terms in Eq.\ \eqref{eq:sme_mbq-qd-qpc} lead to a decay of the off-diagonal components of the density matrix $\hat{\rho}$ in the charge basis, but preserve the diagonal components. The action of the stochastic terms can be understood from the equation of motion for $n(t) \equiv \ev{\hat{n}(t)} = \textrm{tr}[\hat{n} \hat{\rho}_c(t)]$,
\begin{equation}\label{eq:charge_stochastic_equation}
    \frac{\textrm{d}}{\textrm{d}t} n(t) = \sqrt{4k}[n(t) - n^2(t)]\xi(t).
\end{equation}
Its two fixed points $n=0$ and $n=1$ correspond to the two possible measurement outcomes. Conservation of the diagonal components of the ensemble-averaged density matrix ensures that these outcomes occur with the correct probabilities. At the fixed point $n$, the quantum-point-contact current $j(t)$ becomes stationary, $j_{n}(t) = n + \xi(t)/\sqrt{4k}$, and directly reveals the measurement outcome $n$ after an integration time $\tau_m$. The latter is determined by the requirement that the integrated signal dominate over the integrated noise, which happens for $\tau_m \gg \pqty{4k}^{-1}$. We note that we use the steady-state measurement signal when estimating measurement times. While transients may also provide information in principle, we assume that in practice, typical measurement times will exceed the time scale on which the system decoheres. 

\begin{figure}
    \includegraphics[width=0.45\textwidth]{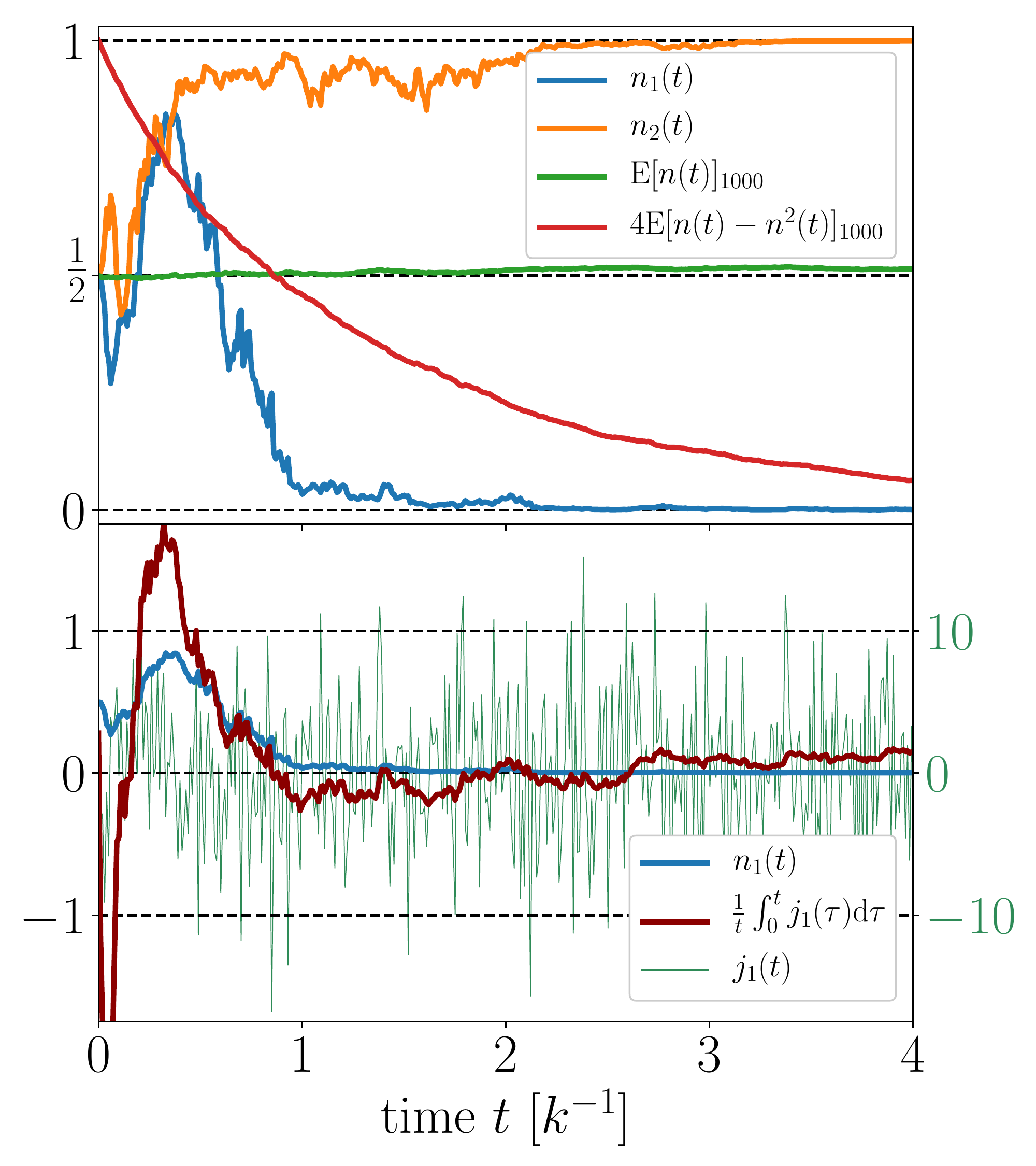}
    \caption{ 
    Continuous measurement of the quantum dot charge for $H_T = 0$, with initial state $(\ket{\uparrow,0} + \ket{\downarrow,1})/\sqrt{2}$ corresponding to $n(0) = 1/2$. Top panel: Two sample trajectories $n_1(t)$ (blue) and $n_2(t)$ (orange) corresponding to measurement results $n=0$ and $n=1$, respectively. The ensemble averaged evolution of the quantum dot charge (green, obtained from $1000$ trajectories) stays near 1/2. The ensemble average of $4\pqty{n(t)-n^2(t)}$ (red) equals unity for uncertain charge and zero when a fixed point has been reached, and therefore quantifies the advance of the measurement process. Bottom panel: Instantaneous (green, right y-axis labels) and time-averaged (dark red, left y-axis labels) measurement current for sample trajectory $n_1(t)$ (blue, same as in top panel).}
     \label{fig:three graphs}
\end{figure}

This time-resolved description of a projective measurement of $\hat{n}$ is illustrated in Fig.\ \ref{fig:three graphs} based on a numerical solution of Eq.\ (\ref{eq:sme_mbq-qd-qpc}). We show sample trajectories of the expectation value of the quantum dot charge (experimentally inaccessible) as well as the corresponding measurement currents through the quantum point contact. While the instantaneous measurement current fluctuates strongly, its time average reveals the quantum dot charge. 

\subsection{Majorana-qubit readout} \label{sec:majorana-qubit_readout}

Including the tunnel coupling $\hat{H}_T$ between quantum dot and Majorana qubit, the dot occupation is no longer a good quantum number. While the measurement tries to project the state of the system into charge eigenstates, tunneling continuously rotates it out of this basis. This can be seen explicitly from the equation of motion for the charge expectation value,  
\begin{equation}\label{eq:charge_stochastic_equation_2}
    \frac{\textrm{d}n}{\textrm{d}t}  = i \ev{\bqty{\hat{H},\hat{n}}} + \sqrt{4k}[n - n^2]\xi.
\end{equation}
As a result of the nonzero commutator on the right hand side, the evolution of $n(t)$ no longer tends towards fixed points. Similarly, $\langle\hat{Z}(t)\rangle$ also does not evolve towards fixed points as $\hat{Z}$ does not commute with $\hat{H}$, as well. At first sight, this seems to imply that quantum dot charge measurements will not suffice to read out the state of the qubit. 
Remarkably, we find that one may still read out  $\hat{Z}$ from a measurement of the quantum dot charge, but the procedure is more subtle.

The key observation is that the evolution governed by Eq.\ \eqref{eq:sme_mbq-qd-qpc} implements a quantum nondemolition measurement of the combined local parity $\hat{\pi}$, which is in one-to-one correspondence with $\hat Z$ for the initial state in Eq.\ (\ref{initialstate}). Under quantum dot charge measurements, the evolution of $\pi(t) = \ev{\hat{\pi}(t)}$ is a bistable process with fixed points $\pi = \pm 1$, which are reached with the correct probabilities $|\alpha|^2$ and $|\beta|^2$, respectively. This does not yet guarantee a projective measurement of $\hat Z$. First, as a consequence of the tunneling Hamiltonian, the measurement does not properly project the state of the system, but leaves it in an equal mixture of the two eigenstates with combined local parity $\pi$. However, once the measurement outcome $\pi$ is determined, the readout device can be decoupled and the state of the quantum dot-Majorana qubit system appropriately reset (see App.\ \ref{app:qd_reset}). Second, one needs to specify how to read out $\pi$ and thus $ Z$ from the measurement current. Unlike for pure quantum dot charge measurements, the average measurement current in general no longer distinguishes between the two measurement outcomes. Instead, a measurement readout generally requires one to analyze the frequency-dependent noise of the measurement current. 

We now discuss these claims in more detail. We first show that under quantum dot charge measurements, the unconditioned evolution of the density matrix $\hat \rho(t)$ generically tends towards
\begin{equation}
\hat{\rho}^{\infty} = \frac{1}{2} \textrm{diag}(\abs{\alpha}^2,\abs{\alpha}^2,\abs{\beta}^2,\abs{\beta}^2)
\end{equation}
for the initial state in Eq.\ (\ref{initialstate}). This follows because the evolution preserves the weight of the two $\pi$ subspaces and the set of steady states of $\mathcal{L}$ is spanned by $\hat{\rho}^\infty_{+} = \textrm{diag}(1,1,0,0)/2$ and $\hat{\rho}^\infty_{-} = \textrm{diag}(0,0,1,1)/2$. To see this, we decompose $\hat\rho$  into $2\times 2$ blocks ${\rho}_{\pi,\pi'}$ according to the combined parity eigenvalues. Since $\hat\pi$ is a good quantum number and commutes with the quantum dot charge operator, the evolution equation for $\hat\rho$ decouples into independent equations
\begin{subequations}
\begin{align}
\dot{\rho}_{\pi,\pi} =&\ -i\bqty{h_\pi,\rho_{\pi,\pi}} + k\mathcal{D}
\bqty{ n }\rho_{\pi,\pi} \nonumber  \\
=&\ \mathcal{L}_{\pi,\pi}\rho_{\pi,\pi}, \label{eq:++_block_meq}\\
\dot{\rho}_{+-} =&\ -i\pqty{h_{+}\rho_{+-} -\rho_{+-} h_{-}} + k\mathcal{D}\bqty{n} \rho_{+-} \nonumber \\
=&\ \mathcal{L}_{+-}\rho_{+-}. \label{eq:+-_block_meq}
\end{align}
\end{subequations}
for the diagonal and off-diagonal blocks of $\hat\rho$. 

\begin{figure}[t]
    \includegraphics[width=0.9\columnwidth]{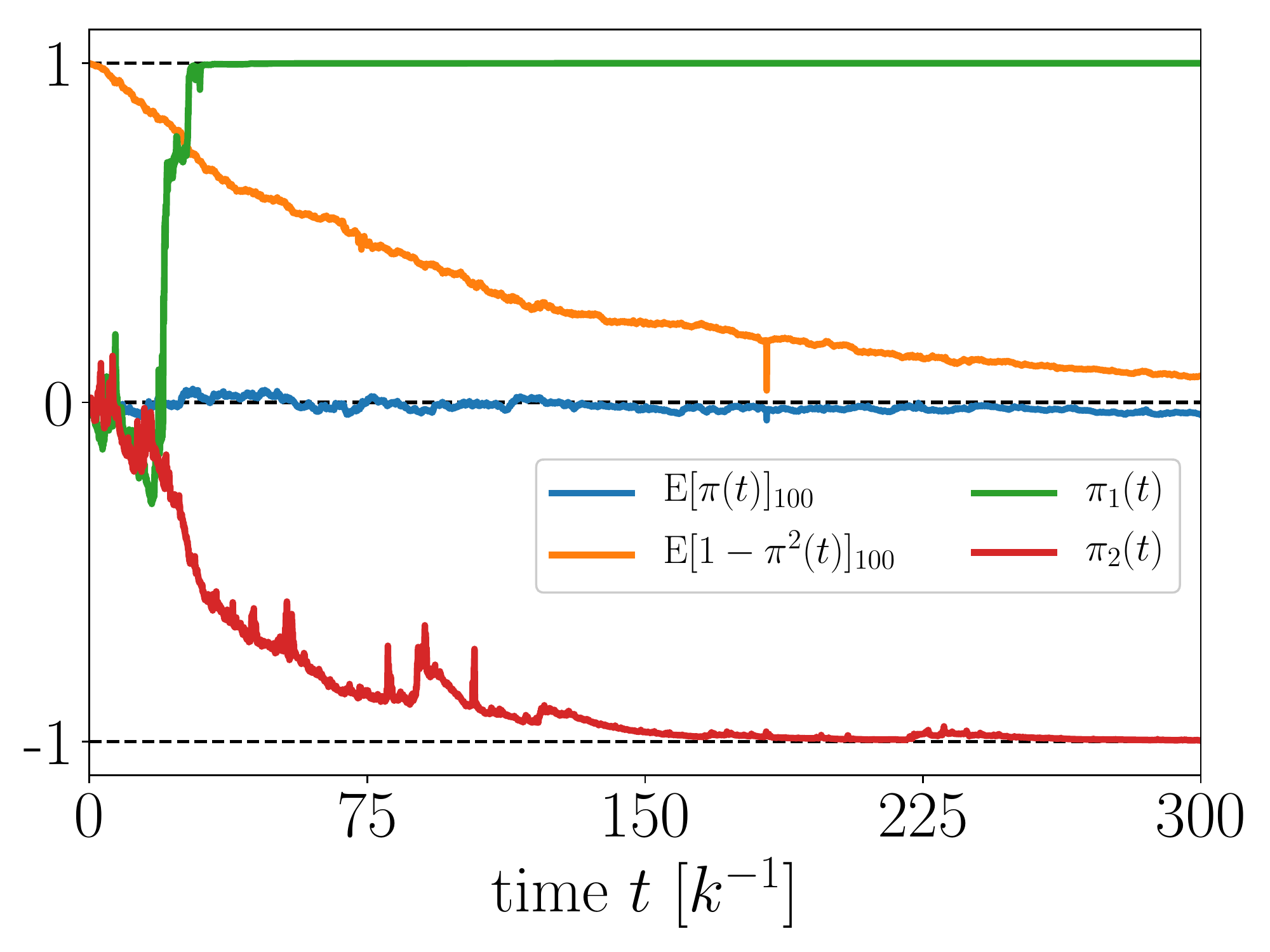}
    \caption{ 
    Continuous readout of Majorana qubit, with initial state $(\ket{\uparrow,0} + \ket{\downarrow,0})/\sqrt{2}$ corresponding to $n(0) = 0$. Two sample trajectories $\pi_1(t)$ (green) and $\pi_2(t)$ (red) show different measurement outcomes $\pi=1$ and $\pi=-1$, respectively. The ensemble average of $\pi(t)$ (blue, computed for 100 trajectories) remains close to zero for all times. The ensemble average of $1-\pi^2(t)$ (orange) quantifies the distance from the fixed points $\pi=\pm1$.  Parameters: $\varepsilon = 20k, t_1 = e^{-i\varphi} t_2 = 2k$ with $\varphi = \pi/4$. }
     \label{fig:majorana_measurement_generic}
\end{figure}

The equations for the diagonal blocks have themselves Lindblad form and preserve the trace. As $h_\pi$ does not commute with the quantum dot charge $n$ (unless $t_1 = \pm i t_2$ in which case the tunneling Hamiltonian vanishes for one of the blocks; we will comment on this case below) and as $n$ is hermitian, their only zero mode is the completely mixed state. Then, preservation of the trace implies that the diagonal blocks of the density matrix do indeed tend towards the fixed points 
\begin{equation}
\rho_{++}^{\infty} = \frac{\abs{\alpha}^2}{2}\mathds{1} \,\,\,\,\,\,\,\, {\rm and} \,\,\,\,\,\,\,\, \rho_{--}^{\infty} = \frac{\abs{\beta}^2}{2}\mathds{1},
\label{asymprho}
\end{equation} 
respectively. (We analyze the complete set of eigenvalues $\lambda_{\pi,n}$ and eigenmodes of $\mathcal{L}_{\pi,\pi}$ in App.\ \ref{app:liouvillian_evs_diag}.) Anticipating that the off-diagonal blocks generically decay to zero, we obtain the correct Born-rule probabilities for $\hat{Z}$. The final state has weight $\abs{\alpha}^2$ in the $\pi = +1$ subspace and $\abs{\beta}^2$ in the $\pi = -1$ subspace, which just corresponds to the probabilities of finding $Z = +1$ or $Z=-1$, as required.

\begin{figure*}
    \centering
    \includegraphics[width=1.0\textwidth]{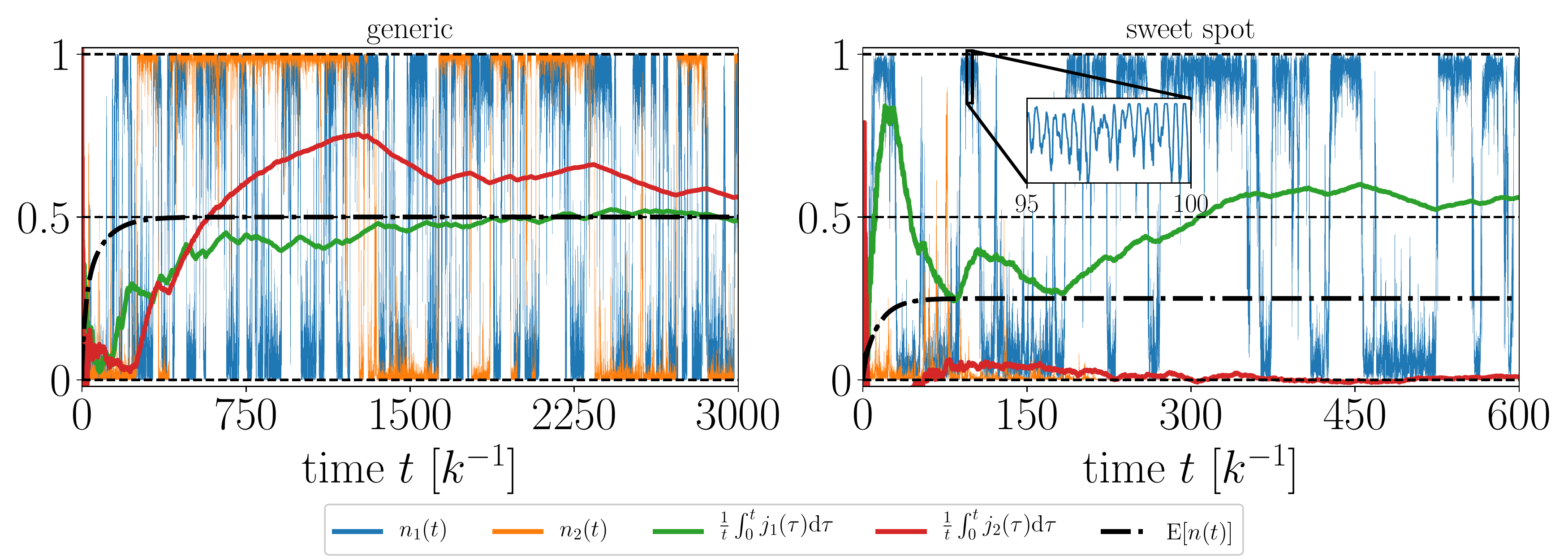}
    \caption{\label{fig:majorana_readout_compare_generic_and_sweet_spot}
    Continuous readout of Majorana qubit, with initial state $(\ket{\uparrow,0} + \ket{\downarrow,0})/\sqrt{2}$ corresponding to $n(0) = 0$. The panels show the (experimentally accessible) time-averaged measurement currents for two different combined-parity outcomes $\pi = 1$ (index $1$) and $\pi = -1$ (index $2$) and the (experimentally inaccessible) expectation value of the quantum dot charge. Left panel: Generic parameters as in Fig.\ \ref{fig:majorana_measurement_generic} ($\varphi=\pi/4$). Average measurement currents converge to $1/2$ for both $\pi=+1$ (dark green) and $\pi = -1$ (dark red), reflecting $\mathbb{E}[n(t)]\to 1/2$ (dash-dotted line), and the measurement outcome cannot be deduced from the time-averaged measurement current in the long-time limit. The charge expectation value $n_i(t)$ remains close to the charge eigenvalues, with occasional transitions occurring more frequently for outcome $\pi = +1$ (blue trace) than for outcome $\pi = -1$ (orange trace), reminiscent of telegraph noise. For suitable parameters, it may be feasible to base readout on extracting the transition rate between $n \simeq 0$ and  $n \simeq 1$ from the measurement current. Moreover, the integrated current relaxes more slowly to $1/2$ in the low-frequency sector (here $\pi = -1$) which may also help readout in some instances.
    Right panel: Same for the sweet spot $t_1 = -it_2$ ($\varphi = \pi/2$), where the system Hamiltonian $h_\pi$ commutes with the quantum dot charge $\hat n$ for $\pi=-1$. The time-averaged measurement current now converges to $1/2$ for outcome $\pi= +1$ (dark green, $\hat n$ not conserved) and to zero for outcome $\pi = -1$ (dark red, $\hat n$ conserved), so that $\pi$ is accessible from the time-averaged measurement current. This is a limiting case of the readout based on transition rates between $n\simeq 0$ and $n\simeq1$, which become increasing unequal as parameters approach the sweet spot. The ensemble average of $n(t)$ (dash-dotted line) now converges to $1/4$. The inset shows the coherent charge oscillations at frequency $\Omega_+$.
} 
\end{figure*}

The equation for the off-diagonal block deviates from Lindblad form since the first term on the right hand side involves both Hamiltonians $h_+$ and $h_-$. We analyze the eigenvalues $\tilde{\lambda}_n$ of $\mathcal{L}_{+-}$ in App.\ \ref{app:liouvillian_evs_offdiag} and find that they generically correspond to decaying modes. Then, the off-diagonal blocks decay to zero and the two $\pi$ subspaces decohere, $\rho_{+-}^{\infty} = 0$. The only exception occurs when $\Im{t_1t_2^*} = 0$. In this case, the off-diagonal block supports a nondecaying mode since the characteristic frequencies coincide for the $\pi=+1$ and $\pi = -1$ eigenspaces and we find $\rho_{+-}^{\infty} = {\alpha^*\beta}\mathds{1} / {2}$.

Consistent with these results, the conditional evolution of $\hat\rho$ is a bistable process for $\pi(t)$ and a projective measurement of $\hat{\pi}$. We can readily derive the stochastic evolution equation for $\pi(t)$,
\begin{equation}
    \dot{\pi} = \sqrt{4k}\xi \pqty{ \ev{\hat{n} \hat{\pi}} - n\pi }. 
\end{equation}
Clearly, $\pi = \pm 1$ are fixed points of this equation. This is illustrated in Fig.\ \ref{fig:majorana_measurement_generic}, where we show two representative sample trajectories of $\pi(t)$ for measurement outcomes $\pi = +1$ and $\pi = -1$. We find numerically that provided $\Im{t_1 t_2^*} \neq 0$, these are the only fixed points, cp.\ $\mathbb{E}[1-\pi^2(t)] \to 0$ in Fig.\ \ref{fig:majorana_measurement_generic}. We analyze the stochastic evolution in more detail in App.\ \ref{app:stochastic_evolution}. 

The decay of the off-diagonal block to zero implies that $\pi$ can be extracted from the measurement current $j(t)$. Once the measurement current becomes stationary, $j(t) = j_\pi (t)$, the system explores the full subspace with fixed $\pi$ in an ergodic manner. This follows from the fact that the ensemble average over all trajectories with outcome $\pi$ yields a completely mixed state on the subspace. Importantly, this implies that the ensemble (or time) average of $j_\pi(t)$ does not carry information on $\pi$,
\begin{align}
\label{eq:enavj}
    \mathbb{E}[j_\pi (t)] =&\ \textrm{tr}\bqty{\hat{n}\mathbb{E}[\hat{\rho}_c (t)\vert_{\pi}]} + \frac{\mathbb{E}[\xi(t)]}{\sqrt{4k}}  = n^{\infty}_\pi = \frac{1}{2}.
\end{align}
In agreement with Eq.\ (\ref{eq:enavj}) and in contrast to the simple quantum dot charge readout (see Sec.\ \ref{sec:simple_charge_readout}), the integrated measurement currents converge to $1/2$ irrespective of the measurement outcome. This is shown in Fig.\ \ref{fig:majorana_readout_compare_generic_and_sweet_spot} (left panel). 

\begin{figure}
    \includegraphics[width=0.9\columnwidth]{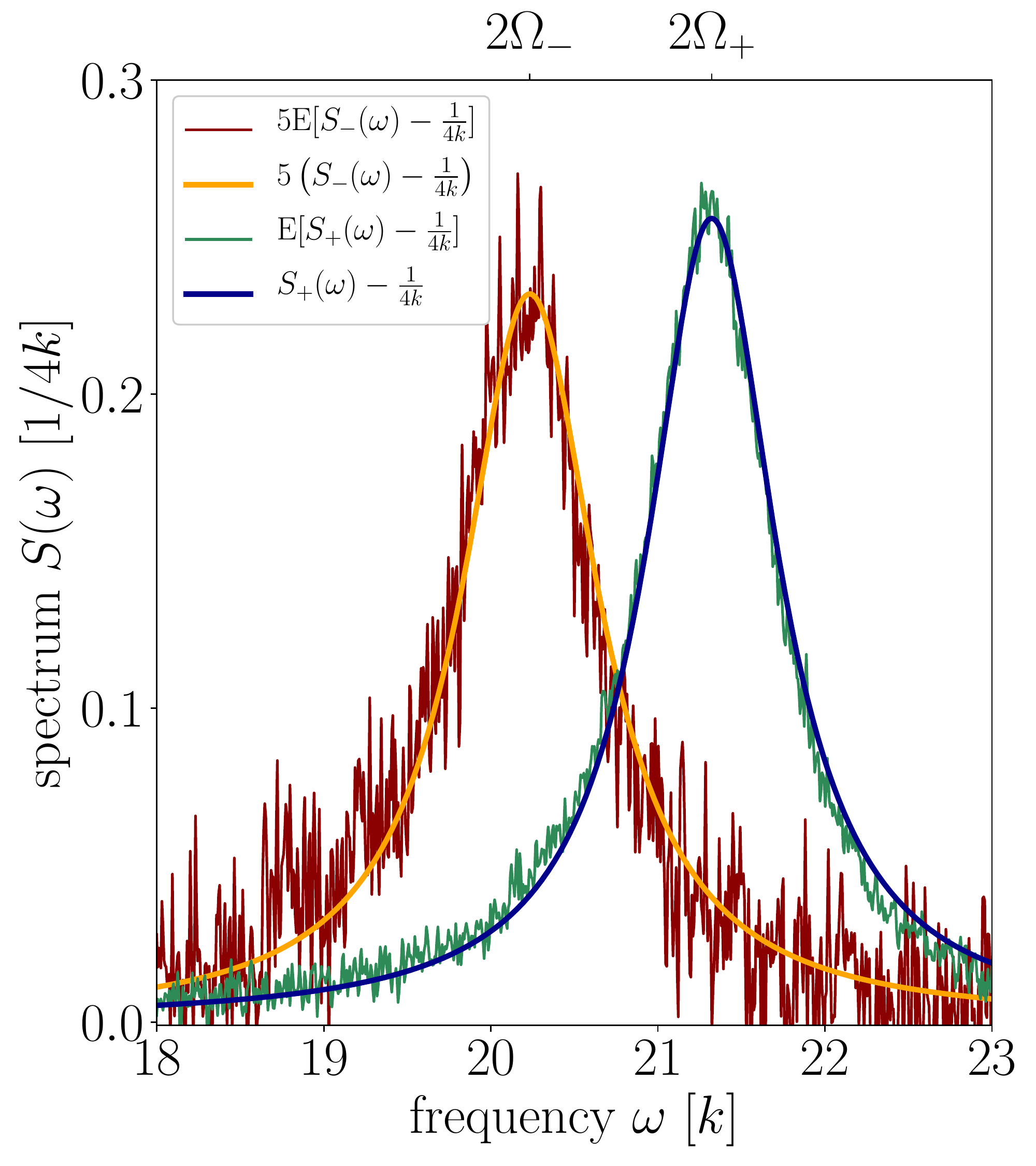}
    \caption{
    Power spectra of the measurement signal corresponding to outcomes $\pi = +1$ and $\pi = -1$. Numerical simulations based on Eq.\ (\ref{eq:sme_mbq-qd-qpc}) (green and red traces) are in excellent agreement with the analytical expression in Eq.\ (\ref{eq:spectrum_main_text}) (blue and orange traces). The numerical power spectra were obtained by generating measurement signals for long time intervals of $T \sim 10^7 k^{-1}$. The long integration time is necessitated by the small weight of the finite-frequency Lorentzians when $\sin\theta_\pi \ll 1$. Other parameters as in Fig.\ \ref{fig:majorana_measurement_generic}. The power spectrum for $\pi=-1$ is scaled up by a factor of five for better visibility.}
     \label{fig:spectrum}
\end{figure}

Nonetheless, information on $\pi$ is generically encoded in the noise correlations of the measurement current,
\begin{equation}\label{eq:autocorrelation_main}
     S_\pi(\tau) = \mathbb{E}\bqty{j_\pi(t)j_\pi(t+\tau)}. 
\end{equation}
In the stationary (long-time) limit, the corresponding power spectrum $S_\pi(\omega)$ can be readily computed for the two values of $\pi$ (see App.\ \ref{app:spectrum}). In the limit of weak measurements, $k \ll \Omega_\pi$, we find that in addition to a white-noise background, which is just the shot noise power of the quantum point contact, the power spectrum exhibits Lorentzian peaks at $\omega = 0$ and $\omega=\pm 2\Omega_\pi$, which reflect the dynamics of the Majorana qubit-quantum dot system. Explicitly, we find
\begin{multline}\label{eq:spectrum_main_text}
    S_\pi(\omega) = \frac{1}{4k} + \frac{\cos^2\theta_\pi}{2} \frac{ \kappa_\pi }{\omega^2 + \kappa_\pi^2} \\ + 
		\frac{\sin^2\theta_\pi}{4} \sum_\pm 
		\frac{ \tilde{\kappa}_\pi }{(\omega \pm 2\Omega_\pi)^2 +\tilde{\kappa}_\pi^2 }  ,
\end{multline}
where we introduced the widths 
\begin{subequations}
\begin{align}
    \kappa_\pi =&\ \frac{\sin^2\theta_\pi}{2} k, \\
    \tilde{\kappa}_\pi =&\  \frac{\pqty{1+\cos^2\theta_\pi}}{4} k
\end{align}
\end{subequations}
of the Lorentzians. The measurement result for $\pi$ can be read off in particular from the location of the Lorentzians in frequency.  This is illustrated in Fig.\ \ref{fig:spectrum} which shows the power spectrum for a numerically generated measurement current and compares it to Eq.\ (\ref{eq:spectrum_main_text}).

Physically, the zero-frequency peak in Eq.\ (\ref{eq:spectrum_main_text}) is associated with the telegraph noise of the charge expectation value shown in Fig.\ \ref{fig:majorana_readout_compare_generic_and_sweet_spot}. In principle, the width of this peak also encodes the measurement outcome. Indeed, Fig.\ \ref{fig:majorana_readout_compare_generic_and_sweet_spot} illustrates that the dwell time near $n=0$ and $n=1$ depends on the $\pi$ subspace. For suitable parameters, it may also be possible to monitor the transition rate directly by means of an appropriately smoothened measurement current, or to extract information on the measurement outcome from associated transients in the time-averaged measurement current. The peak at finite frequency originates from Rabi oscillations, which are seen in Fig.\ \ref{fig:majorana_readout_compare_generic_and_sweet_spot} as the small deviations of the quantum dot charge from its eigenvalues $n=0$ and $n=1$ (see inset of right panel). The measurement pushes the system into a charge eigenstate and thus a superposition of energy eigenstates, which then leads to oscillations as a result of the Hamiltonian dynamics.

It is interesting to consider two special parameter choices. First, for $\Im{t_1 t_2^*} = 0$, the off-diagonal block $\rho_{+-}$ of the density matrix does not decay. Indeed, in this case, not only the average measurement signal, but also its noise correlations are independent of $\pi$. More generally, the failure to decohere in the $\hat{\pi}$ basis reflects the fact that the measurement signals for the two subspaces are indistinguishable. 

While the measurement fails for $\Im{t_1 t_2^*} = 0$, it can be simplified at the sweet spot $t_1 = - i t_2$ ($t_1 = i t_2$ is analogous). In this case, the Hamiltonian commutes with $\hat{n}$ in the $\pi = -1$ block and charge conservation in this block makes all density matrices which are diagonal in the charge basis into zero modes of $\mathcal{L}_{--}$. In this fine-tuned situation, it is not necessary to measure noise correlations. Instead, the measurement outcome for $\hat{\pi}$ and hence $\hat{Z}$ can be extracted from the ensemble-averaged charge alone, which yields $\mathbb{E}[n]=1/2$ for $\pi = +1$ and $\mathbb{E}[n]=0$ for $\pi = -1$. For small deviations from the fine-tuned point, $t_1 = - i t_2 + \delta$ with $\delta \ll \abs{t_1}$, charge is no longer conserved in both blocks and $\mathbb{E}[n]=1/2$ regardless of $\pi$. However, the relaxation rate to the stationary state will be smaller in the $\pi=-1$ block by a factor $\abs{\delta}^2/\abs{t_1}^2$. For a sufficiently high measurement efficiency, it might then be possible to resolve $\pi$ from transient differences in the average charge. Figure \ref{fig:majorana_readout_compare_generic_and_sweet_spot} (right panel) shows corresponding  simulations (with $\delta = 0$), which confirm that in principle, the integrated measurement signals suffice to identify the measurement outcome in a measurement time  $\tau_m \sim k^{-1}$.

\begin{figure}
    \centering
    \includegraphics[width=\columnwidth]{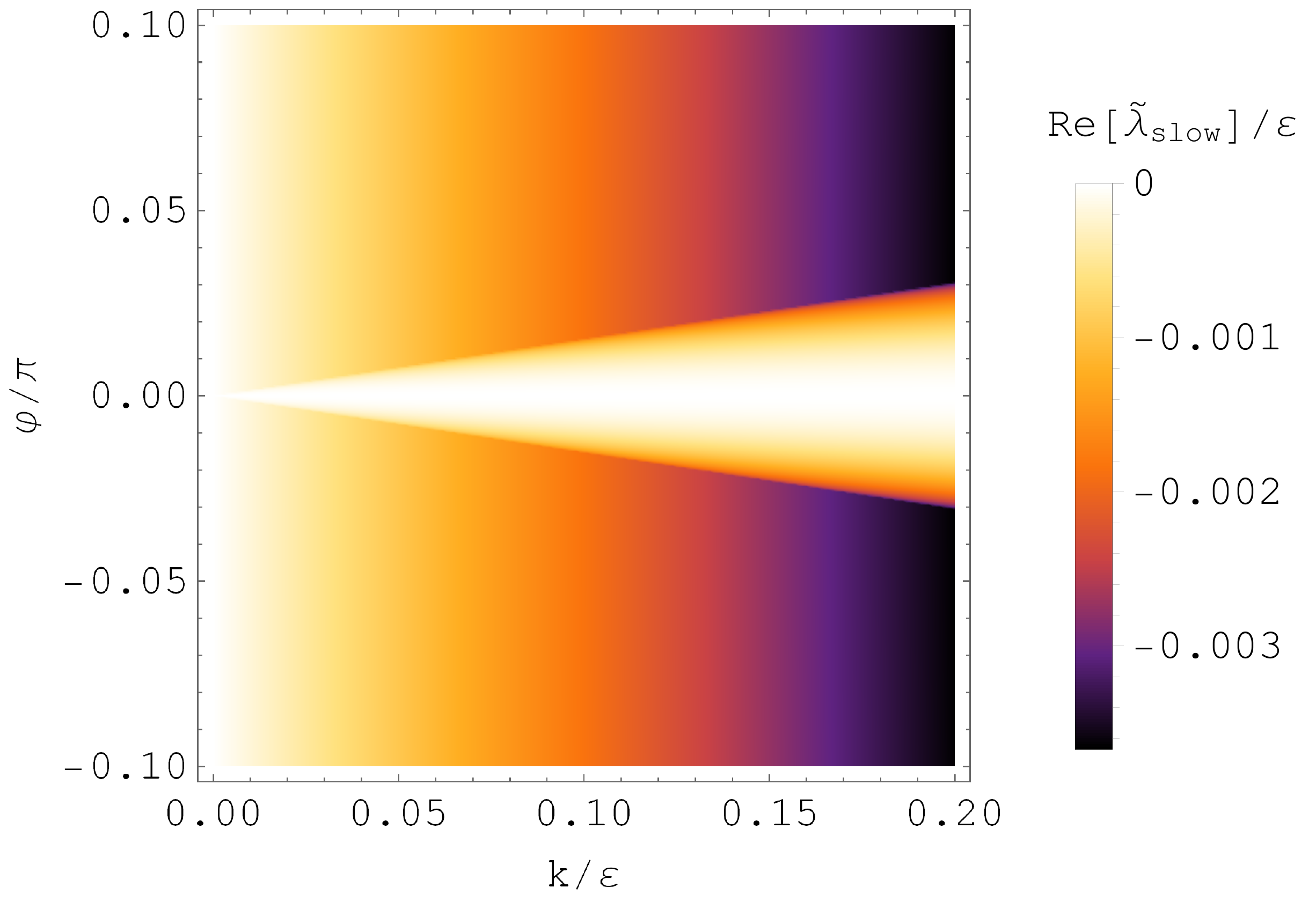}
    \caption{Decoherence rate as characterized by the real part of the slowest decaying eigenvalue of $\mathcal{L}_{+-}$, $\Re\{\tilde{\lambda}_{\textrm{slow}}\}$ as a function of the measurement strength $k$ and $\varphi$ defined through $t_1 = e^{-i\varphi} t_2 = 0.1 \varepsilon$.}
    \label{fig:slowest_decaying_ev}
\end{figure}

An important characteristic of the measurement is the measurement-induced decoherence time (see Refs.\ \cite{Goldstein2011,Rainis2012,Schmidt2012,Hu2015,Knapp2018,Li2018,Bauer2020} for discussions of Majorana qubit decoherence unrelated to measurements). The decoherence time is closely related to (and in fact upper bounded by) the inverse of the real part of the slowest decaying eigenvalue $\tilde{\lambda}_{\textrm{slow}}$ of $\mathcal{L}_{+-}$. Figure \ref{fig:slowest_decaying_ev} shows numerical results for $\Re\{\tilde{\lambda}_{\textrm{slow}}\}$. We focus on weak measurements, $k, |t_1|,|t_2| \ll \varepsilon$, and nonzero but possibly small $\varphi$ defined through $t_1 = e^{-i\varphi}t_2$. We observe that there is no decoherence along the line $\varphi=0$ where $\Im{t_1t_2^*} = 0$. At fixed $k$, $\Re\{\tilde{\lambda}_{\textrm{slow}}\}$ grows quadratically in $\varphi$ up until discontinuous lines, where $\tilde{\lambda}_{\textrm{slow}}$ and the corresponding eigenmatrix coalesce with another eigenvalue-eigenmatrix pair. Then, the eigenmatrices of $\mathcal{L}_{+-}$ fail to span the space of $2 \times 2$ complex matrices along these \textit{exceptional} lines  emanating from $\pqty{\varphi,k}=\pqty{0,0}$. At fixed $\varphi$ and to leading order in $k$, we observe a linear decrease as $k$ increases towards the discontinuity. The measurement is more efficient when tuning to the large-$|\varphi|$ side of the exceptional lines. In this region, $\Re\{\tilde{\lambda}_{\textrm{slow}}\}$ depends only weakly on $\varphi$, which is why we only display the region of small $\varphi$ in Fig.\ \ref{fig:slowest_decaying_ev}. Note that the discontinuities are regularized by relaxation terms, such as those discussed in the following Sec.\ \ref{sec:with_relaxation}. 

\subsection{Majorana qubit readout in the presence of relaxation}\label{sec:with_relaxation}

The results of the previous section may be surprising in that the quantum point contact does not detect the dependence of the ground-state expectation value of the quantum dot charge on the parity sector $\pi$. Instead, the information on $\pi$ can generically only be extracted from noise correlations of the measurement current $j(t)$. Experimentally, however, it would be preferable if $\pi$ could be extracted from the average measurement signal.

The underlying reason for the insensitivity to the quantum dot charge in the ground state is that under the continuous measurement, the density matrix generically becomes proportional to the unit matrix within the subspace with fixed $\pi$ [see Eq.\ (\ref{asymprho})]. Then, the expectation value of the quantum dot charge is just equal to 1/2, independent of the parity $\pi$. The difference in quantum dot charge between the ground states is compensated by the opposite difference between the excited states. 

It is then natural to expect that the average measurement current distinguishes between the two subspaces once one includes additional relaxation processes from the excited state $\ket{e_\pi}$ to the ground state $\ket{g_\pi}$. Unlike the measurement which leads to relaxation in the basis of the quantum dot charge, this additional relaxation should operate within the eigenbasis of $h_\pi$. We now show that this expectation is indeed correct. 

There are various processes which induce relaxation within the eigenstate basis. One relevant example are effective measurements of the Majorana-qubit charge by the environment. For definiteness, we effect relaxation within the eigenstate basis by coupling to the electromagnetic environment. Within the Born-Markov approximation and focusing on $T=0$ for simplicity, this leads to an additional dissipation term in Eq.\ (\ref{eq:sme_mbq-qd-qpc}) (see App.\ \ref{app:dissipation} for details),
\begin{equation}\label{eq:dissipator_main}
    \frac{\textrm{d}}{\textrm{dt}}\hat{\rho}\vert_{\textrm{relax}} 
    = \Gamma_- \sum_\pi \mathcal{D}\bqty{\frac{\sin\theta_\pi}{2} \hat{\tau}^\pi_-}\hat{\rho} \equiv \mathcal{L}'\hat{\rho}.
\end{equation}
Here, we defined the lowering operator in the energy basis for subspace $\pi$, $\hat{\tau}^\pi_- = \ketbra{g_\pi}{e_\pi}$ and the zero-temperature relaxation rate $\Gamma_- = 2\pi J(2\Omega_\pi)$ governed by the spectral density $J(\omega)$ of the electromagnetic environment. While $\Gamma_-$ may depend on $\pi$ in principle, we assume $J(2\Omega_+) \simeq J(2\Omega_-)$ for simplicity. Note that at finite temperatures, there is an additional dephasing term $\Gamma_0 \mathcal{D}\bqty{\cos\theta_\pi \hat{\tau}^\pi_z / 2}\hat{\rho}$ in the eigenstate basis, where $\Gamma_0 = 2\pi \lim_{\omega \to 0} J(\omega)b(\omega)$ with the Bose distribution $b(\omega)$. However, for $k, \Gamma_-, \Gamma_0 \ll \Omega_\pi$, this term does not affect the results,  cp. App.\ \ref{app:liouvillian_evs_diag}. 

\begin{figure}
    \centering
    \includegraphics[width=0.9\columnwidth]{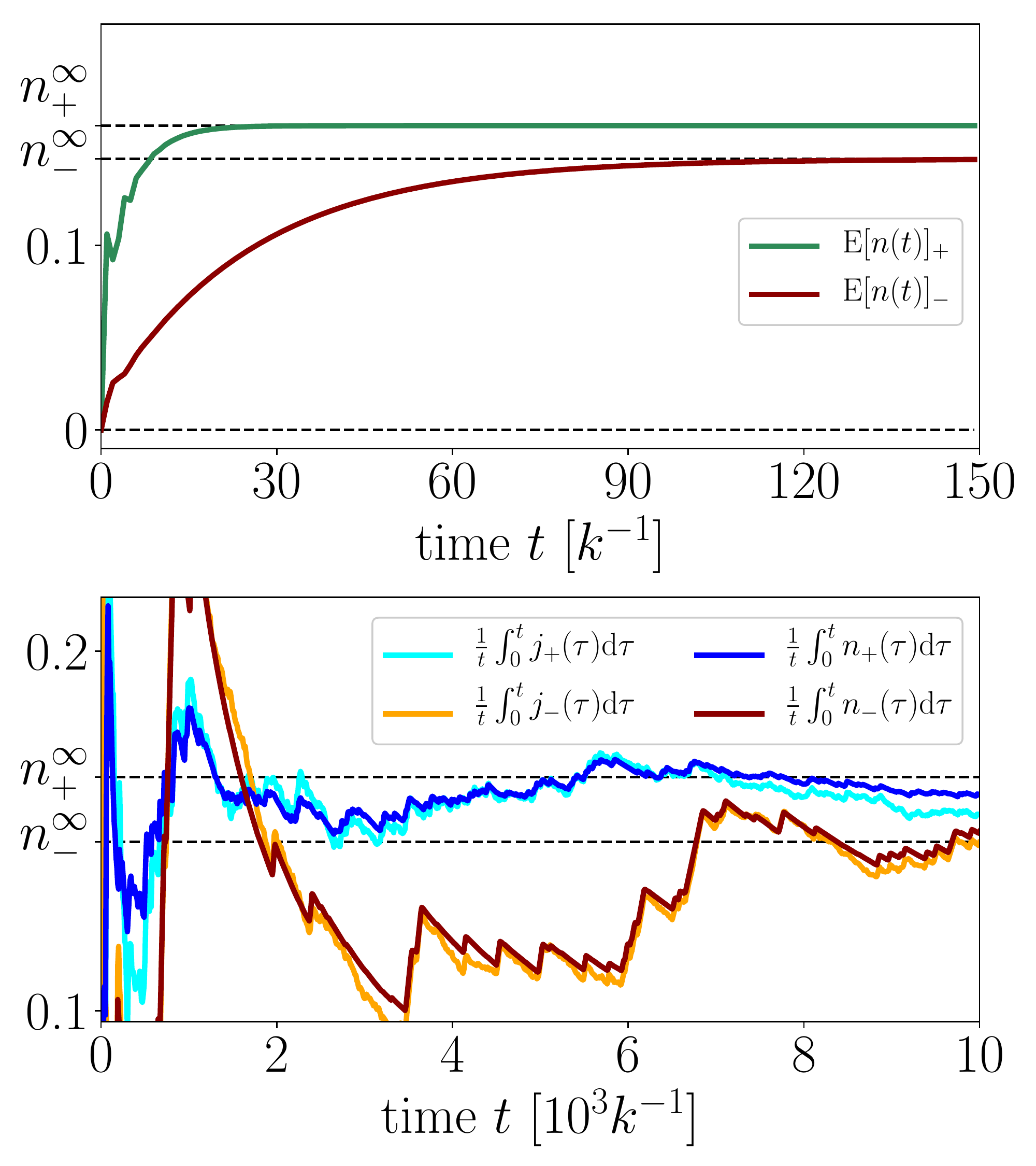}
    \caption{Continuous readout of Majorana qubit in the presence of relaxation with rate $\Gamma_- = 5k$ (other parameters as in Fig.\ \ref{fig:majorana_measurement_generic}). Top panel: Ensemble-averaged quantum dot occupations $n(t)$, restricted to the subspaces $\pi=+1$ (green trace) and $\pi=-1$ (red trace), as obtained from the unconditioned master equation (starting in the ground state with $n=0$). $\mathbb{E}[n(t)]_\pi$ converges to distinct expectation values $n_\pi^\infty$ for the two subspaces, so that $\pi$ can be read out from the time-averaged measurement current of a quantum dot charge measurement. This is shown in the bottom panel: Time averaged measurement signals for trajectories in the $\pi = +1$ (cyan trace) and $\pi = -1$ (orange) sectors, converging to $n_\pi^\infty$. Readout requires integration times which are significantly longer than convergence times of the ensemble-averaged quantum dot occupations in the top panel. From Eq.\ \eqref{eq:meas_time_relax}, the measurement time can be estimated as $\tau_m \sim 10^4 /k$ for the given parameters. This estimate is based on the fluctuations in the $\pi = -1$ sector, where fluctuations are larger since $\cot \theta_- \gg \cot\theta_+$.  We also show the time average of the charge expectation value (not accessible in experiment). The similarity of the curves indicates that fluctuations of the measurement current are dominated by fluctuations of the charge expectation value. } 
    \label{fig:majorana_readout_with_dissipation}
\end{figure}

Importantly, Eq.\ \eqref{eq:dissipator_main} also conserves $\hat{\pi}$ and the total unconditioned master equation [obtained by incorporating Eq.\ (\ref{eq:dissipator_main}) into Eq.\ \eqref{eq:sme_mbq-qd-qpc}] still decouples into blocks. The off-diagonal block obeys 
\begin{align}
     \dot{\rho}_{+-} =&\  \pqty{ \mathcal{L}_{+-} + \mathcal{L}'_{+-} } \rho_{+-}. 
\end{align}
Here, $\mathcal{L}'_{+-}  = \frac{1}{4}\Gamma_- \sin\theta_+\sin\theta_-\mathcal{D}\bqty{\tau_-}$ has Lindblad form and a negative semidefinite real part. Since $\mathcal{L}_{+-}$ generically has only decaying eigenvalues, $\mathcal{L}_{+-} + \mathcal{L}'_{+-}$ is also decaying. The only exception occurs when $\Im{t_1 t_2^*} = 0$, as in the absence of relaxation. In this special case, $\mathcal{L}_{+-}$ and thus also $\mathcal{L}_{+-} + \mathcal{L}'_{+-}$ as a whole have Lindblad form and preserve the trace. Thus, the condition for the measurement to work remains unchanged in the presence of relaxation.

The evolution of the diagonal blocks tends to 
\begin{equation}
    \rho_{\pi,\pi}^{\infty} = \frac{1}{2}\pqty{\tau_0^\pi + R \tau^{\pi}_z }
\end{equation}
in the energy basis to leading order in $\Omega_\pi \gg k, \Gamma_-$. Here we defined the ratio
\begin{equation}\label{eq:measurement_relaxation_ratio}
    R = \frac{\Gamma_-}{\Gamma_- + 2k}
\end{equation}
characterizing the strength of the additional relaxation. The associated ensemble average of the quantum dot charge becomes \begin{equation}
    n_{\pi}^{\infty} = \frac{1}{2}\pqty{1 +  R\cos\theta_\pi} .
\end{equation}
When the measurement is stronger than dissipation, we have $R\ll 1$ and the quantum dot charge is close to $1/2$, independent of $\pi$. However, in the opposite limit, when dissipation is stronger than the measurement and $R$ approaches unity, the average charge is approximately given by the ground-state expectation value of the charge in the respective sector, $\bra{g_\pi} \hat{n} \ket{g_\pi} = (1+\cos \theta_\pi)/2$. In this limit, the time-averaged measurement signal depends on $\pi$, in agreement with the heuristic arguments given above. 

We illustrate these considerations by the numerical simulations shown in Fig.\ \ref{fig:majorana_readout_with_dissipation}. Including the relaxation process in Eq.\ \eqref{eq:dissipator_main} in the simulations of the unconditioned master equation, we compute the time-averaged measurement currents and find that indeed, they converge towards $n_{\pi}^{\infty}$, albeit slowly.

The corresponding measurement time is determined by the requirement to resolve the difference 
\begin{equation}
    \abs{n_{+}^{\infty} - n_{-}^{\infty}} = 4R\abs{\frac{\Im{t_1 t^*_2}}{\varepsilon^2}}
\end{equation}
in quantum dot occupations. The time-averaged measurement current 
\begin{equation}
    j_{\textrm{int},\pi}(T) = \frac{1}{T}\int_0^T\textrm{d}t\ j_\pi(t) 
\end{equation}
fluctuates around $n_{\pi}^{\infty}$, with decreasing magnitude of the fluctuations as $T$ grows. The fluctuations can be estimated via the variance
\begin{equation}\label{eq:int_curr_variance_main}
    \mathbb{V}[j_{\textrm{int},\pi}(T)] = \frac{1}{4kT} +   \cot^2\theta_\pi R^2\  \frac{4k}{\Gamma_-^2 T}.
\end{equation}
The first term reflects the white noise background, whereas the second originates from fluctuations of $n(t)$. The latter depends on the sector $\pi$. The measurement time can then be estimated by comparing the variance with the resolution necessary to distinguish the two possible measurement outcomes. This gives
\begin{equation}\label{eq:meas_time_relax}
    \tau_{m} \sim \frac{\varepsilon^4}{16 \abs{\Im{t_1 t^*_2}}^2} \pqty{ \frac{1}{4kR^2} +  C  \frac{4k}{\Gamma_-^2}},
\end{equation}
where we use the variance of the sector with larger fluctuations, defining
\begin{equation}
    C = \max_\pi{ \cot^2\theta_\pi}.
\end{equation}
We observe that a large splitting resulting from a large $\abs{\Im{t_1 t^*_2}}$ and a small $\varepsilon$ are advantageous for a fast measurement. (Note, however, that there is a tradeoff since a small $\varepsilon$ enhances quasiparticle poisoning rates.) Moreover, the terms in the brackets have interesting structure. They diverge for both $k \to 0$ and $k\to\infty$. Thus, at a given $\Gamma_-$, there is an optimal measurement strength $k_{\textrm{opt}} = \Gamma_-/2\sqrt{1 + 4C}$ to identify the measurement outcome based on the time-averaged signal. The corresponding optimal measurement time becomes
\begin{equation}
    \tau_{m,\textrm{opt}} \sim \frac{\varepsilon^4\pqty{1 + \sqrt{1+4C }}}{16 \Gamma_- \abs{\Im{t_1 t^*_2}}^2}.
\end{equation}

\begin{figure}
    \centering
    \includegraphics[width=0.9\columnwidth]{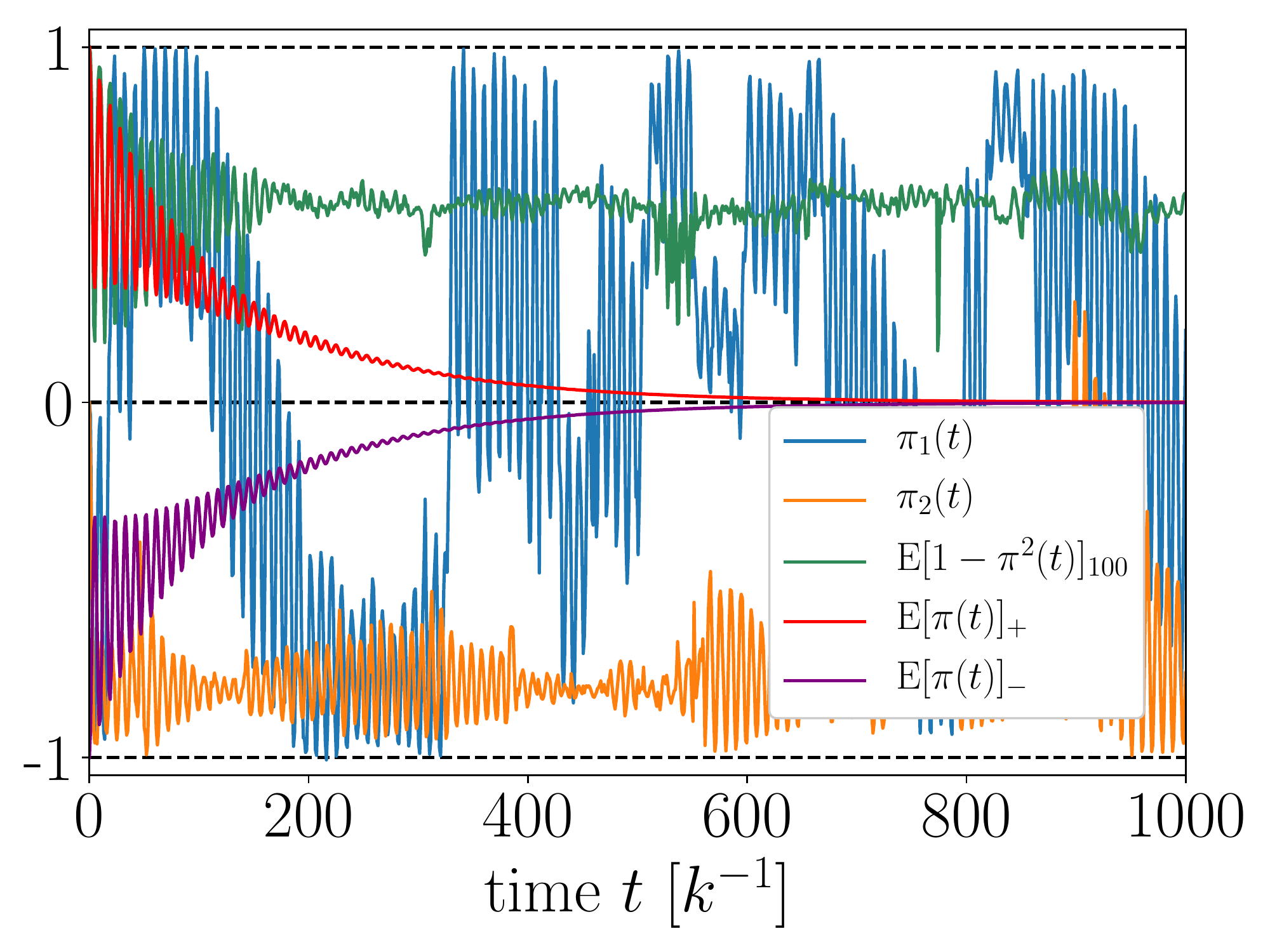}
    \caption{Effect of Majorana hybridization $\epsilon_{23} = 0.2k$ on Majorana-qubit readout (in the absence of relaxation, $\Gamma_- =0$). Blue and orange traces: Sample trajectories of $\pi(t)$, clearly not reaching a fixed point. Red and purple traces: Ensemble-averaged evolution of $\pi(t)$ for states initialized in the $\pi = +1$ (red) and $\pi = -1$ (purple) sectors. Both curves converge towards $\mathbb{E}[\pi] = 0$, corresponding to an equal mixture of the two sectors, regardless of initial condition. Similarly, $\mathbb{E}[1-\pi^2(t)]$ (green) does not approach $0$, so that $\pi \neq \pm 1$. Other parameters as in Fig.\ \ref{fig:majorana_measurement_generic}.   }
    \label{fig:majorana_readout_with_hybridization_no_dissipation_fast}
\end{figure}

\subsection{Charge nonconservation and Majorana hybridizations}
\label{sec:imperfect}

An essential assumption underlying the readout of the Majorana qubit is that the combined parity $\hat\pi$ is a good quantum number. In practice, there can be processes which do not conserve $\hat\pi$. First, the combined parity does not commute with the residual Majorana hybridizations $\varepsilon_{ij}$ in Eq.\ \eqref{eq:charging_ham} (except for $\varepsilon_{12}$). Second, $\hat\pi$ is no longer conserved in the presence of leakage of the quantum dot charge, say into additional reservoirs.

\begin{figure}
    \centering
    \includegraphics[width=0.9\columnwidth]{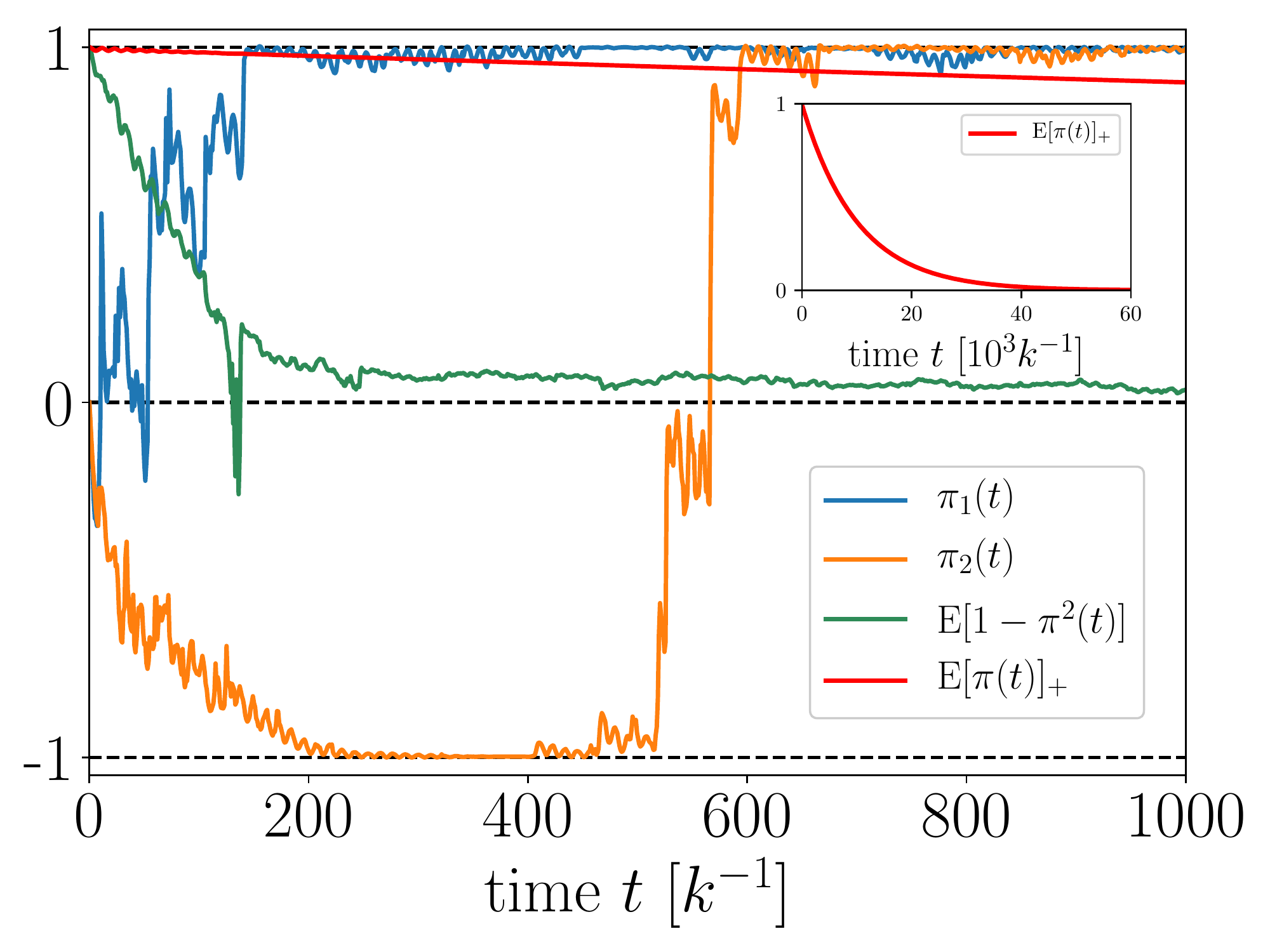}
    \caption{
    Effect of a weak Majorana hybridization $\epsilon_{23} = 0.02k$ on Majorana-qubit readout. For this hybridization strength, the measurement evolution, which tries to project $\pi$ onto an eigenvalue of $\hat{\pi}$, is stronger than the evolution due to $\hat{H}_{23}$. Thus, in contrast to Fig.\ \ref{fig:majorana_readout_with_hybridization_no_dissipation_fast}, individual $\pi_i(t)$ traces (blue and orange) remain predominantly near the fixed points $\pi = \pm 1$. This is also reflected in the fact that $\mathbb{E}[1-\pi^2]$ (green trace) reaches a steady state value which is different from but still close to zero. At short times, individual trajectories reach the fixed points with a probability reflecting the initial weights $\abs{\alpha}^2$ and $\abs{\beta}^2$ associated with the $\hat{\pi}$ eigenspaces. Eventually, hybridization flips $\pi(t)$ between $1$ to $-1$, as illustrated by trajectory $\pi_2(t)$. These jumps cause a decay of the ensemble average of $\pi(t)$ over trajectories initialized within one fixed point, see red trace for $\mathbb{E}[\pi]_+$ (enlarged in inset). Correspondingly, the initial weights are lost and in the long-time limit, trajectories are close to either fixed point with equal probability (as quantified by the decay of $\mathbb{E}[\pi]_+$) . 
    For good readout fidelity, the measurement outcome must be identifiable as long as $\mathbb{E}[\pi]_+ \simeq 1$. 
    Other parameters as in Fig. \ref{fig:majorana_measurement_generic}.}
    \label{fig:majorana_readout_with_hybridization_no_dissipation_slow}
\end{figure}

It is natural to expect that these processes spoil the measurement by allowing weight to move between the $\pi$ subspaces and thereby scrambling the probabilities associated with the measurement outcomes. We analyze this in more detail for the Majorana hybridizations. For definiteness, we focus on $\varepsilon_{23}$ with the corresponding contributions
\begin{align}
    \hat{H}_{23} =&\ -i\varepsilon_{23} \hat{\gamma}_2\hat{\gamma}_3  \\
    =&\ \varepsilon_{23} \sum_n \pqty{ \ketbra{\uparrow,n}{\downarrow,n}+\ketbra{\downarrow,n}{\uparrow,n}}
\end{align}
to the Hamiltonian and $\mathcal{L}_{23} \hat{\rho} = -i[\hat{H}_{23},\hat{\rho}]$ to the Liouvillian. In the absence of relaxation and for $\Im{t_1 t_2^*} \neq 0$, the new total Liouvillian $ \mathcal{L}+ \mathcal{L}_{23}$ has $\hat{\rho}^{\infty} = \textrm{diag}(1,1,1,1)/4$ as the only zero mode and, consequently, does not preserve information on the weights $\abs{\alpha}^2$ and $\abs{\beta}^2$ of the initial Majorana-qubit state in the long-time limit (see App.\ \ref{app:steady_state_w_hybridizations}).  

This is illustrated in Fig.\ \ref{fig:majorana_readout_with_hybridization_no_dissipation_fast} which shows $\mathbb{E}[\pi(t)]_\pm$, the ensemble-averaged evolution of $\pi(t)$ for initial states $\pi(0) = \pm 1$. For significant values of $\epsilon_{23}$, $\mathbb{E}[\pi(t)]_\pm$ relaxes to $0$ faster than the measurement can project $\hat{\pi}$, as indicated by the fact that $\mathbb{E}[1-\pi^2(t)]$ remains large for all times $t$. This implies that the system forgets the weights associated with the $\pi$ eigensectors too fast to perform a measurement. In contrast, Fig.\ \ref{fig:majorana_readout_with_hybridization_no_dissipation_slow} shows data for a much smaller value of $\epsilon_{23}$. Here, $\varepsilon_{23} \ll \tau_m^{-1}$ and the information on the weights is retained transiently. Still, in the long-time limit, this information is lost and Majorana hybridizations set an upper limit for the time a measurement may take. Including relaxation does not change this qualitatively. In this case, the steady state will no longer be completely mixed, but importantly, there is only one steady state and information on the qubit state is lost in the long time limit.

\subsection{Readout via double quantum dot}
\label{sec:DQDreadout}

\subsubsection{Readout of two-Majorana parities}

\begin{figure}
    \centering
    \includegraphics[width=0.7\columnwidth]{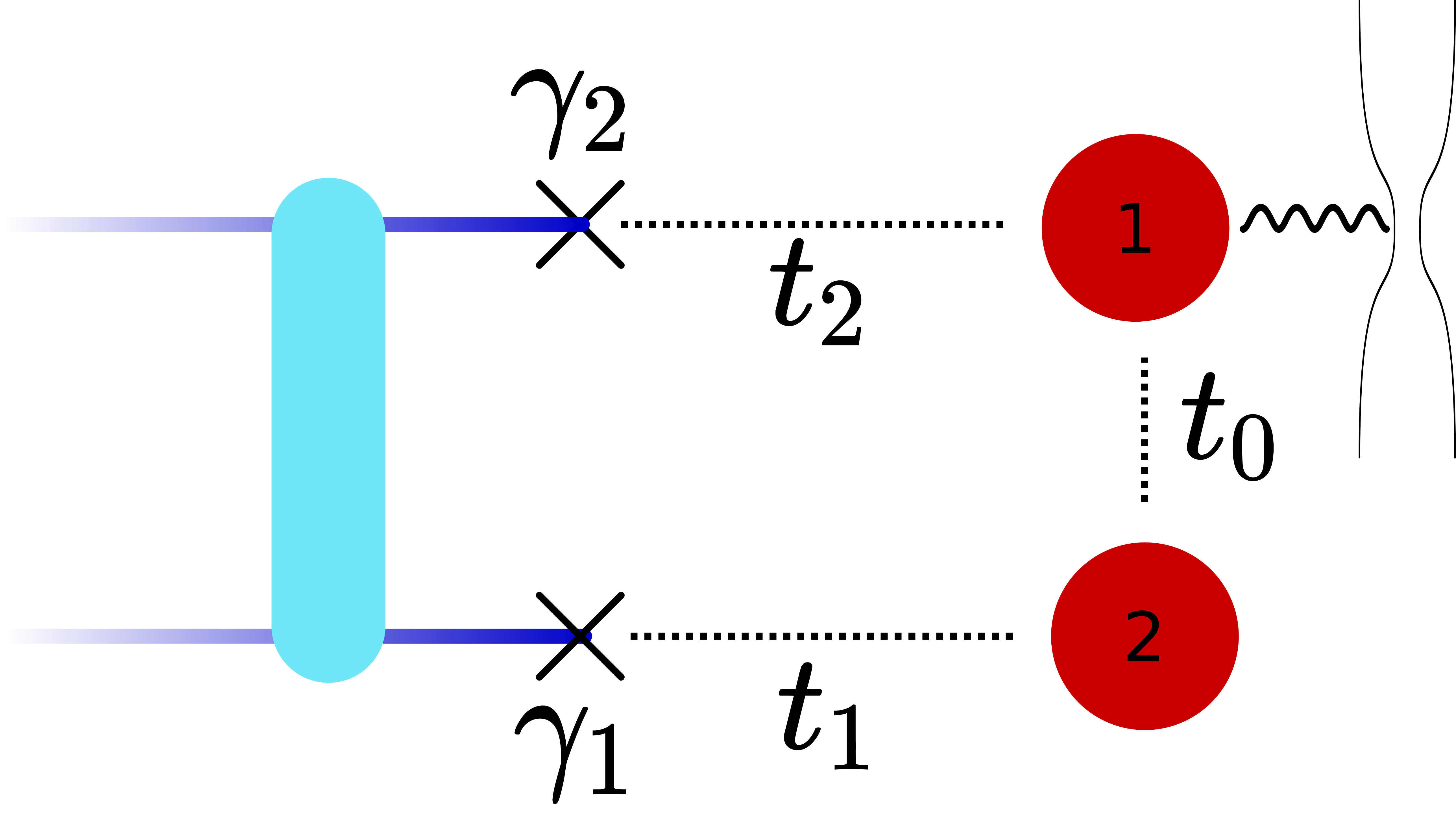}
    \caption{Majorana qubit readout by means of a double quantum dot (with inter-dot tunneling $t_0$), with charge monitoring by a quantum point contact of one (as shown) or both quantum dots. Symbols as in Fig.\ \ref{fig:tetron_quantum_dot_measurement_device}.}
    \label{fig:tetron_double_dot_with_readout}
\end{figure}
It is interesting to compare the scheme discussed so far with a modified readout setup which couples the Majorana qubit to a double quantum dot, such that Majoranas $\hat\gamma_1$ and $\hat\gamma_2$ entering into $\hat Z$ are coupled to one quantum dot each, see Fig.\ \ref{fig:tetron_double_dot_with_readout}. In this case, the effective hopping amplitude between the quantum dots equals 
\begin{equation}
t_{\hat Z} = t_0+\frac{it_1t_2^*}{E_c}\hat Z.    
\end{equation}
Here, $t_0$ denotes direct hopping, while the second term originates from indirect hopping via the Majorana qubit. We consider the subspace in which a single electron in the double quantum dot can reside in either of the two quantum dots, with basis states $\ket{1,0}$ and $\ket{0,1}$. The Hamiltonian of the system, written in the basis $\{\ket{1,0;\uparrow},\ket{0,1;\uparrow},\ket{1,0;\downarrow},\ket{0,1;\downarrow}\}$ becomes block-diagonal, 
\begin{equation}
    {\hat H}=\left(\begin{array}{cc}  h_\uparrow & 0 \\ 0 & h_\downarrow 
    \end{array}\right),
\end{equation}
where the $2\times 2$ blocks take the form 
\begin{equation}
\label{eq:hamDQD}
    h_Z= \left(\begin{array}{cc} \epsilon/2 & t_Z \\
    t_Z^* & -\epsilon/2  \end{array}\right).
\end{equation}
Unlike the single-dot case, the block-diagonal structure is now directly related to the operator of interest, $\hat Z$. At first sight, this may seem to simplify readout based on monitoring the charge of one of the quantum dots.

However, this is not the case and our analysis of the readout via a single quantum dot carries over to the present case with only small changes. In particular, the time-averaged measurement signal of the quantum point contact does not distinguish between the two $Z$ values, unless there is relaxation in the energy eigenbasis. This is because the measurement attempts to project the quantum dot into a charge eigenstate of one of the quantum dots, which is not an eigenstate, thus causing Rabi oscillations of the charge between the quantum dots. In the stationary limit, the system explores both charge states, $\ket{1,0}$ and $\ket{0,1}$, with equal probability and the ensemble-averaged charge becomes equal to 1/2, independent of $Z$. The similarities with the single-dot setup are, of course, rooted in the fact that the Hamiltonians (\ref{eq:sector_hams}) and (\ref{eq:hamDQD}) for the single and double-dot setups, respectively, are closely analogous. 

Despite these similarities, the present setup may have some advantages which could compensate for the additional effort. First, the diagonal elements of the Hamiltonian (\ref{eq:hamDQD}) can now be tuned by a gate, making a wider parameter range accessible. Second, the double quantum dot presumably couples efficiently to the electromagnetic environment, which induces relaxation in the energy basis and enables readout of the qubit via the average measurement current. Third,
the setup obviates the need for resetting the qubit as electrons enter the Majorana qubit only virtually. 

\subsubsection{Readout of four-Majorana parities}
\label{sec:four_majorana_readout}

\begin{figure}
    \centering
    \includegraphics[width=0.9\columnwidth]{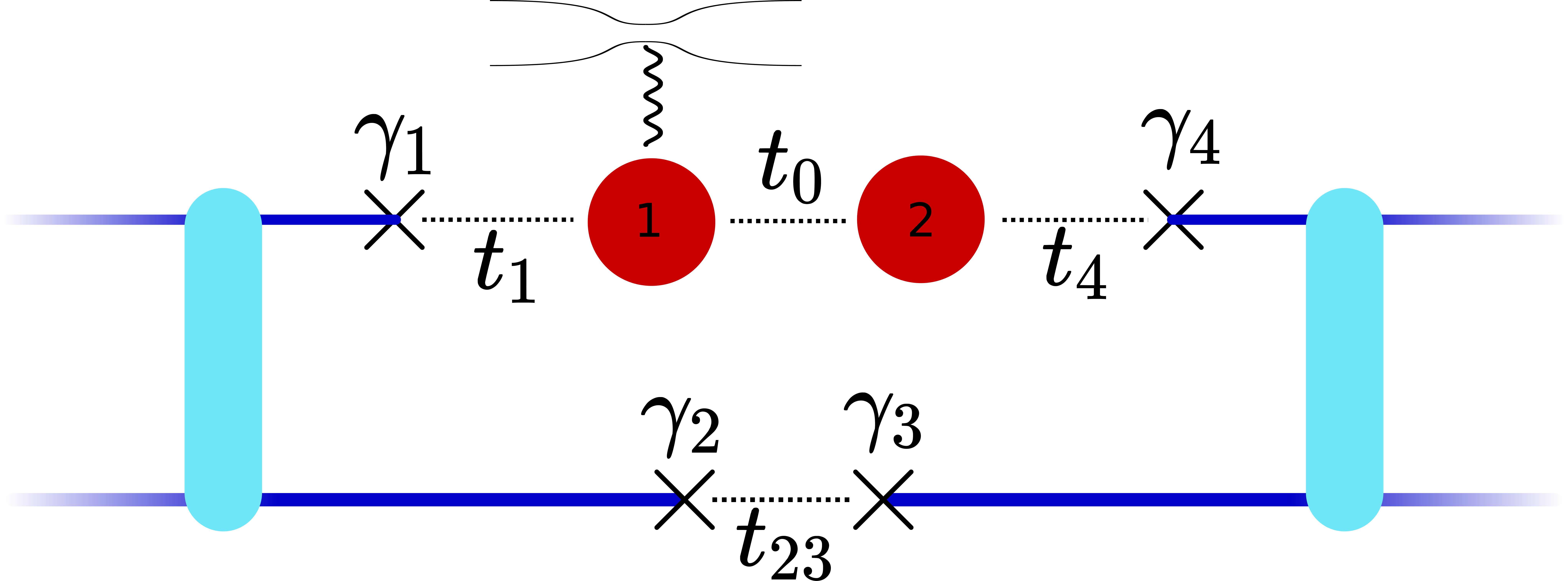}
    \caption{Four-Majorana readout by charge measurements on a double quantum dot. The Majorana bound states $\hat{\gamma}_2$ and $\hat{\gamma}_3$ are tunnel coupled directly via the tunneling link $t_{23}$. Symbols as in Fig.\ 1.}
    \label{fig:four_majorana_measurement}
\end{figure}

Universal quantum computing requires a gate which entangles qubits such as the controlled NOT. For Majorana qubits, the entangling gate can be implemented using measurements of two-qubit Pauli operators \cite{Zilberberg2008,karzig_scalable_2017,litinski_quantum_2018}, say $\hat{Z}_1 \hat{Z}_2$, where $\hat{Z}_1 = -i \hat{\gamma}_1\hat{\gamma}_2$ and $\hat{Z}_2 = -i \hat{\gamma}_3\hat{\gamma}_4$, cf.\ Fig.\ \ref{fig:four_majorana_measurement}. This requires measurements of products of four Majorana operators. Measurements of Majorana parities with even more operators are required to read out stabilizer operators of various topological error correcting codes \cite{Oreg2020}. 

Measurements of four-Majorana parities can be implemented using double quantum dots as in Sec.\ \ref{sec:DQDreadout}, replacing the tunneling path through a single Majorana qubit in Fig.\ \ref{fig:tetron_double_dot_with_readout} by a tunneling path through a sequence of two Majorana qubits, as shown in Fig.\ \ref{fig:four_majorana_measurement}. If the path involves all four Majoranas included in $\hat{Z}_1 \hat{Z}_2$, the corresponding tunneling amplitude becomes
\begin{equation}
    t_{\hat{Z}_1\hat{Z}_2} = t_0+\frac{t_1 t_{23} t^*_4}{E^2_c} \hat{Z}_1\hat{Z}_2. 
\end{equation}
By analogy with our discussion in Sec.\ \ref{sec:DQDreadout}, the quantum dot charge measurement leads to decoherence in the eigenbasis of $\hat{Z}_1\hat{Z}_2$. At the same time, the density matrix remains unaffected within the diagonal blocks of fixed two-qubit parity $\hat{Z}_1\hat{Z}_2$, so that no information is gained on ${\hat Z}_1$ or ${\hat Z}_2$. Clearly, this can, at least in principle, be extended to the measurement of larger products of Majorana operators. 

\section{Symmetry protected readout}\label{sec:general_readout_analysis}

We found in Sec.\ \ref{sec:majorana-qubit_readout} that even though $\hat{Z}$ was not a conserved quantity and the measurement device was coupled to $\hat{n}$, we could read out $\hat Z$ by effectively extracting the combined local parity $\hat{\pi}$ which is a symmetry of both the system and the measurement Hamiltonian. This is a special case of a more general result (see, e.g., \cite{baumgartner_analysis_2008,Albert2014}). If an operator $\hat\Pi$ commutes with both, the Hamiltonian, $[\hat{H},\hat{\Pi}] = 0$, and the full set of jump operators describing the measurement and decoherence channels, $[\hat{L}_\alpha,\hat{\Pi}] = 0$, the system generically decoheres in the $\hat{\Pi}$ basis. In particular, decoherence occurs as long as the measurement current distinguishes between the eigenspaces of $\hat{\Pi}$ \cite{molmer_hypothesis_2015}. Before justifying the validity of this statement, we further illustrate its usefulness by additional applications to Majorana qubits. 

It was shown by Akhmerov \cite{akhmerov_topological_2010} that coupling Majorana zero modes $\hat\gamma_i$ to other fermionic quasiparticles $\hat\alpha_{i,k}$ localized in their vicinity is not detrimental to topological protection. Due to their localized nature, the quasiparticles do not couple distant Majoranas and the operators
\begin{equation}
    \hat{\gamma}'_i = \hat{\gamma}_i (-1)^{\hat{N}_i}
\end{equation}  
with $\hat{N}_i = \sum_{k} \hat{\alpha}^{\dagger}_{i,k} \hat{\alpha}_{i,k}$ are dressed but protected zero modes of the system which commute with the Hamiltonian. This was recently studied further for a specific model in Ref.\ \cite{munk_fidelity_2019}.

For these dressed zero modes to be useful for topological quantum computation, we need to be able to use them in Majorana qubits and to perform projective measurements of corresponding qubit operators such as $\hat{Z}' = -i\hat{\gamma}'_1 \hat{\gamma}'_2$ \cite{litinski_quantum_2018}. The general statement mentioned above implies that this is indeed possible. Consider a measurement of $\hat{Z}$ by coupling $\hat{\gamma}_1$ and $\hat{\gamma}_2$ to a quantum dot as before. We can define a modified combined local parity 
\begin{equation}
    \hat{\pi}' = \hat{Z}(-1)^{\hat{n}+ \hat{N}_1 + \hat{N}_2} =  \hat{Z}' (-1)^{\hat{n}},
\end{equation}
which includes the localized quasiparticles. Unlike $\hat\pi$, the modified combined parity $\hat{\pi}'$ is a symmetry of the system in the absence of processes coupling to other Majorana bound states or changing the charge $\hat{n}+ \hat{N}_1 + \hat{N}_2$. A measurement which distinguishes between the two eigenspaces of $\hat{\pi}'$ will then no longer decohere the system in the eigenbasis of $\hat{\pi}$, but in the eigenbasis of $\hat{\pi}'$, as required for a projective readout of a qubit based on the dressed zero modes. There may, however, be a reduction in the readout speed, as the coupling to other localized modes reduces the hybridization of the zero mode with the quantum dot. 

Our description of the measurement process in terms of the stochastic master equation (\ref{eq:sme_mbq-qd-qpc}) assumes a large bias applied to the quantum point contact, which might cause unnecessary heating of the quantum dot-Majorana qubit system as a consequence of the measurement. The general statement above implies that this assumption, although technically convenient, is unnecessary. Inspecting the derivation of the stochastic master equation in App.\ \ref{app:derivation}, we see that relaxing this assumption will change the argument of the decoherence operator $\mathcal{D}[\hat{n}]$ in Eq.\ (\ref{eq:sme_mbq-qd-qpc}). Nevertheless, $\hat{\pi}$ is conserved by all interactions and thus necessarily by the argument of $\mathcal{D}$, as well. Then, the system still decoheres in the $\hat{\pi}$ eigenbasis. It is worthwhile noting, however, that for smaller bias voltages the argument of $\mathcal{D}$ will in general no longer be hermitian and the associated steady state will not be completely mixed within each $\pi$ subspace.

Now we turn to justifying the general statement. If, for simplicity, the symmetry squares to one, $\hat{\Pi}^2 = \mathds{1}$, the unconditional master equation decouples into blocks labeled by the eigenvalues of $\hat{\Pi}$ (cp.\ Sec.\ \ref{sec:majorana-qubit_readout}),
\begin{subequations}
\begin{align}
\dot{\rho}_{\pi\pi} =&\ -i\bqty{h_\pi,\rho_{\pi\pi}} + \sum_{\alpha} k_{\alpha} \mathcal{D}
\bqty{ l^{\alpha}_\pi }\rho_{\pi\pi} \nonumber \\ 
=&\ \mathcal{L}_{\pi\pi}\rho_{\pi\pi}, \label{eq:++_block_meq_general}\\
    \dot{\rho}_{+-} =&\ -i\pqty{h_{+}\rho_{+-} -\rho_{+-} h_{-}} + \sum_{\alpha} \tilde{\mathcal{D}}\bqty{l^{\alpha}_+,l^{\alpha}_-} \rho_{+-}  \nonumber \\ 
=&\ \mathcal{L}_{+-}\rho_{+-}. \label{eq:+-_block_meq_general}
\end{align}
\end{subequations}
Here, we use the notation $\tilde{\mathcal{D}}\bqty{A,B} \rho = A\rho B^{\dagger} - (A^{\dagger}A\rho + \rho B^{\dagger}B)/2$ and decompose $\hat{L}_{\alpha} = \textrm{diag}[ l^{\alpha}_+ , l^{\alpha}_- ]$ as well as $\hat{H} = \textrm{diag}[ h_+ , h_- ]$. Just as in Sec.\ \ref{sec:majorana-qubit_readout}, the diagonal blocks have Lindblad form and the evolution preserves the weights in the respective blocks. 

We then need to understand when $\mathcal{L}_{+-}$ leads to a decay of $\rho_{+-}$.  Baumgartner and Narnhofer \cite{baumgartner_analysis_2008} show that nontrivial off-diagonal steady states exist if and only if there is a unitary \begin{equation}
    \hat{U}= \begin{pmatrix}
    0 & u^{\dagger} \\ u & 0
    \end{pmatrix}
\end{equation}
connecting the two subspaces, $\hat{U}\hat{P}_+ = \hat{P}_-\hat{U}^{\dagger}$, which commutes with the Hamiltonian and all the $\hat{L}^{\alpha}$. Here, $\hat{P}_\pm$ denotes the projectors onto the two eigenspaces of $\hat\Pi$ and $u$ is a unitary acting on the $\hat\Pi$ eigenspaces.
Then, one has 
\begin{equation}
    h_- = u^{\dagger} h_+ u\,\, , \,\, l^{\alpha}_- = u^{\dagger} l^{\alpha}_+ u,
\end{equation}
so that both the spectra of the Hamiltonians and the algebras formed by $\Bqty{h_{\pm},l^{\alpha}_{\pm}}$ are identical. This implies that the two sectors are unitarily equivalent and the associated measurement currents are indistinguishable. The existence of such a unitary $U$ requires finetuning. Generically, the subspaces are not related in this manner and the measurement signals distinguish between the two sectors. Then, decoherence occurs in the eigenbasis of $\hat{\Pi}$. 

This holds true regardless of the details of the measurement procedure. For instance, one could alternatively base the charge measurement on circuit-QED reflecto\-metry, where the coupling to the quantum dot charge takes the form 
\begin{equation}
    \hat{H}_{\textrm{cQED}} = g\ \hat{n} \pqty{\hat{a}^{\dagger}_0 + \hat{a}_0}.
\end{equation}
Here, $\hat{a}_0$ annihilates a bosonic resonator mode and $g$ quantifies the coupling strength. Since the coupling respects the symmetry $[\hat{H}_{\textrm{cQED}},\hat{\pi}] = 0$, this generically decoheres the system in the eigenbasis of $\hat{\pi}$. This emphasizes that it is really the symmetry that counts, not the details of the measurement, and we refer to this mechanism as symmetry protected decoherence or symmetry protected readout. 

While decoherence generically occurs in the eigenbasis of $\hat{\pi}$, the decoherence rates depend on the specifics of the measurement and can be linked to the rate at which it is possible to distinguish the measurement signals of the two sectors \cite{molmer_hypothesis_2015}. In particular, $\abs{\textrm{tr} \rho_{+-}(t)}^2$ is closely related to the probability to correctly identify the measurement outcome from the measurement signal up to time $t$. Since this assumes an ideal measurement, the decay of $\rho_{+-}$ generally provides only bounds on the measurement time.

\section{Discussion}\label{sec:discussion}

Readout of Majorana-based topological qubits is an important problem and has attracted much attention in the literature for qubit designs based on Coulomb-blockaded superconducting islands \cite{karzig_scalable_2017,plugge_majorana_2017,grimsmo_majorana_2019,Qin2019} or alternative settings \cite{flensberg_non-abelian_2011,yavilberg_fermion_2015,ohm_microwave_2015,aasen_milestones_2016,gharavi_readout_2016,zhou2020doubledot}. Its importance is rooted in the fact that promising schemes for Majorana-based quantum computation \cite{karzig_scalable_2017,litinski_quantum_2018,Oreg2020} rely on measurements as an integral part of quantum information processing. This implies that the measurements must not only provide the measurement outcome, but also reliably project the qubit into the corresponding eigenstate. 

A variety of techniques have been proposed to read out Majorana qubits, including interferometry of transport currents passed through the Majorana qubit \cite{plugge_majorana_2017,Qin2019,zhou2020doubledot}, techniques borrowed from circuit quantum electrodynamics \cite{yavilberg_fermion_2015, ohm_microwave_2015, plugge_majorana_2017, karzig_scalable_2017, grimsmo_majorana_2019}, or measurements relying on charge sensing \cite{flensberg_non-abelian_2011,aasen_milestones_2016,gharavi_readout_2016,plugge_majorana_2017,karzig_scalable_2017,szechenyi_parity_2019}. Techniques borrowed from circuit quantum electrodynamics can frequently be treated theoretically in close analogy to the description for (nontopological) superconducting qubits \cite{blais_cavity_2004}. At the same time, these schemes involve a substantial hardware overhead and may significantly increase the effective dimensions of each qubit. 
Therefore, we focused here on readout of Majorana qubits based on coupling to a quantum dot whose charge is measured by means of a quantum point contact. This approach combines suitability to the basic design of Majorana qubits with conceptual simplicity, and thus relevance for near-term devices with accessibility of a thorough theoretical analysis at an analytical level.

Despite its apparent simplicity, this scheme poses nontrivial questions. In particular, we discuss charge-based readout protocols of parity-protected Majorana qubits which are distinctly different from charge-based readout protocols of other types of qubits. Spin qubits (by spin-charge conversion) \cite{petta_coherent_2005} or Majorana qubits without parity protection \cite{aasen_milestones_2016} (by parity-to-charge conversion) can also be effected by charge measurements. In these cases, the computational basis of the qubit is robustly brought into one-to-one correspondence with the charge basis. In contrast, the charge-based readout of parity-protected Majorana qubits projects in the charge basis, while the qubit operator enters through a tunneling Hamiltonian which does not commute with the charge. Generically, this makes a single projective charge measurement insufficient to identify the qubit state. Moreover, readout by repeated charge measurements would necessitate very high levels of control. Instead, the readout process is a weak continuous measurement and its theoretical description requires a time-resolved description of the measurement. 

A systematic measurement theory of this readout scheme for (parity-protected) Majorana qubits is the central contribution of this paper. Our theory reveals under which conditions a measurement of the quantum dot charge constitutes a projective measurement of the Majorana parity of the qubit, describes the time it takes to decohere the system in the measurement basis, and includes the noisy measurement signal which can be analyzed to estimate required measurement times. 

Our central insight is that generically, one does not directly measure the Majorana parity but rather a combined parity which includes the quantum dot charge in addition to the Majorana parity of the qubit. We find that this is not detrimental to the readout as the combined parity can eventually be converted into the desired Majorana parity. Importantly, this observation generalizes and our theory also applies more generally. In particular, this implies that other local charges which the Majorana might couple to are not a hindrance to topological protection. Topological quantum computation, including qubit readout, can be based on dressed zero modes which include these additional local charges. In its general form, the underlying result states that decoherence generically occurs in the eigenbasis of operators which commute with the system and readout Hamiltonians, which we refer to as symmetry protected decoherence or readout.

The theory also describes how to extract the measurement outcome from the measurement current through the quantum point contact. We find that generically, the measurement outcome cannot be reconstructed from the average measurement current, but only from its noise correlations. This can only be avoided when including additional dissipative processes, or by exploiting transient signals in fine-tuned situations. This surprising result can be traced back to the fact that the quantity to be read out enters into a tunneling amplitude which does not commute with the measured quantity, namely the quantum dot charge. We emphasize that for readout based on a single quantum dot, the Majorana qubit is in an excited state for a significant fraction of the measurement time. This may make the procedure susceptible to qubit errors by uncontrolled electron tunneling. This can be avoided in a measurement setup using a double quantum dot, which may thus promise better readout fidelities. 

Finally, the theory naturally provides estimates of the measurement time. We find that the measurement times for Majorana qubit readout based on setups with a single quantum dot are consistently considerably larger than those for a conventional quantum dot charge readout at the same measurement strength. The reason for this is twofold. First, the decoherence rates are no longer simply controlled by the measurement strength $\sim k$ but involve the small tunnel couplings between quantum dot and Majorana qubit, $\sim \abs{t}^2 k / \varepsilon^2 \ll k$. Second, one generically cannot access the entire information contained in the measurement signal. Instead, typical experiments will only have access to its mean and two-point correlations. Thus, actual measurement times can be large compared to the decoherence time. However, slow readout should not be a generic feature of Majorana qubit readout. In fact, readout setups with two quantum dots can access an increased parameter range and should be less restricted. Investigating the double-dot setup in greater detail represents an interesting direction for future work.

{\em Note added:} Recently, we became aware of related unpublished work \cite{munk_parity_2020}, which contains a partly complementary analysis and reaches similar conclusions where overlapping.

\acknowledgments{We thank Francesco Albarelli, Andrew Doherty, Reinhold Egger, Karsten Flensberg, Marco Genoni, Daniel Litinski, and Morten Munk for useful discussions. We also thank Reinhold Egger and Karsten Flensberg for sharing their manuscript prior to submission. We gratefully acknowledge financial support by QuantERA project {\em Topoquant} as well as by Deutsche Forschungsgemeinschaft through CRC 910 and CRC 183.}

\bibliography{refs}

\appendix
\section{Majorana qubits and quantum dot -- notation and definitions}\label{app:definitions}

In this appendix, we summarize details of our definitions and conventions for the quantum states of the coupled system of Majorana qubit and quantum dot. 

\subsection{Choice of basis states}

A set of $2m$ Majorana bound states labeled by $i=1,\ldots,2m$ is described by hermitian fermionic operators $\hat{\gamma}_i = \hat{\gamma}_i^{\dagger}$ which satisfy  $\{\hat{\gamma}_i,\hat\gamma_j\} = 2\delta_{ij}$. The associated $2^m$-dimensional  Hilbert space is spanned by the Fock occupations $\hat{n}_{ij} = \hat{f}_{ij}^{\dagger} \hat{f}_{ij}$ of complex fermions $\hat{f}_{ij}= \pqty{\hat{\gamma}_i + i\hat{\gamma}_j}/2$.  

The total fermion parity of these basis states is given by the operator
\begin{equation}\label{eq:parity_op}
   \hat{P} = \pqty{-i}^m\prod_{j=1}^{m}\hat{\gamma}_{2j-1}\hat{\gamma}_{2j} 
\end{equation}
As quantum superpositions exist only for states of the same fermion parity, a qubit requires at least four Majorana bound states. States with even and odd fermion parity can be split energetically by a charging energy, Eq.\ \eqref{eq:charging_ham}, as they have different charge $N=2N_C + N_{M}$, where $N_C$ is the number of Cooper pairs and $N_M$ the charge in the Majorana sector. 
For definiteness, we choose the ground (excited) states of the Majorana qubit to have $N=0$ ($N=-1$). The Hilbert space is spanned by $\Bqty{\ket{N,n_{12},n_{34}}}$. Due to the parity constraint
\begin{equation}\label{eq:parity_constraint}
    P = \pqty{-1}^{N} = \pqty{-1}^{n_{12}+n_{34}},
\end{equation}
it is sufficient to specify the state as $\ket{N,n_{12}}$. 

In the main text, we choose a slightly different labeling of the basis states. First, instead of using the label $n_{12}$, we specify $n_{12}$ via the eigenvalue $z \in \Bqty{1,-1} \equiv \Bqty{\uparrow,\downarrow}$ of the Pauli-$Z$ operator
\[
\hat{Z} = -i\hat{\gamma}_1\hat{\gamma}_2 = 1-2\hat{n}_{12} = (-1)^{\hat{n}_{12}}.
\]
Second, the Majorana qubit only exchanges charge with the quantum dot. We always initialize the system to have even total parity with Majorana-qubit charge $N=0$ and quantum dot charge $n=0$, so that these charges are related as $n=-N$ in general. We then label the basis states by the quantum dot charge $n$ and the eigenvalue $z$ of the Pauli-$Z$ operator of the Majorana qubit, 
\begin{equation}
    \ket{z,n} \equiv \ket{N =-n, n_{12} = \frac{1-z}{2} , n_{34} = \frac{1-z(-1)^n}{2} }.
\end{equation}
With this definition, the constraint \eqref{eq:parity_constraint} is automatically satisfied. Furthermore, with our choices, all participating states have even total fermion parity $P_{\textrm{tot}} = P(-1)^n$.  

The four basis states $\ket{z,n}$ differ in the local parity $\hat{\pi} = \hat{Z} (-1)^{ \hat{n}}$. We choose a basis $\{\ket{\uparrow,0},\ket{ \downarrow,1},\ket{\downarrow,0},\ket{\uparrow,1}\}$, where the first (last) two states have local parity $\pi=+1$ ($\pi=-1$). We can then define Pauli matrices acting in a subspace of fixed local parity, 
\begin{subequations}
\begin{eqnarray}
    \sigma^\pi_z &=& \ketbra{\pi,0}{\pi,0} - \ketbra{-\pi,1}{-\pi,1}, \\
    \sigma^\pi_x &=& \ketbra{\pi,0}{-\pi,1} + \ketbra{-\pi,1}{\pi,0}, \\
    \sigma^\pi_y &=& i\sigma^\pi_x \sigma^\pi_z. 
\end{eqnarray}
\end{subequations}
In the main text, we drop the $\pi$ label whenever this does not lead to confusion. In terms of these Pauli matrices, the quantum dot charge is
\begin{equation}
    n = \frac{1}{2}\pqty{\sigma_0 - \sigma_z}.
\end{equation}

\subsection{Coupled Majorana qubit-quantum dot system}

In addition, we define a set of Pauli matrices using the eigenbasis of the Hamiltonian in Eq.\ \eqref{eq:sector_hams} which also act within the subspaces with fixed $\pi$, 
\begin{subequations}
\begin{eqnarray}
    \tau^\pi_z &=& \ketbra{g_\pi}{g_\pi} - \ketbra{e_\pi}{e_\pi}, \\
    \tau^\pi_x &=& \ketbra{e_\pi}{g_\pi} + \ketbra{g_\pi}{e_\pi} = \tau^{\pi}_+ + \tau^{\pi}_-, \\
    \tau^\pi_y &=& i\tau^\pi_x\tau^\pi_z. 
\end{eqnarray}
\end{subequations}
Here, the ground states are given by Eq.\ (\ref{eq:ground_states}), while the excited states are 
\begin{subequations}
\begin{eqnarray}
    \ket{e_+} &=& \cos\frac{\theta_+}{2} \ket{\uparrow,0}  +\sin\frac{\theta_+}{2}e^{i\phi_+} 
    \ket{\downarrow,1},
    \\
    \ket{e_-} &=& \cos\frac{\theta_-}{2} \ket{\downarrow,0}  +\sin\frac{\theta_-}{2}e^{i\phi_-} 
    \ket{\uparrow,1}.
\end{eqnarray}
\end{subequations}
In terms of these Pauli matrices, $h_{\pi} = \Omega_\pi \tau^{\pi}_z$ and 
\begin{align*}
    \hat{n} =&\ \sum_{z=\pm 1} \ketbra{z,1}{z,1} \\
    =&\ \sum_{\pi=\pm 1}\Big[ \cos^2\frac{\theta_\pi}{2}\ketbra{g_\pi}{g_\pi} + \sin^2\frac{\theta_\pi}{2}\ketbra{e_\pi}{e_\pi} \nonumber \\
    &\ \ \ \ \ \ \ - 2\cos\frac{\theta_\pi}{2}\sin\frac{\theta_\pi}{2}\pqty{ \ketbra{e_\pi}{g_\pi} + \ketbra{g_\pi}{e_\pi}   } \Big] \nonumber \\
    =&\ \frac{1}{2} \sum_{\pi=\pm 1} \bqty{\tau^\pi_0 + \cos\theta_\pi \tau^\pi_z - \sin\theta_\pi \tau^\pi_x}.
\end{align*}
In the Heisenberg picture, the quantum dot charge $\hat{n}(t)=\textrm{exp} (i\hat{H}t)\  \hat{n}\ \textrm{exp}(-i\hat{H} t)$ becomes 
\begin{multline}\label{eq:n_heisenberg}
    \hat{n}(t) = \sum_{\pi=\pm 1}\Big[ \frac{\tau^\pi_0 + \cos\theta_\pi \tau^\pi_z}{2} \\
    - \frac{\sin\theta_\pi}{2} \pqty{ e^{2i\Omega_{\pi}t} \tau^{\pi}_+ + e^{-2i\Omega_{\pi}t} \tau^{\pi}_- } \Big].
\end{multline}
We find it convenient to define the time-independent part
\begin{equation}\label{eq:n_time_independent_part}
    \hat{c} = \sum_{\pi=\pm 1} \frac{\tau^\pi_0 + \cos\theta_\pi \tau^\pi_z}{2}.
\end{equation}
Finally, since we are performing our simulations in the charge basis, we need $\tau_{\pm}^{\pi}$ in the charge basis. Within a given $\pi$-block, it takes the form 
\begin{equation}
    \tau_-^\pi = \frac{1}{2} \begin{pmatrix}
    \sin\theta_\pi & \pqty{1-\cos\theta_\pi}e^{-i\phi_\pi} \\
    \pqty{-1-\cos\theta_\pi}e^{i\phi_\pi} & -\sin\theta_\pi,
    \end{pmatrix}
\end{equation}
with $\tau_+^\pi$ given by hermitian conjugation.

\section{Stroboscopic protocol for $\hat{Z}$ readout}
\label{app:repeated_measurements}

\subsection{Measurement protocol}

Section \ref{sec:majorana_qubits_and_quantum_dots} discusses projective measurements of $\hat{Z}$ based on a single projective measurement of $\hat{n}$. In addition to assuming instantaneous charge readout, this requires fine-tuned parameters. In subsequent sections of the main text, we relax both of these conditions. Here, we briefly discuss schemes which assume instantaneous charge readout, but allow for general system parameters $t_i$ and $\varepsilon$. For general parameters, the measurement outcome for the charge of the quantum dot is no longer perfectly correlated with eigenstates of $\hat{Z}$. However, unless $\Im{t_1t_2^*} = 0$, the charge measurement still provides partial information on the qubit state. 

Once coupled, quantum dot and Majorana qubit evolve unitarily from initial state $\ket{\psi} = \pqty{\alpha\ket{\uparrow} + \beta \ket{\downarrow}}\ket{0}$ into 
\begin{align}
    \hat{\mathcal{U}} \ket{\psi}  = &\ 
     \alpha\pqty{ c_{\uparrow ,0} \ket{\uparrow, 0} + c_{\downarrow, 1} \ket{\downarrow, 1} } \nonumber \\
     &+ \beta\pqty{ c_{\downarrow, 0} \ket{\downarrow, 0} + c_{\uparrow, 1} \ket{\uparrow ,1} }.
\end{align}
The unitary time evolution $\hat{\mathcal{U}}$, entangling Majorana qubit and quantum dot, satisfies $[\hat{\mathcal{U}},\hat{\pi}] = 0$, but otherwise depends on details of the protocol. If the charge readout can be effected instantaneously (on the time scale of the hybridization between dot and qubit), or more realistically, qubit and quantum dot are rapidly decoupled following the unitary evolution, the quantum dot charge becomes a good quantum number during charge readout.
Then, the measurement leaves the system in the state 
\begin{align*}
\ket{\psi'_0}=& \frac{1}{\sqrt{p_0}}\pqty{ \alpha c_{\uparrow, 0} \ket{\uparrow} + \beta c_{\downarrow, 0} \ket{\downarrow} } \ket{0},
\end{align*}
or
\begin{align*}
\ket{\psi'_1}=& \frac{1}{\sqrt{p_1}}\pqty{ \alpha c_{\downarrow, 1} \ket{\downarrow} + \beta c_{\uparrow, 1} \ket{\uparrow} } \ket{1},
\end{align*}
with probabilities $p_0 = \vert \alpha c_{\uparrow 0} \vert^2 + \vert \beta c_{\downarrow 0} \vert^2$ and $p_1 = 1- p_0$, respectively. The partial information on $\hat{Z}$ obtained from the measurement transfers weight between qubit states. This can be interpreted in terms of Bayesian inference \cite{korotkov_continuous_1999}, 
\[
p(\uparrow\vert n) = \frac{p (n \vert \uparrow) }{p_n} p(\uparrow),\ p(\downarrow \vert n) = \frac{p(n \vert \downarrow)}{p_n} p(\downarrow),
\]
where we identify the prior probabilities of the qubit states with $p(\uparrow) = \abs{\alpha}^2$ and $p(\downarrow) = \abs{\beta}^2$ and $p( n \vert Z)  = \vert c_{Z n}\vert^2$ with the conditional probabilities to observe measurement outcome $n$. 

Repeating this protocol results in a random walk in the space of qubit and quantum dot states. We find that the random walk has two distinct steady states corresponding to the $\hat{\pi}$ eigensectors. This becomes equivalent to the eigenstates of $\hat{Z}$, if for simplicity, we reset the qubit-quantum dot system in between steps, $\ket{\downarrow, 1} \to \ket{\uparrow, 0}$ and $\ket{\uparrow, 1} \to \ket{\downarrow, 0}$ after every charge readout that gave $n=1$ (for a scheme implementing this reset, see App.\ \ref{app:qd_reset} below). With this reset, we effectively obtain a random walk in the space of qubit states, which can be described by the Kraus operators
\begin{align}\label{eq:kraus}
\hat{M}_0=&
\begin{pmatrix}
c_{\uparrow 0} & 0 \\ 0 & c_{\downarrow 0}
\end{pmatrix},\ 
 \hat{M}_1= \hat{X}
\begin{pmatrix}
0 & c_{\uparrow 1} \\ c_{\downarrow 1} & 0
\end{pmatrix}
\end{align}
acting on a qubit state $\ket{\phi}$, so that the qubit state $\ket{\phi'_n}$ conditioned on the measurement outcome $n$ is given by 
\begin{equation}
    \ket{\phi'_n} = \frac{\hat{M}_n \ket{\phi}}{\sqrt{p_n}},
\end{equation}
with $p_n = \bra{\phi} \hat{M}_n^{\dagger} \hat{M}_n \ket{\phi}$. The Pauli $\hat{X}$ in the definition of $\hat{M}_1$ makes the reset of the qubit-quantum dot system explicit.

\begin{figure}
    \centering
        \includegraphics[width=0.9\columnwidth]{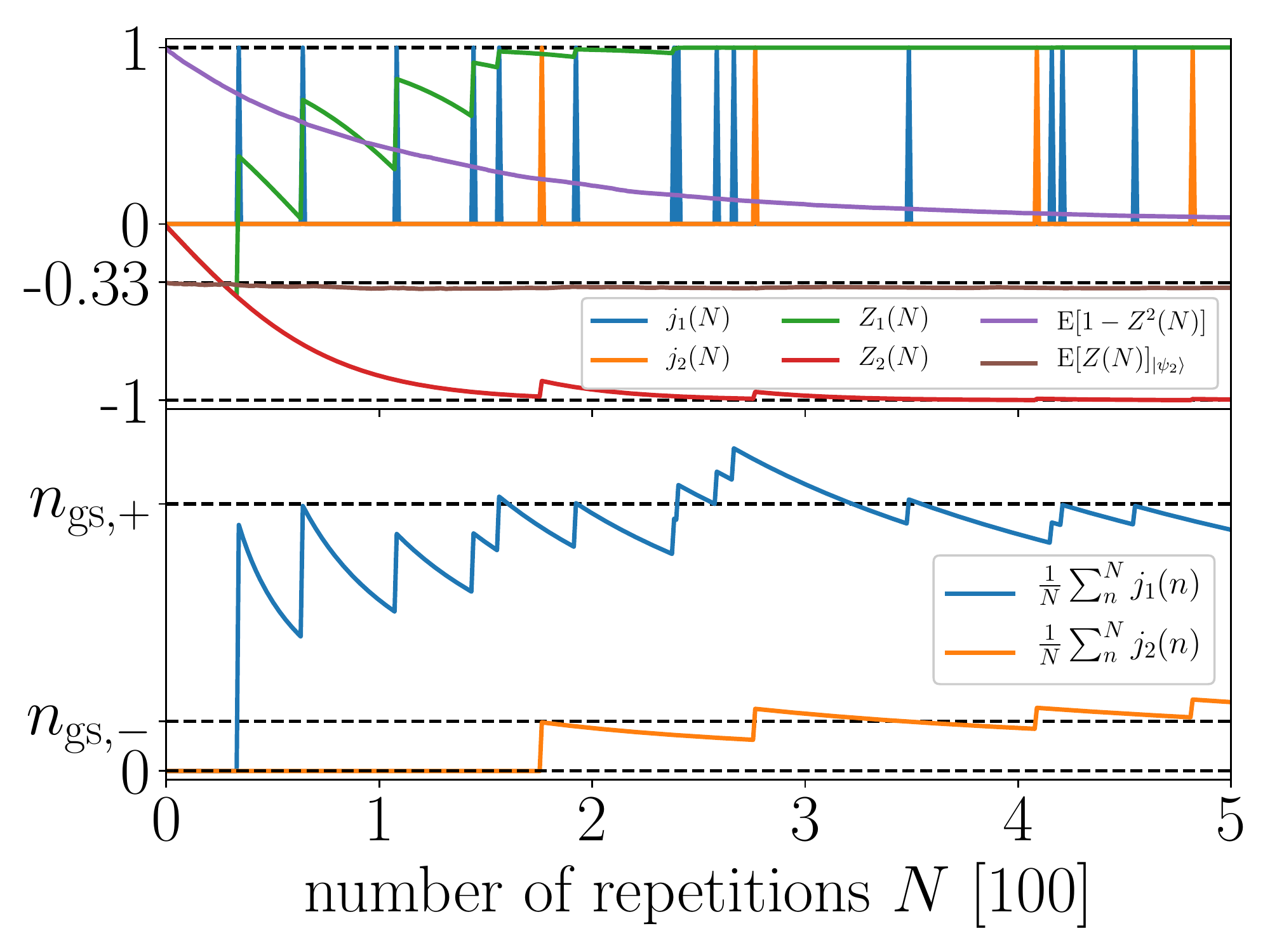}
    \caption{Evolution under the stroboscopic measurement protocol for adiabatic coupling with initial state $\ket{\psi} = (\ket{\uparrow}+\ket{\downarrow})\ket{0}/\sqrt{2}$, as function of time (number of iterations). Sample trajectories of $Z = \langle \hat{Z} \rangle$ (green and red traces) demonstrate the evolution towards the fixed points. The evolution tends slowly towards the $\downarrow$ state (corresponding to likely outcomes), interrupted by jumps (unlikely outcomes) that provide much more information, favoring the $\uparrow$ state and causing greater backaction. The measurement signal for two trajectories is shown by the blue and orange traces. Jumps in the trajectory are correlated with $n=1$ outcomes. The correct statistics of the measurement is illustrated by the constant ensemble average of $Z(N)$ (brown trace), which we computed for the initial state $\ket{\psi_2}  = (\ket{\uparrow}+\sqrt{2}\ket{\downarrow})\ket{0}/\sqrt{3}$. Projection onto the fixed points is indicated by the decay of $\mathbb{E}[1-Z^2(N)]$ (purple trace) to zero. Ensemble averages are over $10^4$ trajectories. Parameters: $t_2 = t_1 e^{i\pi/4},\varepsilon = 10 t_1$.}
    \label{fig:discrete_majorana_readout_adiabatic}
\end{figure}

Figure \ref{fig:discrete_majorana_readout_adiabatic} shows numerical simulations of this protocol for the adiabatic coupling scheme discussed in Sec.\ \ref{sec:basic_protocols}, so that $\mathcal{\hat{U}}$ involves the coefficients $c_{Z,n}$ of the state given in Eq.\ \eqref{eq:adiabatic_turn_on}. The trajectories reach the fixed points $Z=\pm1$ with the correct probabilities. Moreover, the measurement outcome can be extracted from the sequence of $n$ outcomes of a single trajectory by noting that $n$ averages to the ground-state charge corresponding to the respective fixed point. Thus, this protocol implements a projective measurement of $\hat{Z}$. 

We conclude this section with a number of comments. First, resetting the quantum dot charge is not essential. Without intermediate resetting, the protocol projects $\hat{\pi} = \hat{Z}(-1)^{\hat{n}}$ (see discussion in the main text) and one may reset once the outcome of the measurement is determined. However, in this case, the average charge no longer equals the ground state charge and a more involved signal analysis (for instance using the Bayes theorem) is required. Second, qubit and quantum dot can also be entangled through evolution with the Hamiltonian \eqref{eq:sector_hams}. In this case, decoupling qubit and quantum dot in between steps is necessary only if charge readout is slower than tunneling. The resulting evolution is closely related to the continuous evolution discussed in the main text, which arises naturally from sequences of repeated measurements when relaxing more and more assumptions on the strengths of various couplings. 

\begin{figure}
    \centering
    \includegraphics[width=\columnwidth]{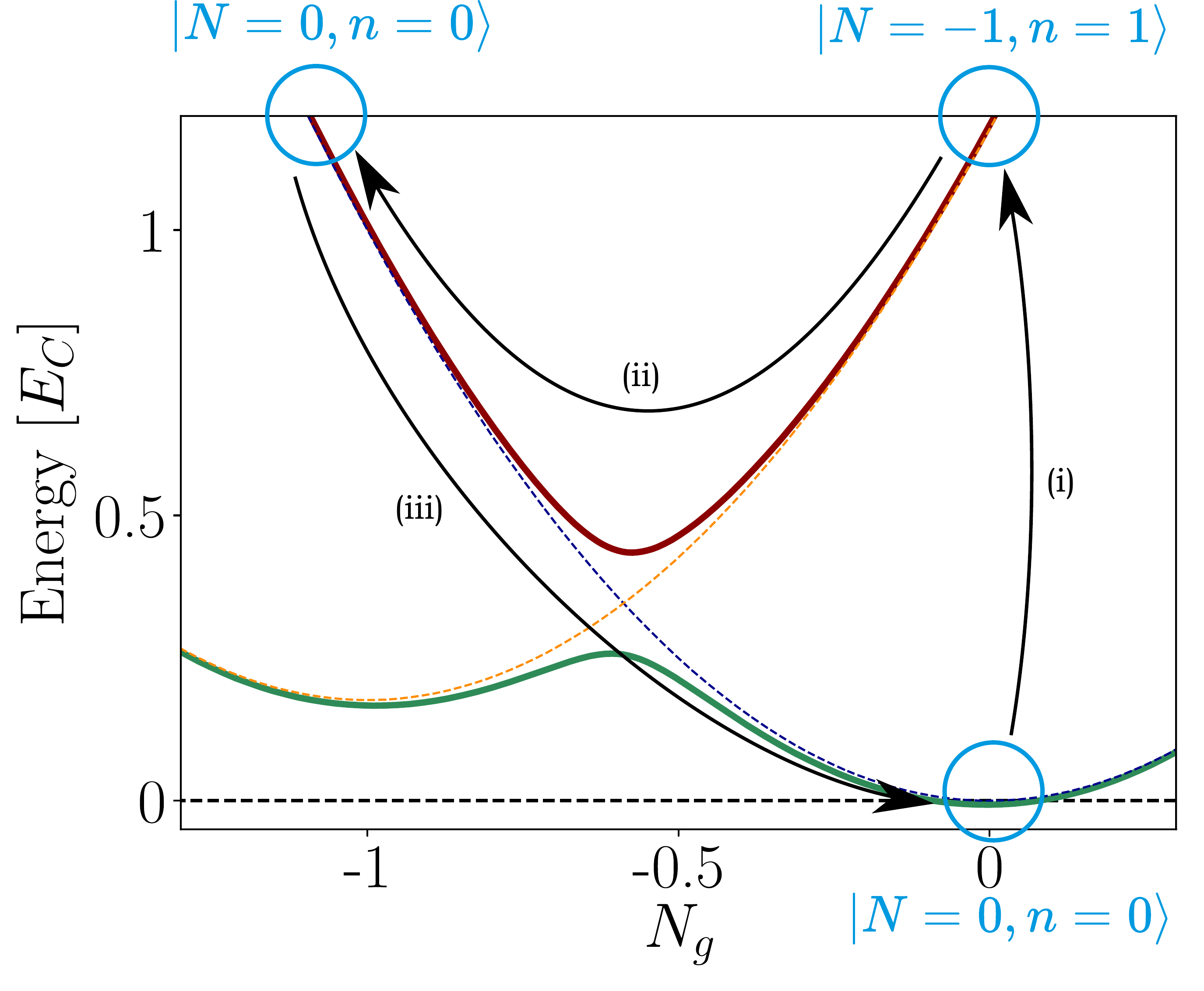}
    \caption{Spectrum of the quantum dot-Majorana qubit system as a function of the qubit gate charge $N_g$. The dashed lines for $N=-1$ (orange) and $N=0$ (blue) correspond to the energies of the system without coupling to the quantum dot. (The vertical shift of the orange curve reflects the energy of the quantum dot.) The full lines (red, green) refer to the coupled system and exhibit an avoided crossing.   Initially, the Majorana qubit gate is tuned to $N_g = 0$ and the system is in the charge ground state  $\ket{\psi} \propto \ket{N=0,n=0}$.  (i) A measurement outcome of $n=1$ transfers the system into the excited charge state $\ket{\psi'} \propto \ket{N=-1,n=1}$. (ii) The charge state is reset by adiabatically changing $N_g \to -1$. The system state is now $\propto \ket{N=0,n=0}$ again. (iii) Suddenly set $N_g \to 0$, so that the system state does not change and the initial situation is restored. Alternatively, one may also decouple the quantum dot and slide down the dashed green curve adiabatically.}
    \label{fig:qd_reset}
\end{figure} 

\subsection{Resetting the qubit-quantum dot system}\label{app:qd_reset}

When the charge measurement yields $n=1$, the Majorana qubit is in the excited charge state $N= - 1$. To avoid uncontrolled charging events, the electron should be swiftly returned from quantum dot to Majorana qubit.

This can be achieved by adiabatic variations of $\varepsilon$ and $t_i$,
transforming  
\begin{equation}
    \ket{\psi} = \pqty{\alpha \ket{\downarrow} + \beta \ket{\uparrow} }\ket{1}
\end{equation}
into
\begin{equation}
    \ket{\psi_{\textrm{reset} }} = \pqty{\alpha \ket{\uparrow} + \beta \ket{\downarrow} }\ket{0}
\end{equation}
for general $\alpha$ and $\beta$. We focus on the subspace $\pi = +1$ for definiteness. 

Without fine tuning of the dynamical phase, $\ket{\downarrow, 1}$ can only be transformed into $\ket{\uparrow, 0}$ if it is an eigenstate of the initial Hamiltonian. Consider an initial Hamiltonian $h_+$ given by $\theta_+ = \theta_0$ and $\phi_+ = \phi_0$, and expand the initial state in eigenstates of $h_+$,
\[
\ket{\downarrow, 1} = e^{-i\phi_0}\pqty{ \sin\frac{\theta_0}{2} \ket{e_+}_0 - \cos\frac{\theta_0}{2} \ket{g_+}_0  }.
\]
Adiabatically changing $\theta_0,\phi_0 \to \theta_1,\phi_1$, the eigenstates evolve as $\ket{e_+}_0 \to e^{i\chi_e} \ket{e_+}_1$ and $\ket{g_+}_0 \to e^{i\chi_g} \ket{g_+}_1$, where the subscripts distinguish eigenstates of the initial ($\theta_0,\phi_0$) and final ($\theta_1,\phi_1$) Hamiltonians. Then 
\begin{equation}
    \ket{\downarrow , 1} \to e^{-i\phi_0}\pqty{ e^{i\chi_e}\sin\frac{\theta_0}{2} \ket{e_+}_1 - e^{i\chi_g} \cos\frac{\theta_0}{2} \ket{g_+}_1 }.
\end{equation}
Writing $\ket{e_+}_1$ and $\ket{g_+}_1$ in the basis $\ket{\uparrow 0}, \ket{\downarrow 1}$ and setting the coefficient of $\ket{\downarrow 1}$ to zero yields the condition
\begin{equation}
    e^{i\chi_g} \sin\frac{\theta_0}{2}\sin\frac{\theta_1}{2} + e^{i\chi_e} \cos\frac{\theta_0}{2}\cos\frac{\theta_1}{2} = 0.
\end{equation}
Without fine tuning, $\chi_g$ and $\chi_e$ are arbitrary phases and the two terms need to vanish separately. This implies $\theta_0 = 0$ and $\theta_1 = \pi$, or vice versa, so that the state $\ket{\downarrow, 1}$ was an eigenstate of $h_+(\theta_0,\phi_0)$ to begin with. 

The charge state of the quantum dot-Majorana qubit system may then be reset from the initial state $\ket{\psi}$ as follows:
\begin{enumerate}
    \item Suddenly decouple quantum dot and Majorana qubit ($t_i = 0$ and $\varepsilon>0$ or $(\theta_{\pm})_0 = \pi$). Then, $\ket{\downarrow, 1} = \ket{e_+}_0$ and $\ket{\uparrow ,1} = \ket{e_-}_0$  are energy eigenstates. The system state $\ket{\psi}$ is now a superposition
    \[
    \ket{\psi} = \alpha \ket{e_+}_0 + \beta\ket{e_-}_0.
    \]
    \item Adiabatically swap $\ket{e_{\pm}}$ and $\ket{g_{\pm}}$, rotating the Hamiltonian $h_{\pm}$ from the south $(\theta_{\pm})_0 = \pi$ to the north pole $(\theta_{\pm})_1 = 0$ of the Bloch sphere, 
    \begin{align*}
         \ket{\psi} \to&\ \alpha \ket{e_+}_1 + \beta e^{i\chi} \ket{e_-}_1 
         \\ &= \alpha \ket{g_+}_0 + \beta e^{i\chi} \ket{g_-}_0,
         \\ &= \pqty{\alpha \ket{\uparrow} + \beta e^{i\chi} \ket{\downarrow}}\ket{0},
    \end{align*}
    where $\chi$ is the relative dynamical phase between the $\hat{\pi}$ eigensectors introduced in the adiabatic evolution. Although uncontrolled, the relative phase does not affect the readout evolution as it preserves the weights in the $\hat{Z}$ eigenbasis. The step requires $t_i \neq 0$ at some point during the evolution to avoid gap closing, but eventually quantum dot and Majorana qubit are again decoupled.
    \item Finally, suddenly reset the gates to their initial values, so that the quantum dot-Majorana qubit system again resides in a superposition of charge ground states. 
\end{enumerate}
This protocol requires a sign flip of $\varepsilon = E_C + \epsilon$, which can be realized by varying the quantum dot energy $\epsilon$ to compensate for the charging energy $E_C$ of the Majorana qubit, or by also varying the spectrum of Majorana-qubit charge states by the gate offset $N_g$, see Fig.\ \ref{fig:qd_reset}. 

The same procedure can be applied at the end of the continuous readout discussed in the main text which may also require a charge reset once the measurement outcome is certain.

\section{Derivation of the stochastic master equation (\ref{eq:sme_mbq-qd-qpc})}\label{app:derivation}

For completeness, we include a derivation of the stochastic master equation  \eqref{eq:sme_mbq-qd-qpc} which describes the evolution of the Majorana qubit-quantum dot system under continuous monitoring of the quantum dot charge by a quantum point contact, see, e.g., Ref.\ \cite{goan_continuous_2001}. 

The Hamiltonian  
\begin{equation}
    \label{eq:qd_ham}
    \hat{H}_{\textrm{readout}} = \hat{H}_{\textrm{leads}} + \hat{H}_{\textrm{jct}} + \hat{n}\ \delta\hat{H}_{\textrm{jct}}.
\end{equation}
of the quantum point contact describes two (left and right) free-fermion leads, 
\[
\hat{H}_{\textrm{leads}} = \sum_{\alpha} \pqty{ \xi_{L\alpha} \hat{c}^{\dagger}_{L\alpha}\hat{c}_{L\alpha} + \xi_{R\alpha} \hat{c}^{\dagger}_{R\alpha}\hat{c}_{R\alpha} },
\]
and a tunneling Hamiltonian 
\begin{equation}
    \hat{\mathcal{V}} = \hat{H}_{\textrm{jct}} + \hat{n}\ \delta\hat{H}_{\textrm{jct}} 
    = \pqty{ \tau + \chi \hat{n} } \hat{\psi}_L^{\dagger} \hat{\psi}_R e^{ieVt}+ \textrm{h.c.},
    \label{Vcoup}
\end{equation}
which includes the coupling to the quantum dot charge $\hat n$. The chemical potentials of the two leads differ by the bias voltage $V$ applied across the quantum point contact, 
\begin{equation}
eV = \mu_L-\mu_R
\end{equation}
The factor $e^{ieVt}$ in the tunneling Hamiltonian (\ref{Vcoup}) already accounts for a time-dependent unitary transformation such that the single-particle energies of the left and right leads are measured from the respective chemical potentials, $\xi_{L\alpha}= \epsilon_{L\alpha} - \mu - eV$ and $\xi_{R\alpha} = \epsilon_{R\alpha} - \mu$, with $\alpha$ labeling the single-particle eigenstates with corresponding electron operators $c_{L\alpha}$ and $c_{R\alpha}$. The electron operators evaluated at the junction position are denoted by $\hat{\psi}_L$ and $\hat{\psi}_R$. Importantly, there is only capacitive coupling, but no charge transfer between quantum dot and quantum point contact. 

At zero temperature, the quantum point contact carries an average current
\begin{equation}\label{eq:current_unocc}
    I_0 = 2\pi\nu_L\nu_R \abs{\tau}^2 eV,
\end{equation}
when the quantum dot is unoccupied, and 
\begin{equation}\label{eq:current_occ}
    I_1 = 2\pi\nu_L\nu_R \abs{\tau+\chi}^2 eV,
\end{equation}
when the quantum dot is occupied. Here, $\nu_{L/R}$ denotes the lead density of states. The sensitivity of the detector depends on $\delta I = I_1 - I_0$. We assume that the quantum dot affects the current weakly, $\delta I \ll I_0$.

\subsection{Unconditional Lindblad master equation}\label{app:derivation_unconditional}

We first derive the unconditional master equation for the quantum dot state following the standard procedure of  tracing over the leads, assuming factorization of the system-lead density matrix at all times (Born approximation), and finally assuming fast decay of the lead correlation functions to obtain a Markovian equation of motion. We will subsequently use the Lindblad equation to identify the Kraus operators, which allows one to derive the conditional master equation which accounts for the monitoring of the quantum-point-contact current.

We start in the interaction picture, with operators and states evolving according to $\hat{H}_0 = \hat{H} + \hat{H}_{\textrm{leads}}$ and $\hat{\mathcal{V}}=\hat{H}_{\textrm{jct}} + \hat{n}\ \delta\hat{H}_{\textrm{jct}}$, respectively. The corresponding density matrix for system and leads, $\hat{\chi}$, satisfies the equation of motion
\begin{multline}\label{eq:IA_pic_eom}
    \frac{\textrm{d}}{\textrm{d}t}\hat{\chi}(t) 
        = -i\  \bqty{ \hat{\mathcal{V}}(t) , \hat{\chi}(t_0) } \\
            -\int_{t_0}^{t}\textrm{d}t'\ \bqty{ \hat{\mathcal{V}}(t) , \bqty{ \hat{\mathcal{V}}(t') , \hat{\chi}(t') } }.
\end{multline}
Weak coupling between system and quantum point contact allows for the Born approximation $\hat{\chi}(t) = \hat{\rho}(t) \otimes e^{-\beta H_{\textrm{leads}}}/Z$ since the effect on the density matrices of the leads remains small at all times. 
We can then trace out the leads, which enter the resulting equation for the (interaction-picture) density matrix $\hat{\rho}(t)$ of the system only through correlation functions. The first term in \eqref{eq:IA_pic_eom} vanishes since cross-lead correlation functions are assumed zero. The second term gives 
\begin{widetext}
\begin{align}
    \frac{\textrm{d}}{\textrm{d}t}\hat{\rho}(t) = 
        -\int_{t_0}^{t}\textrm{d}t'\ 
            \Big\{ & 
                    G^{<}_{L}(t'-t) G^{>}_{R}(t-t') 
                    e^{ieV(t-t')}
                        \bqty{  \hat{m}(t)     \hat{m}^{\dagger}(t') \hat{\rho}(t') 
                            - \hat{m}^{\dagger}(t') \hat{\rho}(t') \hat{m}(t) }  \nonumber \\ 
                    &
                    - G^{>}_L(t'-t) G^{<}_{R}(t-t') 
                    e^{ieV(t-t')}
                        \bqty{  \hat{m}(t) \hat{\rho}(t') \hat{m}^{\dagger}(t')  
                            -  \hat{\rho}(t') \hat{m}^{\dagger}(t') \hat{m}(t)  }\nonumber \\ 
                    &
                    +  G^{>}_{L}(t-t') G^{<}_{R}(t'-t)
                    e^{-ieV(t-t')}
                        \bqty{  \hat{m}^{\dagger}(t) \hat{m}(t') \hat{\rho}(t') 
                            -   \hat{m}(t') \hat{\rho}(t') \hat{m}^{\dagger}(t) }  \phantom{\Big\{}\nonumber \\ 
                    &
                    -  G^{<}_{L}(t-t') G^{>}_{R}(t'-t)
                    e^{-ieV(t-t')}
                        \bqty{ \hat{m}^{\dagger}(t) \hat{\rho}(t') \hat{m}(t')   
                            -  \hat{\rho}(t') \hat{m}(t') \hat{m}^{\dagger}(t)  }
            \Big\},
\label{eomrhoia}
\end{align}
\end{widetext}
where we defined the shorthand $\hat{m}(t) = \tau + \chi\hat{n}(t)$ as well as the greater and lesser lead Green functions  
\begin{align*}
G^{>}_{j}(t-t') =&\ -i \ev{ \hat{\psi}_{j}(t)\hat{\psi}^{\dagger}_{j}(t')  }
   = -\nu_j \frac{\pi/\beta}{\sinh \frac{\pi (t-t'-i\eta)}{\beta}} \\
G^{<}_{j}(t-t') =&\ i \ev{ \hat{\psi}^{\dagger}_{j}(t') \hat{\psi}_{j}(t)}
   = -\nu_j \frac{\pi/\beta}{\sinh \frac{\pi (t-t'+i\eta)}{\beta}}. 
\end{align*} 
Here, $\beta$ denotes the inverse temperature and $\eta$ is a positive infinitesimal. 

In the limit of a large bias voltage, the prefactors of the square brackets in the integrand on the right-hand side of Eq.\ (\ref{eomrhoia}) effectively become sharply peaked functions in $t-t'$. Using the Fourier transform of the function
\begin{equation}
    g(E)= Eb(E) = -\int \frac{\mathrm{d}t}{2\pi} e^{iEt}\left(\frac{\pi/\beta}{\sinh \frac{\pi (t+i\eta)}{\beta}} \right)^2
\end{equation}
with the Bose distribution $b(E)$, we can thus approximate 
\begin{align}
   G^{<}_{L}(t'-t) & G^{>}_{R}(t-t') e^{ieV(t-t')}\nonumber\\
   &\simeq 2\pi\nu_L\nu_Rg(-eV)\delta(t-t') + \textrm{imaginary} \nonumber\\
   G^{>}_{L}(t'-t) & G^{<}_{R}(t-t') e^{ieV(t-t')}\nonumber\\
   &\simeq 2\pi\nu_L\nu_Rg(eV)\delta(t-t') + \textrm{imaginary} \nonumber\\
      G^{>}_{L}(t-t') & G^{<}_{R}(t'-t) e^{-ieV(t-t')}\nonumber\\
   &\simeq 2\pi\nu_L\nu_Rg(eV)\delta(t-t') + \textrm{imaginary} \nonumber\\
      G^{<}_{L}(t-t') & G^{>}_{R}(t'-t) e^{-ieV(t-t')}\nonumber\\
   &\simeq 2\pi\nu_L\nu_Rg(-eV)\delta(t-t') + \textrm{imaginary}. 
\end{align}
Here, we do not specify the imaginary terms as they correspond to perturbative renormalizations of the system Hamiltonian. The approximate $\delta$-functions in time have a width of order $1/eV$. We assume that both the density matrix $\rho(t)$ and the quantum dot occupation $\hat n(t)$ vary slowly within times of order $1/eV$. In particular, this requires that the applied bias is large compared to characteristic system frequencies, $eV\gg \Omega_\pm$. In this limit, the equation of motion for $\hat\rho(t)$ becomes Markovian, and we obtain
\begin{multline}\label{eq:master_eq_system_app_1}
    \frac{\textrm{d}}{\textrm{d}t}\hat{\rho}(t) 
     = -i\bqty{\hat{H}, \hat{\rho}} \\
     + \frac{k}{eV\abs{\chi}^2}\Big\{ g(-eV) \mathcal{D}\bqty{\tau^*+\chi^* \hat{n}}  \\ 
     +  g(eV) \mathcal{D}\bqty{\tau + \chi \hat{n}} \Big\}\hat{\rho}(t).
\end{multline}
Here, we reverted from the interaction to the Schr\"odinger picture and defined the measurement strength $k=2\pi\nu_L\nu_R \abs{\chi}^2 eV$. The $(\tau^*+\chi^* \hat{n})$-term describes a process in which electrons tunnel from the left to the right lead. For positive $eV$, this happens even at $T=0$. The $(\tau+\chi \hat{n})$-term describes a process in which electrons tunnel from the right to the left lead which cannot occur at $T=0$. Absorbing a shift $2\pi \nu_L\nu_R eV \Im\bqty{\chi\tau^*}$ into the quantum dot energy $\epsilon$ in the Hamiltonian and taking the limit of zero temperature with $g(-eV) = eV$ and $g(eV) = 0$, we obtain the unconditional part of Eq.\ \eqref{eq:sme_mbq-qd-qpc}  

\subsection{Stochastic master equation}
\label{app:derivation_conditional_jump}

Since the tunneling amplitude depends only weakly on the quantum dot occupation, the current measurement constitutes a weak measurement of the quantum-dot-qubit system. The change of the system state 
\begin{equation}
    \ket{\psi} \to \frac{1}{\sqrt{p_i}}M_i\ket{\psi},
\end{equation}
due to these weak measurements is described by Kraus operators $M_1$ and $M_0$ which can be respectively associated with transmission or absence of transmission of electrons by the quantum point contact (still assuming $T=0$ so that tunneling is unidirectional). The change in the density matrix $\hat{\rho}_c$ takes the form 
\begin{equation}
    \hat{\rho}_c \to \frac{1}{p_i}M_i\hat{\rho}_c M_i^{\dagger}.
\end{equation}
In these expressions, 
\begin{equation}
    p_i = \bra{\psi}M_i^{\dagger} M_i \ket{\psi} = \textrm{tr}\bqty{M_i^{\dagger} M_i \hat{\rho}_c}
\end{equation}
denotes the probability for outcome $i$. 

The tunneling current through the quantum point contact can be described as a point process
\[
I_c(t) = e\frac{\textrm{d}N_c(t)}{\textrm{d}t}, 
\]
where $\textrm{d}N_c(t) \in \Bqty{0,1}$ is a Poisson element which is not infinitesimal but has an infinitesimal ensemble average
\begin{align}
    \mathbb{E}\bqty{\textrm{d}N_c(t)} 
    = \textrm{tr}\bqty{M_1^{\dagger} M_1 \hat{\rho}_c(t)}
\end{align}
equal to the probability that an electron is transmitted in time $\mathrm{d}t$. To find the Kraus operator $M_1$, we note that the ensemble average of the current  
\begin{equation}
\mathbb{E}\bqty{I_c(t)} = \frac{\textrm{tr}\bqty{M_1^{\dagger} M_1 \hat{\rho}_c(t)}} {\textrm{d}t}
\end{equation}
has to equal $I_0$ for $n=0$ and $I_1$ for $n=1$. This is satisfied for
\[
M_1 = \pqty{\tau^*+\chi^*\hat{n}}\sqrt{\textrm{d}t},
\]
where we rescaled $\sqrt{2\pi\nu_L\nu_R eV}\tau \to \tau$ and $\sqrt{2\pi\nu_L\nu_R eV}\chi \to \chi$ for notational simplicity (so that $I_0 = \abs{\tau}^2$,  $I_1 = \abs{\tau+\chi}^2$, and $k = \abs{\chi}^2$, setting $e=1$). 

To find the Kraus operator $M_0= \mathds{1} + \hat{A} \textrm{d}t$, we equate $\mathbb{E}\bqty{\hat{\rho}_c(t+\textrm{d}t)} = M_0 \hat{\rho}(t) M_0^{\dagger} + M_1 \hat{\rho}(t) M_1^{\dagger}$ and
$\mathbb{E}\bqty{\hat{\rho}_c(t+\textrm{d}t)} = 
\hat{\rho}(t+\textrm{d}t) = \pqty{1+\mathcal{L}\textrm{d}t}\hat{\rho}(t)$. Reading off the Liouvillian from Eq.\ (\ref{eq:master_eq_system_app_1}), this yields
\[
M_0 = \mathds{1} - i \textrm{d}t \hat{H} - \frac{1}{2} M_1^{\dagger} M_1
\]
and thus $\hat{A} = -i\hat{H} - (I_0 + \delta I \hat{n})/2$. As mentioned below Eq.\ (\ref{eq:master_eq_system_app_1}) above, we absorb a shift of the quantum dot energy into the Hamiltonian, 
\begin{align} \label{eq:a_operator}
    \hat{A} = -i\pqty{\hat{H} + \Im\bqty{\tau^*\chi}\hat{n}} - \frac{1}{2}\bqty{I_0 + \pqty{2 \tau\chi^* + k} \hat{n}} .
\end{align}
Below, we will no longer display this shift of $\hat{H}$ explicitly.  

The conditional evolution of $\hat{\rho}_c(t)$ takes the form 
\begin{align}
    \hat{\rho}_c(t+\textrm{d}t) =&\ \pqty{1-\textrm{d}N_c(t)} \frac{ M_0 \hat{\rho}_c(t) M_0^{\dagger} }{ \textrm{tr}\bqty{ M_0 \hat{\rho}_c(t) M_0^{\dagger}}} \nonumber \\
    &+ \textrm{d}N_c(t) \frac{ M_1 \hat{\rho}_c(t) M_1^{\dagger}}{ \textrm{tr}\bqty{ M_1 \hat{\rho}_c(t) M_1^{\dagger}}},
\end{align}
where the denominators ensure normalization. Expanding to linear order in $\textrm{d}t$ and neglecting higher-order terms of the form $\textrm{d}t\ \textrm{d}N_c(t)$, we obtain
\begin{widetext}
\begin{multline}\label{eq:sme_jump}
    \textrm{d}\hat{\rho}_c = \textrm{d}t  
        \Bqty{ -i \bqty{\hat{H},\hat{\rho}_c} 
        - \frac{1}{2}\pqty{ 
            \Bqty{ I_0 + k \hat{n},\hat{\rho}_c} 
            + 2 \chi^*\tau \hat{n}\hat{\rho}
            + 2\tau^*\chi \hat{\rho} \hat{n}
        } 
        + \pqty{ I_0 + \delta I \ev{ \hat{n} } } \hat{\rho}_c } \\ 
        + \textrm{d}N_c
        \Bqty{
        \frac{ \pqty{\tau^*+\chi^*\hat{n}} \hat{\rho}_c  \pqty{\tau+\chi\hat{n}} }{ I_0 + \delta I \ev{ \hat{n}}  } 
        - \hat{\rho}_c }.
\end{multline}
\end{widetext}
This describes the stochastic evolution of $\hat{\rho}_c$ as the quantum dot is monitored by the quantum point contact. The evolution of $\hat{\rho}_c$ is conditioned on the stochastic measurement current $I_c(t)$ of the quantum point contact.

We can alternatively describe the evolution in terms of a stochastic Schr\"odinger equation which takes the form  
\begin{widetext}
\begin{equation}
    \textrm{d}\ket{\psi_c} = 
        \textrm{d}t  \Bqty{ 
            -i \hat{H}  
            - \frac{1}{2}  \bqty{I_0 + \pqty{2\chi^*\tau + k} \hat{n}}  
            + \frac{1}{2}\bqty{ I_0 + \delta I \ev{\hat{n}} }  
        }  \ket{\psi_c} 
        + \textrm{d}N_c \Bqty{ 
            \frac{ \tau^*+\chi^*\hat{n} }{ \sqrt{ I_0 + \delta I\ \ev{ \hat{n}} } } - 1 
        } \ket{\psi_c}.
\end{equation}
\end{widetext}

\subsection{Diffusive approximation}
\label{app:derivation_conditional_diffusive}

The assumption that the average current $I_0$ is much larger than the shift $\delta I$ induced by changes in the quantum-dot occupation allows one to approximate the point process by a Wiener process. 

We consider a time interval $\delta t$ which is short enough that the changes in the density matrix $\hat{\rho}_c$ remain small and the quantum expectation value of $\hat n$ remains approximately constant. Then, the probability distribution of the number $N$ of tunneling events within the time interval $\delta t$ is given by the Poisson distribution
\begin{align}\label{eq:p_n}
    P(N) = \frac{[(I_0+\delta I \langle\hat n\rangle)\delta t]^{N} }{ N ! } e^{-(I_0+\delta I \langle\hat n\rangle)\delta t}
\end{align}
with ensemble average
\begin{align}
    \mathbb{E}\bqty{N} =   [I_0 + {\delta I} \ev{\hat{n}}]\delta t, 
\end{align} 
and variance
\begin{align}
    \mathbb{V}\bqty{N} \simeq I_0\delta t.
\end{align}
The expression for the variance uses that $\delta I\ll I_0$. Assuming that $I_0\delta t$ is large, one can approximate the Poisson by a Gauss distribution with the same average and variance. 

Assuming that $\delta I$ is sufficiently small and the Hamiltonian dynamics sufficiently slow, we can approximate both the unitary dynamics and the changes of the density matrix $\hat{\rho}_c$ induced by the weak measurements of the quantum dot charge to linear order in the time interval $\delta t$, 
\begin{widetext}
\begin{multline}
    \delta\hat{\rho}_c = {\delta}t  
        \Bqty{ -i \bqty{\hat{H},\hat{\rho}_c} 
        - \frac{1}{2}\pqty{ 
            \Bqty{ I_0 + k \hat{n},\hat{\rho}_c} 
            + 2 \chi^*\tau \hat{n}\hat{\rho}
            + 2\tau^*\chi \hat{\rho} \hat{n} } 
        + \pqty{ I_0 + \delta I \ev{ \hat{n} } } \hat{\rho}_c } \\ 
        + {\delta}N_c
        \Bqty{
        \frac{ \pqty{\tau^*+\chi^*\hat{n}} \hat{\rho}_c  \pqty{\tau+\chi\hat{n}} }{ I_0 + \delta I \ev{ \hat{n}}  } 
        - \hat{\rho}_c }.
        \label{eq:c23}
\end{multline}
\end{widetext}
Here, $\delta N_c$ describes the Wiener process
\begin{equation}
   \delta N_c(t) =  [I_0 + {\delta I} \ev{\hat{n}}]\delta t + \sqrt{I_0} \xi(t) \delta t
\end{equation}
with a Gaussian random process $\xi(t)$ with variance $1/\delta t$. Writing this in the continuum limit, we find 
\begin{multline}\label{eq:sme_diffusive_appendix}
    \frac{\textrm{d}}{\textrm{d}t}\hat{\rho}_c = -i \bqty{\hat{H}_{\textrm{eff}} ,\hat{\rho}_c} 
        + k\mathcal{D}\bqty{\hat{n}}\hat{\rho}_c \\
        + \sqrt{k}\xi(t) \mathcal{H}\bqty{\hat{n} e^{i \phi}}\hat{\rho}_c
\end{multline}
with $\tau^*\chi=\abs{\tau\chi}e^{i\phi}$ and the $\delta$-function correlator $\mathbb{E}\bqty{\xi(t)\xi(t')}=\delta(t-t')$. This simplifies to the evolution equation \eqref{eq:sme_mbq-qd-qpc} in the main text if one fixes $\phi = \pi$ (corresponding to a decrease in current through the quantum point contact due to the presence of an electron on the quantum dot). 

The measurement current is obtained by subtracting the background current $I_0$ and normalizing. For $\phi = \pi$, this yields
\begin{align} \label{eq:measurement_current_app}
    j(t)  = \frac{1}{\delta I} \pqty{ \frac{{\delta} N_c(t)}{{\delta}t} - I_0 } 
    = \ev{\hat{n}(t) } + \frac{1}{\sqrt{4 k}} \xi(t)
\end{align} 
in agreement with Eq.\ (\ref{eq:meas_signal}) of the main text.

\subsection{Relaxation by the electromagnetic environment}\label{app:dissipation}

Coupling to the electromagnetic environment leads to relaxation in the eigenbasis of the Majorana qubit-quantum dot system, as described by Eq.\ (\ref{eq:dissipator_main}) in the main text. Here, we sketch its derivation. 

The electrostatic potential of the electromagnetic environment is described as a free bosonic field 
\begin{equation}
   \hat v(\mathbf{r})  = \sum_{\bf q} \left[ \tilde{m}_{\bm{q}}^* \hat{a}_{\bm{q}}^{\dagger} e^{i\mathbf{qr}}+ \tilde{m}_{-\bm{q}}\hat{a}_{\bm{-q}}e^{-i\mathbf{qr}} \right].
\end{equation}
with Hamiltonian 
\begin{equation}
    \hat{H}_{v} = \sum_{\bm{q}} \omega_{\bm{q}}\hat{a}_{\bm{q}}^{\dagger}\hat{a}_{\bm{q}}
\end{equation}
and assumed to be in a thermal state $\hat{\rho}_v \propto \textrm{exp}(-\hat{H}_v/T)$. The potential $\hat v$ is an additional contribution to the gate voltages of Majorana qubit and quantum dot and varies slowly in space compared to the spatial extent of the system. Majorana qubit and quantum dot are then subject to the same potential $\hat v$, and we obtain the interaction
\begin{align}
    \hat{V} =& -\frac{2E_C C_g}{e} \hat{N} \hat{v} - \frac{2\epsilon_C c_g}{e} \hat{n}\hat{v}  \\
    =& \lambda \hat n \hat v \\
    =&  \hat n \sum_{\bf q} \left[ {m}_{\bm{q}}^* \hat{a}_{\bm{q}}^{\dagger} + {m}_{-\bm{q}}\hat{a}_{\bm{-q}} \right],
\end{align}
where we used $\hat{N} = -\hat{n}$ due to charge conservation and absorbed 
\begin{align}
     \lambda = \frac{2E_C C_g}{e} - \frac{2\epsilon_C c_g }{e}  
\end{align}
into the coefficients $m_{\mathbf{q}}$. (Here, $\epsilon_C$ and $c_g$ are charging energy and gate capacitance of the quantum dot, and $C_g$ is the gate capacitance of the Majorana qubit.)

We write the charge operator in the interaction picture with respect to the system Hamiltonian as given in Eq.\ \eqref{eq:n_heisenberg}. Then, the coupling Hamiltonian in rotating wave approximation becomes 
\begin{multline}
    \hat{V}(t) \simeq \hat{c} [\hat{B}(t) + \hat{B}^{\dagger}(t)] \\
        - \frac{\sin\theta}{2} \pqty{ e^{2i\Omega t} \hat{\tau}_+ \hat{B}(t) + e^{-2i\Omega t} \hat{\tau}_- \hat{B}^{\dagger}(t)  }, 
\end{multline}
where we defined the shorthand
\begin{align}\label{eq:bath_ops_environment}
    \hat{B}(t) =&\ \sum_{\bm{q}}  m_{\bm{q}} e^{-i\omega_{\bm{q}}t}\hat{a}_{\bm{q}}.
\end{align}
Notice that we suppressed $\pi$ indices as $\hat{V}(t)$ conserves $\hat{\pi}$ and does not mix the two subspaces. 

Following the same steps as above, the master equation for a general system-bath interaction
\begin{equation}\label{eq:sb_schmidt_decomp}
    \hat{H}_{SB} = \sum_{i} \hat{S}_{i} \otimes \hat{B}_{i}
\end{equation}
with system operators $\hat S_i$, here associated with the Majorana qubit-quantum dot system, and bath operators $\hat B_i$, here associated with the electromagnetic environment, can be written
as 
\begin{multline}\label{eq:reduced_evolution_equation}
    \frac{\textrm{d}}{\textrm{d}t}\hat{\rho}(t) 
     = -i\bqty{\hat{H}, \hat{\rho}} \\ 
     - \sum_{i} \Big\{ \hat{S}_{i}\hat{\mathcal{S}}^+_{i} \hat{\rho}(t) 
     - \hat{\mathcal{S}}^+_{i}\hat{\rho}(t) \hat{S}_{i} \\
     + \hat{\rho}(t) \hat{\mathcal{S}}^-_{i} \hat{S}_{i}
     -  \hat{S}_{i} \hat{\rho}(t) \hat{\mathcal{S}}^-_{i} \Big\}.
\end{multline}
Here, we defined the (calligraphic) operators
\begin{subequations}
\begin{align}
    \hat{\mathcal{S}}^+_{i} =&\ \sum_j \hat{\mathcal{S}}^+_{ij} = \sum_j\int_{0}^{\infty}\textrm{d}\tau\ C_{ij}(\tau) \hat{S}_{j}(-\tau), \\
    \hat{\mathcal{S}}^-_{i} =&\ \sum_j \hat{\mathcal{S}}^-_{ij} =\sum_j\int_{0}^{\infty}\textrm{d}\tau\ C_{ji}(-\tau) \hat{S}_{j}(-\tau)
\end{align}
\end{subequations}
including the bath correlation functions 
\begin{equation}\label{eq:bath_correlators_general}
    C_{ij}(t-t') = \ev{\hat{B}_{i}(t)\hat{B}_{j}(t')}.
\end{equation}
We note that the $\hat{S}_{i}$ and $\hat{B}_{i}$ are not necessarily hermitian. If they are, $C^*_{ij}(\tau) = C_{ji}(-\tau)$ and thus $(\hat{\mathcal{S}}^-_{i})^{\dagger} = \hat{\mathcal{S}}^+_{i} \equiv \hat{\mathcal{S}}_{i}$.

Applying this to the problem at hand, we identify
\begin{subequations}
\begin{align}
    \hat S_1 =&\ \hat c \\
    \hat S_2 =&\ -\sin\theta\ \textrm{exp}(2i\Omega t) \hat{\tau}_+ /2 \\
    \hat S_3 =&\ -\sin\theta\ \textrm{exp}(-2i\Omega t) \hat{\tau}_- /2 
\end{align}
\end{subequations}
and 
\begin{subequations}
\begin{align}
    \hat B_1 =&\ \hat{B} + \hat{B}^{\dagger} \\
    \hat B_2 =&\  \hat{B}  \\
    \hat B_3 =&\  \hat{B}^{\dagger}. 
\end{align}
\end{subequations}
One readily evaluates the basic bath correlation functions (with $\tau=t-t'$)
\begin{subequations}
\begin{align}
    C_{BB^{\dagger}}(\tau) =&\ \langle \hat B(t) \hat B^\dagger(t') \rangle
        = \int_0^{\infty}\textrm{d}\omega\ J(\omega)  e^{-i\omega\tau}(1+ b(\omega) ),  \\
    C_{B^{\dagger} B}(\tau) =&\ \langle \hat B^\dagger(t) \hat B(t') \rangle  
        = \int_0^{\infty}\textrm{d}\omega\ J(\omega)  e^{i\omega\tau} b(\omega).
\end{align}
\end{subequations}

Within the rotating wave approximation, we retain only terms which are slowly varying on the scale of the system dynamics. Moreover, we retain only dissipative terms and drop renormalizations of the system Hamiltonian. This yields the result
\begin{equation} \label{eq:relaxation_appendix}
    \frac{\textrm{d}\hat{\rho}}{\textrm{d}t} = 
    \underbrace{\Bqty{\frac{\cos^2\theta}{4}\Gamma_0 \mathcal{D}\bqty{\hat{\tau}_z} + \frac{\sin^2\theta}{4}\pqty{\Gamma_+ \mathcal{D}\bqty{\hat{\tau}_+} + \Gamma_- \mathcal{D}\bqty{\hat{\tau}_-}} }}_{\equiv \mathcal{L}'}\hat{\rho},
\end{equation}
where we defined 
\begin{subequations}
\begin{align}
    \Gamma_0\, =&\ \pi \lim_{\omega \to 0}  J(\omega)( 1 + 2 b(\omega)), \\
    \Gamma_{+} =&\ 2\pi J(2\Omega)  b(2\Omega) \\
    \Gamma_{-} =&\ 2\pi J(2\Omega)\pqty{ 1 + b(2\Omega)}.
\end{align}
\end{subequations}
The $\Gamma_0$ term causes decoherence in the energy basis, whereas the $\Gamma_\pm$ terms cause transitions between ground and excited states. At low temperatures, $\Gamma_- \gg \Gamma_+$ with $\Gamma_+$ vanishing at $T=0$ and $\Gamma_-$ remaining finite. For this reason, we neglect $\Gamma_+$ relative to $\Gamma_-$ in the main text. 

\section{Spectra and eigenmodes of Liouvillians}\label{app:liouvillian_evs}

\subsection{Eigenvalues and eigenmatrices of diagonal block Liouvillian $\mathcal{L}_{\pi,\pi} + \mathcal{L}'_{\pi,\pi}$}\label{app:liouvillian_evs_diag}

This appendix gives the eigenvalues and eigenvectors of the diagonal blocks of the full Liouvillian including both measurement and relaxation dynamics, $\mathcal{L}_{\pi,\pi} + \mathcal{L}'_{\pi,\pi}$. To this end, we vectorize by columns, $\rho \to \ket{\rho} = (\rho_{11}, \rho_{21} , \rho_{12} , \rho_{22})^T$. In this notation, $\textrm{tr}[AB] = \braket{A^{\dagger}}{B}$ for square matrices $A$ and $B$. The Liouvillian matrix is then given by 
\begin{align}
    \mathcal{L}_{\pi,\pi} + \mathcal{L}_{\pi,\pi}' =&\  -i\pqty{\mathds{1}\otimes h_\pi - h_\pi^T \otimes \mathds{1} } + k \mathrm{D}[n] \nonumber \\ 
    &\ + \frac{\cos^2\theta_\pi}{4} \Gamma_0 \mathrm{D}\bqty{\tau_z} +  \frac{\sin^2\theta_\pi}{4}  \Gamma_- \mathrm{D}\bqty{ \tau_-} \nonumber \\
    =&\ \sum_n  \lambda_{\pi,n}  \ketbra{\psi_{\pi,n}}{\phi_{\pi,n}}, 
\end{align}
where we defined the vectorized decoherence superoperator
\begin{equation}
    \mathrm{D}[L] = L^* \otimes L - \frac{1}{2}\bqty{ \mathds{1}\otimes L^{\dagger} L + \pqty{L^{\dagger} L}^T \otimes \mathds{1} },
\end{equation} 
and expanded in terms of left and right eigenvectors $\ket{\psi_{\pi,n}}$ and $\ket{\phi_{\pi,n}}$ to eigenvalue $\lambda_{\pi,n}$.   
In the following, we work to leading order in $t_i, k, \Gamma_-, \Gamma_0 \ll \Omega_\pi$ and will suppress $\pi$ subscripts of $\theta$ and $\Omega$. The eigenvalues are
\begin{subequations}
\begin{align}
    \lambda_0 =&\ 0, \\
    \lambda_{\textrm{slow}} =&\ -\frac{\sin^2\theta}{4}\pqty{\Gamma_- + 2k}, \\
    \lambda_{\textrm{fast} , \pm} =&\  \pm 2i\Omega - \frac{1+\cos^2\theta}{4}k  \nonumber \\ & -\frac{\sin^2\theta}{8}\Gamma_- - \frac{\cos^2\theta}{2} \Gamma_0 .
\end{align}
\end{subequations}
The corresponding right eigenvectors, written in the energy basis, are
\begin{subequations}
\begin{align}
    &\ket{\psi_0} \simeq  \frac{1}{2}\pqty{ 1 + R , 0  , 0 , 1-R }^T  , \\
    &\ket{\psi_{\textrm{slow}}} \simeq \pqty{ 1 , 0 , 0, -1}^T, \\
    &\ket{\psi_{\textrm{fast}, +}} \simeq \pqty{ 0 , 0, -1   ,  0 }^T, \\
    &\ket{\psi_{\textrm{fast}, -}} \simeq \pqty{ 0 , 1 ,  0 ,  0 }^T,
\end{align}
\end{subequations}
where $R$ was defined in Eq. \eqref{eq:measurement_relaxation_ratio}, and the left eigenvectors are
\begin{subequations}
\begin{align}
    &\bra{\phi_0} = \pqty{ 1 , 0 , 0 , 1 }  ,\\
    &\bra{\phi_{\textrm{slow}}} \simeq \pqty{ \frac{k}{\Gamma_- + 2k} , 0 , 0 ,  -\frac{\Gamma_-+k}{\Gamma_-+2k}}, \\
    &\bra{\phi_{\textrm{fast}, +}} \simeq \pqty{ 0 , 0, -1   ,  0 }, \\
    &\bra{\phi_{\textrm{fast}, -}} \simeq \pqty{ 0 , 1  , 0  ,  0 }.
\end{align}
\end{subequations}
Note that $\ket{\psi_0} = \ket{\rho^{\infty}}$ and $\bra{\phi_0} = \bra{\mathds{1}_2}$. We normalized $\braket{\phi_m}{\psi_n} = \delta_{mn}$ to leading order. 

\subsection{Off-diagonal block Liouvillian $\mathcal{L}_{+-}$ in Eq.~ (\ref{eq:+-_block_meq})}\label{app:liouvillian_evs_offdiag}

In this appendix, we analyze the off-diagonal Liouvillian $\mathcal{L}_{+-}$ in Eq.\ \eqref{eq:+-_block_meq} with $h_{\pm}$ in Eq.\ \eqref{eq:sector_hams} and one decoherence channel given by the quantum-point-contact coupling (i.e., without relaxation).

\subsubsection{Steady state of $\rho_{+-}$}
To understand the long-time behavior of the density matrix, we find the eigenvalues of $ \mathcal{L}_{+-}$ and show that generically, the real parts of all eigenvalues are strictly negative. Thus, the only steady state is $\rho_{+-}=0$. 

We again write the superoperator $ \mathcal{L}_{+-}$ in vectorized form with $\rho_{+-}\to (\rho_{+-}^{11},\rho_{+-}^{21},\rho_{+-}^{12},\rho_{+-}^{22})^T$, such that (in the charge basis)
\begin{align}
    \mathcal{L}_{+-} =&\ -i\pqty{\mathds{1}\otimes h_+ - h_-^T \otimes \mathds{1} } \nonumber \\
    &\ + k\bqty{ n\otimes n - \frac{1}{2}\pqty{\mathds{1}\otimes n + n \otimes \mathds{1}} } \nonumber
    \\
    =&\ 
    \begin{pmatrix}
    0 & -it_1 - t_2 & i t^*_1 + t^*_2 & 0 \\
    -i t^*_1 + t^*_2  & -i\varepsilon -k/2 & 0 & i t^*_1 + t^*_2 \\
    i t_1 - t_2 & 0 & i \varepsilon -k/2  & -it_1 - t_2  \\
    0 & i t_1 - t_2 & -i t_1^* + t^*_2 & 0 
    \end{pmatrix}.
\end{align}
The characteristic polynomial becomes 
\begin{multline}\label{eq:char_pol}
    \chi(\tilde{\lambda}) = \tilde{\lambda}^4 + k\tilde{\lambda}^3 + ({\varepsilon}^2 + 4|t_1|^2+4|t_2|^2 +{k^2}/{4}) \tilde{\lambda}^2 
        \\+2k(|t_1|^2+|t_2|^2)\tilde{\lambda} + 16|t_1|^2|t_2|^2\sin^2\varphi,
\end{multline}
where $\varphi$ denotes the phase difference between the tunneling amplitudes $t_1$ and $t_2$.

Evidently, $\mathcal{L}_{+-}$ has a zero eigenvalue when $\sin\varphi=0$ or, equivalently, $\Im{t_1 t_2^*}=0$. In this case, $\rho_{+-}$ does not decay to zero and neither $\hat{\pi}$ not $\hat{Z}$ are  projectively measured. Physically, the Rabi frequencies of the $\pi = +1$ and the $\pi = -1$ sectors are identical and the steady-state measurement currents of the two sectors are indistinguishable. Thus, the measurement reveals no information on the qubit state and does not decohere the system in the measurement basis. 

Conversely, if $\sin\varphi\neq 0$, there is no zero eigenvalue. We now show that in this case, the eigenvalues have strictly negative real parts. They are non-positive since $\mathcal{L}_{+-}+\mathcal{L}_{+-}^{\dagger}$ is negative semi-definite. To show that the real parts of the eigenvalues are strictly negative, consider the characteristic polynomial. Taking $\sin\varphi \neq 0$, we assume that there is an imaginary eigenvalue $\tilde{\lambda}=iy$ with $y \in \mathds{R}$. This eigenvalue satisfies 
\begin{eqnarray}\label{eq:char_pol_2}
  &&y^4 -i ky^3 - ({\varepsilon}^2 + 4|t_1|^2+4|t_2|^2 +{k^2}/{4}) y^2 
  \nonumber\\
  &&+i2k(|t_1|^2+|t_2|^2)y + 16|t_1|^2|t_2|^2\sin^2\varphi = 0.
\end{eqnarray}
The imaginary part of this equation, $y^3=2(|t_1|^2+|t_2|^2)y$ has solutions $y=0$ and $y=\pm [2(|t_1|^2+|t_2|^2)]^{1/2}$. 
For $y=0$, the real part of Eq.\ \eqref{eq:char_pol_2} implies $\sin^2\varphi=0$, contradicting our assumptions. Similarly, for $y=\pm [2k(|t_1|^2+|t_2|^2)]^{1/2}$, the real part implies 
\begin{equation}
  \frac{4|t_1|^2|t_2|^2}{(|t_1|^2+|t_2|^2)^2}\sin^2\varphi -1 = \frac{\epsilon^2+k^2/4}{2(|t_1|^2+|t_2|^2)}.
\end{equation}
While the left hand side is non-positive, the right-hand side is strictly positive, so that there are no solutions. We conclude that $\mathcal{L}_{+-}$ has only eigenvalues with a strictly negative real part. 

\subsubsection{Decoherence rate}

The decoherence rate is governed by the eigenvalue $\tilde{\lambda}_{\textrm{slow}}$ of $\mathcal{L}_{+-}$ with the largest real part (corresponding to the slowest decay). We perform perturbative analyses for small $k$ as well for small $\sin^2\varphi$. Beyond the perturbative regime, we investigate the behavior of the eigenvalues numerically, see Fig. \ref{fig:slowest_decaying_ev} in the main text. For simplicity, we specify to $t_1 = t_2 e^{-i\varphi}$ with $t_1$ real. We also define the shorthand $\tilde{\varepsilon}^2 = \varepsilon^2+8t_1^2$. Then, the characteristic polynomial takes the form
\begin{equation}
    \chi(\tilde{\lambda}) = \tilde{\lambda}^4 + k\tilde{\lambda}^3 + (\tilde{\varepsilon}^2 + \frac{k^2}{4}) \tilde{\lambda}^2 + 4kt_1^2\tilde{\lambda} + 16t_1^4 \sin^2\varphi.
\end{equation}

\paragraph*{Small $k$:} For weak coupling between quantum dot and quantum point contact, we determine the roots of the characteristic polynomial \eqref{eq:char_pol} to first order in $k$. Expanding $\tilde{\lambda}=\tilde{\lambda}_0 + k \tilde{\lambda}_1 + ...$, we obtain 
\begin{subequations}
\begin{align}
    0 =& \tilde{\lambda}_0^4 + \tilde{\varepsilon}^2 \tilde{\lambda}_0^2 +  16t_1^4\sin^2\varphi, \\
    0 =& 4\tilde{\lambda}_0^3\tilde{\lambda}_1 + \tilde{\lambda}_0^3 + 2\tilde{\varepsilon}^2 \tilde{\lambda}_0 \tilde{\lambda}_1 + 4 t_1^2 \tilde{\lambda}_0
\end{align}
\end{subequations}
with the solutions
\begin{subequations}
\begin{align}
    (\tilde{\lambda}^\pm_0)^2 =& -\frac{\tilde{\varepsilon}^2}{2} \pm \sqrt{\pqty{\frac{\tilde{\varepsilon}^2}{2}}^2-16t_1^4\sin^2\varphi}  \\
    \tilde{\lambda}^\pm_1 =& - \frac{1}{4}\left[1- \frac{\varepsilon^2}{2(\tilde{\lambda}^\pm_0)^2+\tilde\varepsilon^2}\right] \nonumber\\   
    =&  -\frac{1}{4}\bqty{1 \pm \frac{\varepsilon^2}{\sqrt{\varepsilon^4+16\varepsilon^2t_1^2+64t_1^4\cos^2\varphi }}}.
\end{align}
\end{subequations}
While $(\tilde{\lambda}^\pm_0)^2 \leq 0$, so that the $\tilde{\lambda}_0$ are purely imaginary, the first-order correction $k\tilde{\lambda}_1$ is manifestly real and negative. For the slowly decaying eigenvalues, we choose the minus sign in the above expressions. In the limit $t_1 \ll \varepsilon$, the decay of $\rho_{+-}$ is then controlled by 
\begin{equation}
    \Re{\tilde{\lambda}_{\textrm{slow}}} \simeq \Re{\tilde{\lambda}^-_0+ k \tilde{\lambda}^-_{1}} = -2k\frac{t_1^2}{\varepsilon^2}.
\end{equation}
For $\sin^2\varphi=0$, $\tilde{\lambda}_0^-$ vanishes while $\tilde{\lambda}_1^-$ remains finite. Hence, perturbation theory breaks down close to $\sin^2\varphi=0$. Requiring $\vert\tilde{\lambda}^-_1\vert \ll k\vert\tilde{\lambda}^-_2\vert$ yields the condition
\begin{equation}
    \frac{k}{\varepsilon} \ll  \abs{ \abs{\sin\varphi} - \frac{1}{4\abs{\sin\varphi}}}^{-1}.
\end{equation}
The right side of this inequality vanishes for $\varphi = 0$, i.e. the expansion indeed breaks down. 

\paragraph*{Small $\sin^2\varphi$:} The slowest decaying eigenvalue $\tilde{\lambda}_{\textrm{slow}}$ evolves from the zero eigenvalue for $\sin^2\varphi=0$. For small $\sin^2\varphi$, we readily find
\begin{equation}
    \tilde{\lambda}_{\textrm{slow}} = -\frac{4 t_1^2 \sin^2 \varphi}{k} + ...
\end{equation}
Notice that the expansion breaks down for vanishing $k$.

The fact that a perturbative expansion is impossible for small $k$ and $\varphi$ is in accordance with the numerical observation that exceptional lines emanate from this region, see Fig. \ref{fig:slowest_decaying_ev}.

\section{Stochastic evolution of $\pi(t)$} \label{app:stochastic_evolution}

It is instructive to analyze the stochastic evolution of the expectation value of the combined fermion parity $\hat\pi$. For $\hat{H}_T = 0$, Eq.\ (\ref{eq:charge_stochastic_equation}) showed that the evolution of $n(t)$ 
ceases once $n = 0$ or $n = 1$. In the presence of tunneling, $\hat{H}_T \neq 0$, this is no longer the case due to the additional term $\langle i [\hat{H}_T , \hat{n}]\rangle$ in Eq.\ (\ref{eq:charge_stochastic_equation_2}). 

To analyze $\pi$, we consider the set of coupled stochastic differential equations obtained from Eq.\ \eqref{eq:sme_mbq-qd-qpc}. As a result of the transfer of weight between the $\pi$ subspaces due to measurements, the stochastic term couples $\rho_{++}$ and $\rho_{--}$. At the same time, these diagonal blocks of $\hat\rho$ remain uncoupled to $\rho_{+-}$. (However, the time evolution of $\rho_{+-}$ depends on $\mathrm{tr}\rho_{\pi,\pi}$.) We write the coupled equations for the diagonal blocks using the Bloch-vector notation 
\begin{align}
    \rho_{\pi,\pi} 
    =  \frac{p_\pi}{2} + \frac{1}{2}\pqty{\sigma_x x_\pi + \sigma_y y_\pi + \sigma_z z_\pi },
\end{align}
where $p_\pi = \textrm{tr} \rho_{\pi,\pi} = \langle\hat{P}_\pi\rangle$ (with the projector $\hat{P}_\pi$ onto the $\pi$ subspace). With this parametrization, Eq.\ \eqref{eq:sme_mbq-qd-qpc} yields the stochastic differential equations
\begin{subequations}
\begin{align}
    \dot{p}_\pi =& -\sqrt{k} \xi \pqty{z_\pi - \mathcal{Z} p_\pi}  \label{eq:stochastic_projector_plus}  \\
    \dot{z}_\pi =& 2\Omega_\pi \pqty{h_\pi^x y_\pi - h^y_\pi x_\pi} - \sqrt{k} \xi \pqty{p_\pi - \mathcal{Z} z_\pi}, \\
    \dot{x}_\pi =& 2\Omega_\pi \pqty{h_\pi^y z_\pi - h^z_\pi y_\pi} - \frac{k}{2} x_\pi + \sqrt{k} \xi \mathcal{Z}  x_\pi ,\\
    \dot{y}_\pi =& 2\Omega_\pi \pqty{h_\pi^z x_\pi - h^x_\pi z_\pi} - \frac{k}{2} y_\pi + \sqrt{k} \xi \mathcal{Z}  y_\pi,
\end{align}
\end{subequations}
where $\mathcal{Z} = z_+ + z_-$ introduces the coupling between the diagonal blocks.

Fixed points of $\pi(t)=p_+-p_-=2p_+-1$ require that the right hand side of Eq.\ \eqref{eq:stochastic_projector_plus} vanish, $M_\pi = z_\pi - \mathcal{Z} p_\pi = 0$. $p_\pi = 0$ implies $x_\pi = y_\pi = z_\pi =0$. We then find by direct evaluation that $p_+ = 0$ (or similarly $p_-=0$) implies $M_+ = M_- = 0$. We conclude that $p_+=0$ and $p_+=1$ are fixed points of the evolution of $p_+(t)$. These correspond to fixed points $\pi= -1$ and $\pi=+1$ of $\pi(t)$, respectively. We checked numerically that these are the only fixed points provided $\Im{t_1 t_2^*}\neq 0$. (This is indicated, e.g., by the fact that $\mathbb{E}[1-\pi^2] \to 0$ as shown in Fig.\ \ref{fig:majorana_measurement_generic}).

Now consider the case $\Im{t_1 t_2^*} =0$. For simplicity we make the stronger assumption $h_+ = h_-$. To understand this case, it is easiest to consider the evolution in terms of
\begin{align}
    l^\perp_\pi =&\  \pqty{h_\pi^x y_\pi - h^y_\pi x_\pi}, \\
    l^\parallel_\pi =&\  \pqty{h_\pi^x x_\pi + h^y_\pi y_\pi}.
\end{align}
Then, the evolution equations are
\begin{subequations}
\begin{align}
    \dot{p}_\pi =&\ -\sqrt{k} \xi \pqty{z_\pi - \mathcal{Z} p_\pi}\\
    \dot{z}_\pi =&\ 2\Omega_\pi l^\perp_\pi - \sqrt{k} \xi \pqty{p_\pi - \mathcal{Z} z_\pi}, \\
    \dot{l}^\perp_\pi =&\ 2\Omega_\pi \pqty{h^z_\pi l^\parallel_\pi - \sin^2\theta_\pi z_\pi  } - \frac{k}{2} l^\perp_\pi + \sqrt{k} \xi \mathcal{Z}  l^\perp_\pi ,\\
    \dot{l}^\parallel_\pi =&\ - 2\Omega_\pi h^z_\pi l^\perp_\pi -  \frac{k}{2} l^\parallel_\pi + \sqrt{k} \xi \mathcal{Z}  l^\parallel_\pi.
\end{align}
\end{subequations}
We also define $L^\perp = l^\perp_+ + l^\perp_-$ and $L^\parallel = l^\parallel_+ + l^\parallel_-$. For $h_+ = h_- = h = \Omega \bm{h}\cdot \bm{\sigma}$, they satisfy a decoupled set of equations 
\begin{subequations}
\begin{align}
    \dot{\mathcal{Z}} =&\ 2\Omega L^\perp  - \sqrt{k} \xi \pqty{1 - \mathcal{Z}^2 }, \\
    \dot{L}^\perp =&\ 2\Omega  \pqty{h^z  L^\parallel - \sin^2\theta  \mathcal{Z}   } - \frac{k}{2} L^\perp + \sqrt{k} \xi \mathcal{Z}  L^\perp  ,\\
    \dot{L}^\parallel =&\ - 2\Omega h^z L^\perp -  \frac{k}{2} L^\parallel + \sqrt{k} \xi \mathcal{Z}  L^\parallel.
\end{align}
\end{subequations}
Importantly, the equations for the $l$ and $L$ have identical form. We now show that the evolution of $p_+$ becomes frozen if the dynamics of the lower-case variables is locked to that of the upper-case variables, i.e., if 
\begin{subequations}\label{eq:fixed_point_h=h}
\begin{align} 
    0 =&\ z_+ - p_+ \mathcal{Z}, \\
    0 =&\ l^\perp_+ - p_+ L^\perp, \\
    0 =&\ l^\parallel_+ - p_+ L^\parallel, 
\end{align}
\end{subequations}
independently of the value of $p_+$. If $h_+ = u h_- u^\dagger$ similar statements hold for linear combinations of these variables. For the above relations to indeed be fixed points, we need to show that also their Ito differentials vanish. 
Consider first the two lower lines, 
\begin{align}
    \textrm{d}\pqty{l_+ - p_+ L} =&\  \textrm{d} l_+ - p_+ \textrm{d} L - \textrm{d} p_+ \pqty{ { \ldots }} \nonumber \\
    =&\ \ \ \ \pqty{ { \ldots } } \bqty{z_+ - p_+ \mathcal{Z}} \nonumber \\
    &+ \pqty{ { \ldots } } \bqty{l^\perp_+ - p_+ L^\perp} \nonumber \\
    &+ \pqty{ { \ldots } } \bqty{l^\parallel_+ - p_+ L^\parallel}, 
\end{align}
where we used that the evolution equations for $l$ and $L$ have identical form. The terms in square brackets vanish at the fixed point specified by Eq.\ \eqref{eq:fixed_point_h=h}, so that the detailed form of the terms in the parentheses does not matter.
Similarly, 
\begin{align}
    \textrm{d}\pqty{  z_+ - p_+ \mathcal{Z} } =&\   \sqrt{k} \xi \mathcal{Z} \pqty{z_+ - p_+ \mathcal{Z}} \nonumber \\ &+ 2\Omega \pqty{l^\perp_+ - p_+ L^\perp}.
\end{align}
Thus, Eqs.\ \eqref{eq:fixed_point_h=h} indeed describe fixed points with arbitrary $p_+$. If the system is initialized, the dynamics of the two sectors will then tend to lock. Once this has happened, $p_+$ remains constant. The ensemble averaged evolution of course has $p_+(t) = p_+(0)$. 

\section{Noise spectrum of the measurement current}\label{app:spectrum}

\subsection{Autocorrelation function of the steady-state measurement signal}\label{app:auto_correlation}

The measurement outcomes $\pi = +1$ and $\pi = -1$ are distinguished by the noise spectrum of the steady-state measurement current $j_\pi(t)$ for a given $\pi$, 
\begin{align} \label{eq:autocorrelation_expanded}
S(\tau) =&\ \mathbb{E}\bqty{j(t)j(t+\tau)} - \mathbb{E}\bqty{j(t)}\bqty{j(t+\tau)}\nonumber \\
        =&\ \frac{\delta(\tau)}{4k} 
         + \frac{1}{\sqrt{4k}} \left\{\mathbb{E}\bqty{n(t)\xi(t+\tau)} + \mathbb{E}\bqty{n(t+\tau)\xi(t)}\right\} \nonumber \\
         &+  \mathbb{E}\bqty{n(t)n(t+\tau)} - \mathbb{E}\bqty{n(t)}\bqty{n(t+\tau)},
\end{align}
where we suppress all labels indicating the measurement outcome and used Eq.\ (\ref{eq:meas_signal}). Our evaluation of $S(\omega)$ follows Ref.\ \cite{wiseman_quantum_1993} (see App.\ B). 

First consider $\mathbb{E}\bqty{n(t+\tau)\xi(t)}$, which requires one to compute $\mathbb{E}\bqty{\hat\rho_c(t+\tau)\xi(t)}$. From Eqs.\ (\ref{eq:c23}) and (\ref{eq:measurement_current_app}), it is evident that $\xi(t)$ is only correlated with the stochastic contribution to $\delta\rho_c$ for the next time step, from $t$ to $t+\delta t$. We then find
\begin{align}
    \mathbb{E}\bqty{ \hat{\rho}_c(t+{\delta}t) \xi(t)} 
        =& \sqrt{k} \mathbb{E}\bqty{ \mathcal{H}[\hat{n}] \hat{\rho}_c(t) }  \nonumber \\
        =& \sqrt{4k}\pqty{ \frac{\Bqty{\hat{n},\hat{\rho}^{\infty}}}{2} -  \mathbb{E}\bqty{n(t) \hat{\rho}_c(t) }  },
\end{align}
where we write $\mathbb{E}[\hat\rho_c] = \hat{\rho}^{\infty}$. Using the formal solution $\hat{\rho}_c(t) = \mathcal{U}(t,0) \hat{\rho}_c(0)$ of the stochastic master equation (\ref{eq:sme_mbq-qd-qpc}) with
\[
\mathcal{U}(t,t') = \mathcal{T} \exp\Bqty{\mathcal{L}(t-t') + \sqrt{k}\int_{t'}^{t} \textrm{d}t_1 \xi(t_1)\mathcal{H}[\hat{n}] } 
\]
($\mathcal{T}$ is the time ordering operator) and exploiting that $\xi(t)$ is uncorrelated with any of the later $\xi(t_1)$, we conclude that 
\begin{multline}
    \mathbb{E}\bqty{ \hat{\rho}_c(t+\tau) \xi(t)} 
        = \sqrt{4k}\, \theta(\tau) \Big( e^{\mathcal{L}\tau}\frac{\Bqty{\hat{n},\hat{\rho}^{\infty}}}{2}  
        \\
        - \mathbb{E}\bqty{n(t) e^{\mathcal{L}\tau} \hat{\rho}_c(t) }  \Big) ,
\end{multline}
and thus 
\begin{multline}\label{eq:autocorrelation_term_tau_greater_0}
    \frac{1}{\sqrt{4k}}\mathbb{E}\bqty{n(t+\tau)\xi(t)}  \\
      = \theta(\tau) \Big( 
        \textrm{tr}\bqty{\hat{n} e^{\mathcal{L}\tau} \frac{\Bqty{\hat{n},\hat{\rho}^{\infty}}}{2} } 
        - \mathbb{E}\bqty{n(t) \textrm{tr}\bqty{\hat{n}e^{\mathcal{L}\tau} \hat{\rho}_c(t) }} \Big) \\
        =  \theta(\tau) \Big( 
        \textrm{tr}\bqty{\hat{n} e^{\mathcal{L}\tau} \frac{\Bqty{\hat{n},\hat{\rho}^{\infty}}}{2} }  
        - \mathbb{E}\bqty{n(t) n(t+\tau) } \Big) .
\end{multline}
In the last step, we used that
\begin{equation}
 \mathbb{E}\bqty{n(t) \textrm{tr}\bqty{\hat{n}e^{\mathcal{L}\tau} \hat{\rho}_c(t) }} 
   =  \mathbb{E}\bqty{n(t) \textrm{tr}\bqty{\hat{n}\mathcal{U}(t+\tau,t) \hat{\rho}_c(t) }} ,
\end{equation}
since all the additional stochastic terms introduced on the right-hand side average to zero. 

Similarly, $\mathbb{E}\bqty{n(t)\xi(t+\tau)}$ is nonzero for $\tau<0$ only. Then, time translation invariance of the stationary state implies 
\begin{align}
    \mathbb{E}\bqty{n(t)\xi(t+\tau)} &= \theta(-\tau)\mathbb{E}\bqty{n(t)\xi(t-|\tau|)} \\
        &= \theta(-\tau) \mathbb{E}\bqty{n(t+|\tau|)\xi(t)} 
\end{align}
and we conclude
\begin{multline}
      \frac{1}{\sqrt{4k}}\Big(\mathbb{E}\bqty{n(t+\tau)\xi(t)}+\mathbb{E}\bqty{n(t)\xi(t+\tau)}\Big)\\
              =  
        \textrm{tr}\bqty{\hat{n} e^{\mathcal{L}|\tau|} \frac{\Bqty{\hat{n},\hat{\rho}^{\infty}}}{2} }  
        - \mathbb{E}\bqty{n(t) n(t+\tau) }  .
\end{multline}
Inserting this into Eq.\ (\ref{eq:autocorrelation_expanded}) gives
\begin{equation}\label{eq:autocorrelation_almost_final}
    S(\tau) = \frac{\delta(\tau)}{4k} + \textrm{tr}\bqty{\hat{n} e^{\mathcal{L}\abs{\tau}}\frac{\Bqty{\hat{n},\hat{\rho}^{\infty}}}{2} } - \pqty{n^{\infty}}^2
\end{equation}
with $n^\infty = \textrm{tr}[\hat{n} \hat\rho^\infty]$.

\subsection{Explicit evaluation}

We now evaluate Eq.\ \eqref{eq:autocorrelation_almost_final} explicitly for the steady states 
$\hat{\rho}^{\infty}_+ = \textrm{diag}(1+R,1-R,0,0)/2$ and $\hat{\rho}^{\infty}_- = \textrm{diag}(0,0,1+R,1-R)/2$ of $\mathcal{L} + \mathcal{L}'$ [as given in Eq.\ \eqref{eq:sme_mbq-qd-qpc} and Eq.\ \eqref{eq:relaxation_appendix}] corresponding to the two measurement outcomes $\pi=+1$ and $\pi=-1$, respectively. Note that we wrote the $\hat{\rho}^{\infty}_\pi$ in the energy basis here. Since $\mathcal{L} + \mathcal{L}'$ conserves $\hat\pi$, the trace in Eq.\ (\ref{eq:autocorrelation_almost_final}) reduces to a trace in one of the $\pi$ subspaces. We thus have to evaluate
\begin{equation}\label{eq:expression_we_have_to_eval}
    \textrm{tr}\bqty{ n e^{(\mathcal{L}_{\pi,\pi} + \mathcal{L}_{\pi,\pi}')\abs{\tau} }\frac{\Bqty{n,  \rho^{\infty} }}{2}}
\end{equation}
with $\rho^{\infty} = (\tau_0 + R\tau_z)/2$. Clearly, the remaining calculation is identical for the two subspaces. Suppressing $\pi$ labels, we evaluate
\begin{align}
    \frac{\Bqty{n, \rho^{\infty} }}{2} =&\ \frac{n^{\infty}}{2} \pqty{ \tau_0 + \frac{R+ \cos\theta}{2n^{\infty}} \tau_z - \frac{\sin\theta}{2n^{\infty}}\tau_x } \nonumber \\
    =&\ n^{\infty}\bqty{ \rho^\infty + \frac{R+ \cos\theta - 2Rn^{\infty}}{4n^{\infty}}\tau_z - \frac{\sin\theta}{4n^{\infty}}\tau_x }.\nonumber
\end{align}
The first term cancels against the $-(n_\infty)^2$ term in $S(\tau)$. We expand the exponential of the  Liouvillian in eigenmodes, 
\begin{equation}
    e^{(\mathcal{L}_{\pi,\pi} + \mathcal{L}_{\pi,\pi}')\abs{\tau} } = \sum_n e^{\lambda_n \abs{\tau}} \ketbra{\psi_n}{\phi_n},
\end{equation}
where $\ket{\psi_n}$ and $\ket{\phi_n}$ are the right and left eigenmodes of $\mathcal{L}_{\pi,\pi} + \mathcal{L}_{\pi,\pi}'$ to eigenvalue $\lambda_n$, respectively (see App. \ref{app:liouvillian_evs_diag}). Note that we can write $\ket{\tau_z} \simeq \ket{\psi_{\textrm{slow}}} $ and $\ket{\tau_x} \simeq -\ket{\psi_{\textrm{fast},+}} + \ket{\psi_{\textrm{fast},-}} $. With this, we can evaluate expression \eqref{eq:expression_we_have_to_eval} to leading order,
\begin{align}
    \bra{n}  e^{(\mathcal{L}_{\pi,\pi} + \mathcal{L}_{\pi,\pi}')\abs{\tau}} \ket{\tau_z} \simeq&\ \cos\theta e^{\lambda_{\textrm{slow}} \abs{\tau}} , \\ 
    \bra{n}  e^{(\mathcal{L}_{\pi,\pi} + \mathcal{L}_{\pi,\pi}')\abs{\tau} } \ket{\tau_x} \simeq&\ -\sin\theta \cos(2\Omega\tau) e^{ \Re{ \lambda_{\textrm{fast}} } \abs{\tau}} .
\end{align}
Here, we expanded $\bra{n} = (\bra{\phi_0} + \cos\theta \bra{\tau_z} - \sin\theta \bra{\tau_x})/2$ and used the overlaps $\braket{\tau_z}{\psi_{\textrm{slow}}} = 2, \braket{\tau_x}{\psi_{\textrm{slow}}} =0, \braket{\tau_z}{\psi_{\textrm{fast}}} \simeq 0$ and $\braket{\tau_x}{\psi_{\textrm{fast},\pm}} \simeq \mp 1 $. This yields the autocorrelation function
\begin{multline}\label{eq:autocorrelation_detailed}
    S(\tau) \simeq \frac{\delta(\tau)}{4k} + \frac{1}{4}\Bigg[ \frac{2k}{\Gamma_- + 2k} \cos^2\theta e^{\lambda_{\textrm{slow}} \abs{\tau}} \\ 
    + \sin^2\theta \cos\pqty{2\Omega\tau} e^{ \Re{ \lambda_{\textrm{fast}} } \abs{\tau}} \Bigg].
\end{multline}

Finally, we compute the noise spectrum 
\begin{align}
    S (\omega) = \int_{-\infty}^{\infty}\textrm{d}\tau e^{i\omega\tau} S (\tau), 
\end{align}
which becomes
\begin{multline}
	S (\omega) \simeq \frac{1}{4k} + \frac{ \cos^2\theta k}{\Gamma_- + 2k} \frac{\abs{\lambda_{\textrm{slow}}}}{\omega^2 + \abs{\lambda_{\textrm{slow}}}^2} \\
	+ \frac{\sin^2\theta }{4} \sum_\pm  
	\frac{ \abs{\Re{\lambda_{\textrm{fast}} } }}{(\omega \pm 2\Omega )^2 + \abs{\Re{\lambda_{\textrm{fast}}}^2}}.
\end{multline}
Thus, the noise spectrum consists of Lorentzians centered at $\omega=0$ due to $\lambda_{\textrm{slow}}$ and at $\pm 2\Omega$ due to $\lambda_{\textrm{fast},\pm}$. For $\abs{t_i} \ll \varepsilon$, we have $\theta \simeq \pi$ and the zero-frequency peak is higher than the finite-frequency peaks by a factor of order $(\varepsilon / \abs{t_i})^4$. Without relaxation, i.e., for $\Gamma_-= \Gamma_0 = 0$ this reduces to the expression \eqref{eq:spectrum_main_text} in the main text. 

\subsection{Fluctuations of time-averaged measurement signal}\label{app:fluctuations_time_avg}

In the presence of relaxation, readout can be based on the time-averaged measurement signal. To estimate readout times it is necessary to obtain the variance of the time-averaged measurement signal, see Sec.\ \ref{sec:with_relaxation}. The time-averaged measurement signal is
\begin{align}
    j_{\textrm{int},\pi}(T) =&\ \frac{1}{T} \int_0^T \textrm{d}t\ j_\pi(t) \nonumber \\ 
    =&\  \frac{1}{T} \int_0^T \textrm{d}t\ \left\{n_\pi(t) + \frac{X_0}{\sqrt{4kT}}\right\},
\end{align}
where $X_0$ is a Gaussian random variable with zero mean and unit variance. Readout relies on
\begin{equation}
    \mathbb{E}[j_{\textrm{int},\pi}(T)] = n^{\infty}_\pi.
\end{equation}
Here, we evaluate the variance of this quantity (suppressing $\pi$ labels),
\begin{align}
\mathbb{V}[j_{\textrm{int},\pi}(T)] =&\ \frac{1}{T^2} \int_0^T \textrm{d}t \int_0^T  \textrm{d}t'\  \mathbb{E}[j(t) j(t')] - \pqty{n^\infty}^2   \nonumber \\
=&\ \frac{2}{T^2} \int_0^T \textrm{d}t \int_0^t \textrm{d}\tau\ S(\tau).
\end{align}
As $\abs{t_i} \ll \varepsilon$, we can neglect the $\sin^2\theta$ term in Eq. \eqref{eq:autocorrelation_detailed}. Using 
\begin{equation*}
    \frac{2}{T^2} \int_0^T \textrm{d}t \int_0^t \textrm{d}\tau\ e^{\lambda \tau} = 2\pqty{ \frac{e^{\lambda T} - 1}{\lambda^2 T^2} - \frac{1}{\lambda T}} \simeq - \frac{2}{\lambda T}
\end{equation*}
for $\lambda T \ll -1$ (checking a posteriori that the measurement times indeed allow for this simplification), we obtain the final result
\begin{equation}
    \mathbb{V}[j_{\textrm{int}}(T)] = \frac{1}{T} \bqty{ \frac{1}{4k} + \frac{k \cos^2\theta }{\Gamma_- + 2k} \frac{1}{\abs{\lambda_{\textrm{slow}}}}} 
\end{equation}
which may be rewritten as
\begin{equation}
    \mathbb{V}[j_{\textrm{int}}(T)] = \frac{1}{T} \bqty{ \frac{1}{4k} + \frac{1}{\tan^2\theta} \frac{4k}{\pqty{\Gamma_- +2k}^2} }, 
\end{equation}
from which we find Eq.\ (\ref{eq:int_curr_variance_main}).

\section{Steady state in the presence of Majorana hybridizations}\label{app:steady_state_w_hybridizations}
Here, we justify the statement in Sec.\ \ref{sec:imperfect} that $\hat{\rho}^{\infty} = \textrm{diag}(1,1,1,1)/4$ is the only zero mode of $\mathcal{L}+ \mathcal{L}_{23}$ (in the absence of relaxation). For small $\varepsilon_{23}$, this  may be obtained as follows. We decompose $\hat{\rho}$ into the steady states $\hat{\rho}^{\infty}_{\pm}$ of $\mathcal{L}$ plus deviations $\delta\hat{\rho}$,
\begin{equation}
    \hat{\rho} = \abs{\alpha}^2 \hat{\rho}^{\infty}_{+} + \abs{\beta}^2 \hat{\rho}^{\infty}_{-} 
        + \delta\hat{\rho}.
\end{equation}
Thus, $\delta\hat{\rho}$ contains only the traceless parts of the diagonal blocks. Note that 
\begin{subequations}\label{eq:hybrid_liou_maps_y}
\begin{align}
    \mathcal{L}_{23} \hat{\rho}^{\infty}_{\pm} =&\ \mp \frac{\varepsilon_{23}}{2}\begin{pmatrix}
        0 & -i \mathds{1}_2 \\ i \mathds{1}_2 &  0 
    \end{pmatrix}
    = \mp \frac{\varepsilon_{23}}{2} \ket{Y}, \label{eq:hybrid_liou_maps_y_1} \\
    \mathcal{L}_{23} \ket{Y} =&\ 4 \varepsilon_{23} ( \hat{\rho}^{\infty}_{+} -  \hat{\rho}^{\infty}_{-} ) . \label{eq:hybrid_liou_maps_y_2} 
\end{align}
\end{subequations}
We also define the projector $\hat{P}$ onto the non-decaying subspace $\textrm{span}(\hat{\rho}^{\infty}_+ , \hat{\rho}^{\infty}_-)$, as well as its complement $\hat{P}_\perp = 1 - \hat{P}$ which projects onto the fast decaying subspace. We then project the eigenvalue equation 
\begin{equation}
    (\mathcal{L}+\mathcal{L}_{23}) \hat{\rho} = \lambda \hat{\rho}
\end{equation}
onto the two subspaces, 
\begin{align}
   & \lambda\pqty{\abs{\alpha}^2 \hat{\rho}^{\infty}_{+} + \abs{\beta}^2 \hat{\rho}^{\infty}_{-} } = \hat{P} \mathcal{L}_{23} \delta \hat{\rho},   \\
   & \lambda \delta \hat{\rho} = \mathcal{L} \delta \hat{\rho} + \mathcal{L}_{23} \pqty{ \abs{\alpha}^2 \hat{\rho}^{\infty}_{+} + \abs{\beta}^2 \hat{\rho}^{\infty}_{-}  }  + \hat{P}_\perp \mathcal{L}_{23} \delta \hat{\rho}.
\end{align}
We formally solve the second equation for $\delta \hat{\rho}$ and insert it into the first equation. Using Eqs.\ \eqref{eq:hybrid_liou_maps_y} and the fact that only $\ket{Y}$ is mapped onto $\textrm{span}(\hat{\rho}^{\infty}_+ , \hat{\rho}^{\infty}_-)$, i.e., $\hat{P} \mathcal{L}_{23}\ ...  = \mathcal{L}_{23} \ketbra{Y}{Y}\ ... /4$ (the factor of $1/4$ stems from the fact that $\braket{Y}{Y} = 4$), we find, after tracing $\tr (\hat{\rho}^\infty_+ ... )$ and using $\abs{\beta}^2 = 1- \abs{\alpha}^2$,
\begin{equation}
    \lambda \abs{\alpha}^2   = \frac{1}{2} \varepsilon^2_{23} \mathcal{G}_{Y} (\lambda) \pqty{ 2\abs{\alpha}^2 - 1 }.
\end{equation}
Here we defined the ``propagator''
\begin{equation}
    \mathcal{G}_{Y} (\lambda) = \bra{Y} \frac{1}{ \lambda - \mathcal{L} - \hat{P}_\perp \mathcal{L}_{23} } \ket{Y}.
\end{equation}
We are interested in $\lambda = 0$. If $\mathcal{G}_{Y} (0) \neq 0$, it follows that a zero mode necessarily has $\abs{\alpha}^2 = 1/2$. Inserting this into the eigenvalue equations, we obtain the relations $\hat{P} \mathcal{L}_{23} \delta \hat{\rho} = 0$ and $(\mathcal{L}+ \hat{P}_\perp \mathcal{L}_{23}) \delta \hat{\rho} = 0$. The first equation gives $\braket{Y}{ \delta \hat{\rho}}=0$, i.e., the steady state has no weight in the span of $\ket{Y}$. Then, the second equation becomes 
\[
(\mathcal{L} +  \mathcal{L}_{23}) \delta \hat{\rho} = 0.
\]
$\mathcal{L}$ has no zero modes acting on the traceless $ \delta \hat{\rho}$. Hence, for weak perturbations $\varepsilon_{23} \ll \vert\Re\{\tilde{\lambda}_{\textrm{slow}}\} \vert , \vert\lambda_{\textrm{slow}, \pi} \vert$, it follows that $\delta \hat{\rho} = 0$. Finally, it is straightforward to check numerically that $\mathcal{G}_{Y} (0)$ is indeed non-vanishing within the relevant parameter range by expanding in left and right eigenvectors of $\mathcal{L} + \hat{P}_\perp \mathcal{L}_{23}$. Thus, the completely mixed state is indeed the only steady state.

\end{document}